\documentclass{emulateapj}
\usepackage{graphicx}
\usepackage{natbib}
\usepackage{epsfig}
\usepackage{subfigure}

\newcommand{\avg}[1]{\ensuremath{\langle #1 \rangle}}
\newcommand{\bma}{\begin{math}}
\newcommand{\ema}{\end{math}}
\newcommand{\beq}{\begin{equation}}
\newcommand{\eeq}{\end{equation}}
\newcommand{\beqa}{\begin{eqnarray}}
\newcommand{\eeqa}{\end{eqnarray}}
\newcommand{\bc}{\begin{center}}
\newcommand{\ec}{\end{center}} 
\newcommand{\bit}{\begin{itemize}}
\newcommand{\eit}{\end{itemize}}

\font\BFd=cmmib10
\font\BFt=cmmib10
\font\BFs=cmmib10 scaled 700
\font\BFss=cmmib10 scaled 500

\def\bbox#1{%
\relax\ifmmode
\mathchoice
{{\hbox{\BFd #1}}}
{{\hbox{\BFt #1}}}
{{\hbox{\BFs #1}}}
{{\hbox{\BFss #1}}}
\else \mbox{#1} \fi }

\def\rvec{{\bbox{r}}}

\def\p{{\bbox{p}}}
\def\pz{{\bbox{p}^0}}

\begin{document}



 
\submitted{\today. To be submitted to \apj.} 

\title{A Measurement of Small Scale Structure in the $2.2 \leq z \leq 4.2$ Lyman-alpha Forest}
\author{Adam Lidz\altaffilmark{1,2}, Claude-Andr\'e Faucher-Gigu\`ere\altaffilmark{2}, Aldo Dall'Aglio\altaffilmark{3},
Matthew McQuinn\altaffilmark{2}, Cora Fechner\altaffilmark{4}, Matias Zaldarriaga\altaffilmark{5,2,6}, 
Lars Hernquist\altaffilmark{2}, Suvendra Dutta\altaffilmark{2}}
\altaffiltext{1}{Department of Physics \& Astronomy, University of Pennsylvania, 209 South 33rd Street, Philadelphia, PA 19104, USA}
\altaffiltext{2}{Harvard-Smithsonian Center for Astrophysics, 60 Garden Street,
Cambridge, MA 02138, USA}
\altaffiltext{3}{Astrophysikalisches Institut Potsdam, An der Sternwarte 16, 14482 Potsdam, Germany}
\altaffiltext{4}{Institut f\"{u}r Physik and Astronomie, Universit\"{a}t Potsdam, Haus 28, Karl-Liebknecht-Str. 24/25. 14476 Potsdam, Germany}
\altaffiltext{5}{School of Natural Sciences, The Institute for Advanced Study, 1 Einstein Drive, Princeton, NJ 08540, USA}
\altaffiltext{6}{Jefferson Laboratory of Physics; Harvard University; Cambridge, MA 02138, USA}
\email{alidz@sas.upenn.edu}

\begin{abstract}
The amplitude of fluctuations in the Lyman-alpha (Ly-$\alpha$) forest on small spatial scales is 
sensitive to the temperature of
the intergalactic medium (IGM) and its spatial fluctuations. 
The temperature
of the IGM and its spatial variations contain important information about hydrogen and helium 
reionization. We present a new measurement of the small-scale 
structure in the Ly-$\alpha$ forest from 40 high
resolution, high signal-to-noise, VLT spectra for absorbing gas at redshifts 
between $2.2 \leq z \leq 4.2$.
We convolve each Ly-$\alpha$ forest spectrum with a suitably chosen Morlet wavelet filter, which 
allows us to extract
the amount of small-scale structure in the forest as a function of position across each spectrum. We 
monitor contamination from metal line absorbers. We present a first comparison of these
measurements with high resolution hydrodynamic simulations of the Ly-$\alpha$ forest which track more
than 2 billion particles. This comparison suggests that the IGM temperature close to the 
cosmic mean density ($T_0$) peaks at a redshift near $z=3.4$, at which point it is greater
than $20,000$ K at $\gtrsim 2-\sigma$ confidence. The temperature at lower redshift is consistent
with the fall-off expected from adiabatic cooling ($T_0 \propto (1+z)^2$), after the peak temperature
is reached near $z=3.4$. In our highest redshift bin,
centered around $z=4.2$, the results favor a temperature of $T_0 = 15-20,000$ K. However, owing 
mostly to uncertainties in the
mean transmitted flux at this redshift, a cooler IGM model with $T_0 =10,000$ K is only disfavored at 
the 2-$\sigma$ level here,
although such cool IGM models are strongly discrepant with the $z \sim 3-3.4$ measurement.  
We do not detect large spatial fluctuations in the IGM temperature at any redshift covered
by our data set.  The simplest interpretation of our measurements is that HeII reionization completes sometime near $z \sim 3.4$,
although statistical uncertainties are still large.  
Our method can be fruitfully combined with future HeII Ly-$\alpha$ forest
measurements.
 \end{abstract}


\section{Introduction} \label{sec:intro}

A key characteristic in our description of the baryonic matter in the Universe is the
thermal state of the gas in the intergalactic medium (IGM). As such, detailed constraints
on the temperature of the gas in the IGM, its spatial variation, density dependence, and
redshift evolution, are of fundamental importance to observational cosmology. During
the Epoch of Reionization (EoR), essentially the entire volume of the IGM becomes filled
with hot ionized gas. The thermal state of the IGM
subsequently retains some memory of when and how the intergalactic gas
was ionized (Miralda-Escude \& Rees 1994, Hui \& Gnedin 1997), owing to the long cooling times
for this low density gas. Measurements of the thermal history of the IGM hence translate into valuable 
constraints
on the reionization history of the Universe (e.g. Theuns et al. 2002a, Hui \& Haiman 2003). 

Current observations suggest
that there may in fact be two separate EoRs: an early Epoch of Hydrogen Reionization during which
hydrogen is ionized, and helium is singly-ionized, by star-forming galaxies, 
followed by a later Epoch of Helium Reionization during which helium is doubly ionized 
by bright quasars (e.g. Madau et al. 1999). Recent measurements of the quasar luminosity function
(Hopkins et al. 2007), combined with estimates of the quasar spectral shape and the clumpiness of the IGM,
suggest that HeII reionization may complete somewhere near $z \sim 3$ 
(Furlanetto \& Oh 2008, Faucher-Gigu\`ere et al. 2008a, McQuinn et al. 2008). 
Indeed, there are some observational indications
that helium is doubly-ionized close to $z \sim 3$ (see e.g., Schaye et al. 2000, Furlanetto \& Oh 2008, Faucher-Gigu\`ere et al. 2008a, 
McQuinn et al. 2008 for a discussion), although the evidence is generally weak and controversial.

Further detailed studies of the HI Ly-$\alpha$ forest near $z \sim 3$ offer promise to pin-point when HeII reionization occurs and can
potentially constrain properties of HeII reionization, such as the filling factor and size distribution
of HeIII regions at different stages of reionization.
Photoheating during HeII reionization impacts the thermal state of the IGM 
(e.g., Miralda-Escude \& Rees 1994, 
Abel \& Haehnelt 1999, McQuinn et al. 2008, Bolton et al. 2009), and in turn influences the statistics of the HI
Ly-$\alpha$ forest. In the midst of HeII reionization, the temperature of the IGM should be inhomogeneous (e.g. McQuinn et al. 2008): 
there are hot regions where HeII recently reionized, and cooler regions where helium is only singly-ionized. Additionally, regions reionized
by nearby sources will typically be cooler than regions reionized by far away sources. Regions reionized by distant sources receive a heavily
filtered and hardened spectrum, and experience more photoheating than gas elements that are close to an ionizing source. The average temperature,
as well as the amplitude of temperature fluctuations and the scale dependence of these fluctuations, are hence closely related to the filling
factor and size distribution of HeIII regions during reionization. Detailed studies of the HI Ly-$\alpha$ forest may
allow us to detect these temperature inhomogeneities, and thereby constrain details of HeII reionization with existing data. 
In principle, additional processes including heating by large scale structure shocks, heating from 
galactic winds, cosmic-ray heating, 
Compton-heating from the hard X-ray background, photo-electric heating from dust grains, or even heat injection
from annihilating or decaying dark matter, may also impact the
temperature of the IGM (see e.g. Hui \& Haiman 2003 for references and a discussion). Sufficiently 
detailed constraints should help determine the relative importance of photo-heating
and these additional effects.

The aim of the present paper is to make a new measurement of small-scale structure in the Ly-$\alpha$ forest,
which can be used to constrain the thermal history of the IGM, and to search for signatures of HeII reionization
in the HI Ly-$\alpha$ forest.
There have been several previous measurements
of the thermal history 
from the Ly-$\alpha$ forest (Schaye et al. 2000, Ricotti et al. 2000, McDonald et al. 2001, 
Zaldarriaga et al. 2001,
Theuns et al. 2002b, Zaldarriaga 2002). However, the agreement between these studies is somewhat marginal, and
the different authors reach
differing conclusions regarding the thermal history of the IGM. 
Note that it has been almost
a decade since many of these measurements were made. In the meantime, better Ly-$\alpha$ forest data sets have become available,
and we now have better numerical simulations to help interpret and calibrate the observational measurements. It is hence timely to revisit
these issues. 

Of particular interest from the theoretical side is the work of McQuinn et al. (2008), who 
performed the
first detailed, three-dimensional radiative transfer simulations of HeII reionization which self-consistently
track the thermal state of the IGM during HeII reionization (see also Paschos et al. 2007). Recent analytic (Furlanetto \& Oh 2008) and
one-dimensional radiative transfer calculations (Tittley \& Meiksin 2007, Bolton et al. 2009) are also refining
our understanding of HeII reionization. In this paper we use improved observational data, along with a 
somewhat
refined methodology, to make a new measurement of small-scale structure in the Ly-$\alpha$ forest. We also
make a first comparison of the results with high resolution hydrodynamic simulations of the forest, in 
order to
explore broad implications of our measurements for the thermal history of the IGM. In future work,
we will use HeII reionization simulations to obtain more detailed constraints. 

The small-scale
power in the Ly-$\alpha$ forest is very sensitive to the temperature of the IGM (e.g. Zaldarriaga et al. 2001): a hotter IGM leads to more Doppler
broadening, and Jeans-smoothing, which in turn leads to less small-scale structure in the Ly-$\alpha$ forest. The amplitude of the transmission
power spectrum on small-scales hence provides an IGM thermometer. In addition to the {\em average temperature}, we aim to measure or constrain
{\em temperature inhomogeneities}, i.e., we would like to be sensitive to variations in the small-scale power across each quasar spectrum.
In order to accomplish this, we convolve each spectrum with a filter 
that is localized in both Fourier space and configuration space, i.e., a `wavelet' filter. For a 
suitable choice of 
smoothing scale, 
this provides a measurement of the IGM temperature as a function of position across each quasar spectrum. 
Although our basic method 
closely resembles that of Theuns \& Zaroubi (2000) and Zaldarriaga (2002), there
are some differences in the details of our implementation. For instance, we employ a different filter than these authors.

The outline of this paper is as follows. In \S \ref{sec:method} we detail our methodology for constraining
the thermal history of the IGM. In \S \ref{sec:data_analysis}, we describe the data set used in our analysis,
and present measurements. \S \ref{sec:theor} focuses on the theoretical interpretation of the measurements. Here
we describe cosmological simulations which we compare with the observations, present preliminary constraints on the
thermal history of the IGM, comment on the implications for our understanding of the reionization history of the Universe,
and compare with previous measurements. 
\S \ref{sec:cross_heii} discusses cross-correlating temperature measurements from the HI Ly-$\alpha$ forest
with HeII Ly-$\alpha$ forest spectra.
In \S \ref{sec:conclusion} we conclude, mentioning plans and possibilities for related future work. 
Several appendicies explore shot-noise bias, metal line contamination, and the convergence of our numerical
simulations. 

\section{Methodology} \label{sec:method}

In this section, we present our method for constraining the temperature of the IGM, and
illustrate its utility with cosmological simulations. First, we introduce some notation and 
briefly mention a few
relevant facts regarding the thermal history of the IGM and the Ly-$\alpha$ forest.

\subsection{The thermal history of the IGM and the Ly-$\alpha$ forest}
\label{sec:review}

After a low-density gas element is photo-heated during reionization, it will subsequently cool
and gas elements with similar photo-heating histories generally land on 
a `temperature-density relation' (Hui \& Gnedin 1997):
\beqa
T(\rvec) = T_0(\rvec) \left[1 + \delta_\rho(\rvec)\right]^{\gamma(\rvec)-1}.
\label{eq:tdelta}
\eeqa
Here $\delta_\rho(\rvec)$ denotes the fractional gas over-density (implicitly smoothed on the Jeans scale)
at spatial position $\rvec$. $T_0$ is the temperature of a gas element at the cosmic mean density, and the power-law index $\gamma$
approximates the density-dependence of the temperature field. The temperature that a gas element reaches
at say, $z=3$, depends on the temperature that it reaches during reionization, and on its subsequent cooling and heating. 
The temperature attained by each gas element during reionization depends mostly on the shape of the 
spectrum of
the sources that ionize it. The relevant spectrum is generally modified from the intrinsic spectral 
shape of an ionizing source, owing to intervening material between a source and the gas element in 
question, which
tends to harden the ionizing spectrum. After a gas element is photoheated during reionization, adiabatic cooling owing to the
expansion of the Universe is the dominant cooling mechanism (for the bulk of the low density gas that makes up the Ly-$\alpha$
forest).\footnote{Compton cooling off of the CMB is efficient only at higher redshifts than
considered here. Specifically, the Compton cooling
time for gas at the cosmic mean density is equal to the age of the Universe at $z = 6$. Gas reionized sufficiently before
this redshift will lose memory of its initial temperature -- i.e., its temperature 
at reionization -- by $z \leq 6$ (Hui \& Gnedin 1997).}
When a gas element is significantly ionized during reionization it reaches photoionization equilibrium and 
receives only a small amount of additional photoheating as low levels of residual neutral material are 
ionized.
During reionization, gas elements gain heat as hydrogen is ionized, as helium is singly ionized, and when helium is doubly ionized. 
If helium is doubly-ionized significantly after hydrogen is ionized, two separate `reionization events' may be important in determining
the thermal history of the IGM. As both hydrogen and helium reionization are extended, inhomogeneous processes, $T_0$
and $\gamma$ may be strong functions of spatial position following reionization events. 
However, once a sufficiently long time passes
after reionization, gas elements
reach a `thermal asymptote' and lose memory of the initial photoheating during reionization (Hui \& Gnedin 1997). At this point the
inhomogeneities in $T_0$ and $\gamma$ should again be small. 

In the absence of HeII photoheating, one expects the temperature of the IGM at $z \sim 3$ to 
be $T_0 \lesssim 10,000$ K, with the precise temperature depending on the timing of hydrogen reionization and the nature of the 
ionizing sources (Hui \& Haiman 2003). Sufficiently long after a reionization event, the slope of the temperature density relation,
$\gamma$, tends to $\gamma \sim 1.6$, owing to the competition between adiabatic cooling and residual photoionization heating
(Hui \& Gnedin 1997). HeII reionization likely raises the temperature of the IGM by roughly $10,000$ K, with the precise increase
depending on the spectrum of the ionizing sources and other factors. HeII photoheating and the spread in timing of HeII reionization 
flatten the temperature-density relation to $\gamma \sim 1.3$ (McQuinn et al. 2008).

The temperature of the IGM has three separate effects on Ly-$\alpha$ forest spectra. First, increasing the temperature of the 
absorbing gas increases the amount of Doppler broadening: thermal motions spread the absorption of a gas element 
out over a length (in velocity units) of $b = \sqrt{2 k T/m_p} \sim 13$ km/s for $T=10^4$ K gas. 
Second, the gas pressure and Jeans smoothing scale increase with increasing temperature. Since 
it takes some time for the gas to move around and the 
gas pressure to adjust to prior heating, this effect is sensitive not to the 
{\em instantaneous temperature}, but to {\em prior heating} (Gnedin \& Hui 1998). This effect is more challenging for simulators to 
capture,
because properly accounting for it requires re-running entire simulations after adjusting the simulated ionization/reheating 
history. The
Jeans smoothing effect 
is not completely degenerate, however,
with the Doppler broadening one because Jeans-smoothing smooths {\em the gas distribution in three dimensions}, while Doppler broadening
smooths {\em the optical depth in one dimension} (Zaldarriaga et al. 2001). Finally, the recombination coefficient is temperature 
dependent,
scaling as $T^{-0.7}$: hotter gas recombines more slowly, and reaches a lower neutral fraction than
cooler gas. 

The first two of these effects mostly impact the amplitude of small-scale fluctuations in the Ly-$\alpha$ forest (e.g.
Zaldarriaga et al. 2001). For the range
of models we are interested in presently, the first effect (Doppler broadening) should be the dominant influence on the 
small-scale power. 
At a given redshift, the small-scale structure in the Ly-$\alpha$ forest is most sensitive to the temperature of absorbing gas at 
some characteristic density,
with less dense gas giving very little absorption and more dense gas giving rise to mostly saturated absorption. At $z \sim 3$ the
forest is sensitive mostly to the temperature of gas a little more dense than 
the cosmic mean (McDonald et al. 2001, Zaldarriaga et al. 2001). At higher redshifts, the absorption
is sensitive to the temperature of somewhat less dense gas, while at lower redshifts the absorption depends 
on more dense gas (Dav\'e et al. 1999).

\subsection{Data Filtering and Constraining the Temperature}
\label{sec:filter}

Next we describe our method for constraining $T(\rvec)$ (Equation \ref{eq:tdelta}) from absorption spectra. 
Following earlier work (Theuns \& Zaroubi 2000, Zaldarriaga 2002, Theuns et al. 2002b), we convolve Ly-$\alpha$ transmission
spectra with a filter that pulls out high-$k$ modes across each spectrum. As mentioned above, Doppler broadening convolves
the optical depth field with a Gaussian filter with a -- temperature-dependent -- width of tens of km/s. 
We hence desire
a filter that extracts Fourier modes with wavelengths of tens of km/s across each spectrum.

We have found that a very simple choice of filter accomplishes this task. In configuration space, the filter we use may be written as
\beqa
\Psi_n(x) = A \rm{exp}(i k_0 x) \rm{exp}\left[-\frac{x^2}{2 s_n^2}\right].
\label{eq:filt_real}
\eeqa
We fix the normalization, $A$, by requiring the filter to have unit power -- i.e., after
filtering a white-noise field with noise power spectrum $P_N(k) = \Delta u \sigma^2$, the
filtered field has variance $\sigma^2$. ($\Delta u$ denotes the size of a spectral pixel in velocity 
units.)\footnote{The variance is $\sigma^2 = \int_{-\infty}^{\infty} dk/(2 \pi) |\Psi_n(k)|^2 P_N(k)$ for our Fourier
convention.} With 
this normalization, the filter's Fourier transform in $k$-space is
\beqa
\Psi_n(k) = \pi^{-1/4} \sqrt{\frac{2 \pi s_n}{\Delta u}} \rm{exp}\left[-\frac{(k - k_0)^2 s_n^2}{2}\right].
\label{eq:filt_four}
\eeqa

\begin{figure}
\bc
\includegraphics[width=9.2cm]{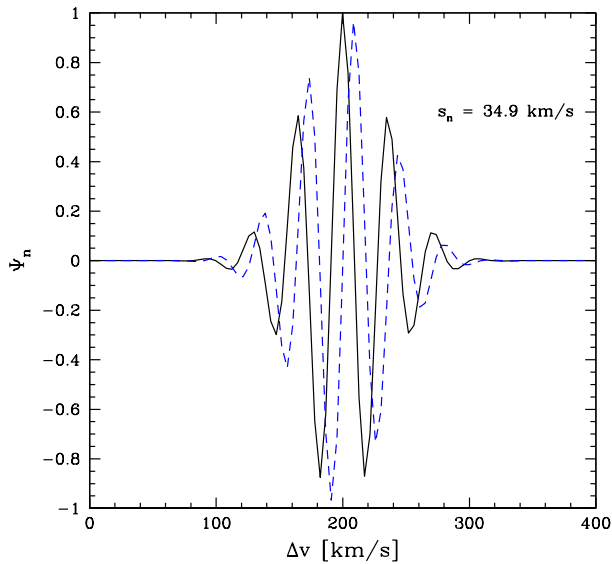}
\caption{The Morlet Wavelet filter in configuration space. The black solid line is the real part of the filter, while the blue dashed line 
is the imaginary part. The filter shown adopts one of the two choices of smoothing scale considered in this work, 
$s_n = 34.9$ km/s. The filter for alternate choices of smoothing scale 
are simply compressed or expanded versions of this fiducial filter. The center of the horizontal scale is arbitrary.}
\label{fig:examp_wavelet}
\ec
\end{figure}

In configuration space this filter is simply a plane-wave, damped by a Gaussian. In Fourier space, the filter is a Gaussian
centered around $k = k_0$. We would like the filter to have zero mean. Throughout this work we choose $k_0 s_n =6$,
in which case Equation \ref{eq:filt_four} shows that the zero mode of the filter $\Psi_n(k=0)$ is extremely close to zero, satisfying closely the
zero mean requirement. This filter clearly has the properties of being localized in both configuration space and 
Fourier space.
These are among the defining properties of a `wavelet filter', and the filter of Equations (\ref{eq:filt_real}) and 
(\ref{eq:filt_four}) is known as a `Morlet Wavelet' in the wavelet 
literature.\footnote{http://en.wikipedia.org/wiki/Wavelets and references therein.}
We plot its form in Figure \ref{fig:examp_wavelet} for $s_n = 34.9$ km/s, which, as we discuss further below, turns out to be one convenient
choice. Note that the filters $\Psi_n$ (Equation \ref{eq:filt_real}) do not form an orthogonal set, but this is unnecessary for 
our present purposes. We do not expand the entire spectrum in terms of a wavelet basis in this work -- the 
Morlet wavelet, with locality in real
and configuration space, is simply a convenient filter. 

We then convolve each observed (or simulated) spectrum with the above filter. 
In this paper, we consider throughout the fractional Ly-$\alpha$ transmission field, $\delta_F = (F - \avg{F})/\avg{F}$. Here $F = e^{-\tau}$
is the Ly-$\alpha$ transmission, and $\avg{F}$ is the global average Ly-$\alpha$ transmission. 
We label the flux field, $\delta_F$, convolved with the filter $\Psi_n$ as $a_n$:
\beqa
a_n(x) = \int dx^\prime \Psi_n(x - x^\prime) \delta_F(x^\prime),
\label{eq:field_filt}
\eeqa
and compute the convolution using Fast Fourier Transforms (FFTs). 
Note that $a_n(x)$ is a complex number for our choice of filter, $\Psi_n(x)$. A measure of
small-scale power is then 
\beqa
A(x) = |a_n(x)|^2,
\label{eq:waveamp}
\eeqa
which for brevity of notation we sometimes refer to as `the wavelet-filtered field' or as `the wavelet amplitudes' (even though
it is proportional to the transmission field squared). It is also useful to note that the average wavelet amplitude is just
\beqa
\avg{|a_n(x)|^2} = \int_{-\infty}^{\infty} \frac{dk^\prime}{2 \pi} \left[\Psi_n(k^\prime)\right]^2 P_F(k^\prime),
\label{eq:meanamp}
\eeqa
with $P_F(k)$ denoting the power spectrum of $\delta_F$. Hence, the 
mean wavelet amplitude is nothing more than the 
usual flux power spectrum for some convenient `band' of wavenumbers (see Figure \ref{fig:pauto_mean_wave} for further illustration). Additional statistics of $A(x)$, beyond the
mean, characterize the spatial variations in the small-scale transmission power. 

We frequently find it convenient to smooth $A(x)$ using a top-hat filter of smoothing length $L$:
\beqa
A_L(x) = \frac{1}{L} \int_{-\infty}^{\infty} dx^\prime \Theta(|x - x^\prime|; L/2) A(x^\prime).
\label{eq:smoothed_waveamp}
\eeqa
Here $\Theta(|x - x^\prime|; L/2) = 1$ for $|x - x^\prime| \leq L/2$ and is zero otherwise. 
Smoothing the wavelet filtered field is desirable since the small-scale power is not a perfect indicator of the local
temperature, and smoothing reduces the noisy excursions that the wavelet amplitudes can take. 
Since the hot regions are expected to
be rather large during HeII reionization (McQuinn et al. 2008), we can smooth considerably without diluting any temperature inhomogeneities.
We generally adopt $L = 1,000$ km/s, corresponding to roughly $\sim 10$ co-moving Mpc/$h$ at $z=3$. We discuss this choice further below.

\begin{figure}
\bc
\includegraphics[width=9.2cm]{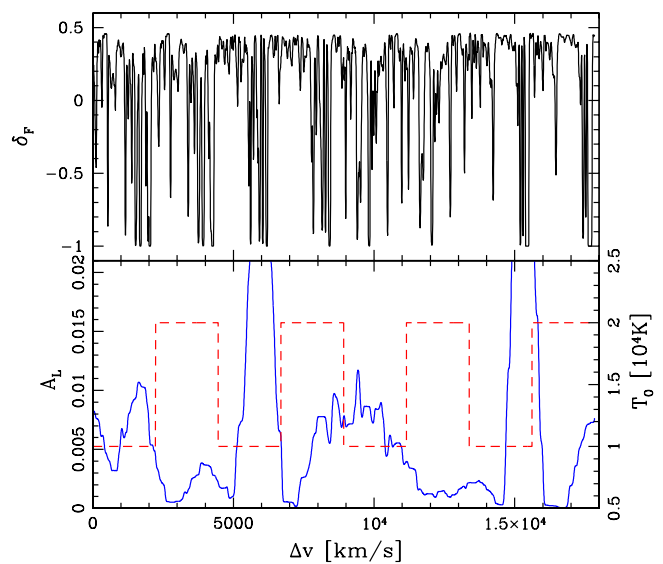}
\caption{Illustration of our filtering method. {\em Top panel}: A simulated spectrum, with some
portions of the spectrum drawn from a simulated `hot' model with $T_0 = 2 \times 10^4$ K and
$\gamma=1.3$, and other regions drawn from a `cold' model with $T_0 = 1 \times 10^4$ K and $\gamma=1.3$.
The hot and cold regions are alternating and are each of length $20$ co-moving Mpc/$h$ ($2230$ km/s). 
{\em Bottom panel}: The red dashed lines and the tick marks on the right hand side of the panel indicate the
temperature of the corresponding regions in the upper panel. The solid blue line shows the wavelet-amplitudes
(for $s_n=34.9$ km/s), top-hat
filtered with a $L = 1,000$ km/s filter. The smoothed wavelet amplitudes are a good tracer of the 
temperature of each region.  
}
\label{fig:wavelet_illust}.
\ec
\end{figure}

Since thermal broadening smooths the optical depth field on tens of km/s scales, $A_L(x)$ should be a good tracer
of the temperature for suitable choices of $s_n$. In order to illustrate this concretely, we apply the filter to a simulated spectrum 
from a simple
toy inhomogeneous temperature model, following a similar example from Theuns \& Zaroubi (2000). Specifically, we splice together simulated
lines of sight (see \S \ref{sec:sims}) with alternating portions of spectrum drawn from each of a `hot' temperature model with 
$T_0 = 2 \times 10^4$ K, and $\gamma=1.3$, and a `cold' temperature model with $T_0 = 1 \times 10^4$ K, and $\gamma=1.3$. We refer
the reader to \S \ref{sec:sims} and \S \ref{sec:box_res} for details regarding the simulated spectra.
If the wavelet filtered field
provides a good indicator of the temperature, regions with hot temperatures should tend to produce low wavelet 
amplitudes, while the cold
regions should produce high wavelet amplitudes. The results of this test are shown in Figure \ref{fig:wavelet_illust}, for smoothing
scales of $s_n = 34.9$ km/s and $L=1,000$ km/s. Cold regions tend to contain several narrow lines, and produce a large response
after filtering: the regions near $\Delta v = 6,000$ km/s and $15,000$ km/s have $A_L \gtrsim 0.02$. The hot regions typically have
$A_L \lesssim 0.005$ and never reach the large amplitudes found in the cold regions. There is some 
variance in the wavelet amplitude
from region to region -- for example, $A_L$ is not as large in the cold region 
near $\Delta v =10,000$ km/s as it is at $\Delta v = 6,000$ km/s and $15,000$ km/s. 
Nonetheless, the smoothed wavelet amplitude is a fairly good tracer of the underlying temperature field.

\begin{figure}
\bc
\includegraphics[width=9.2cm]{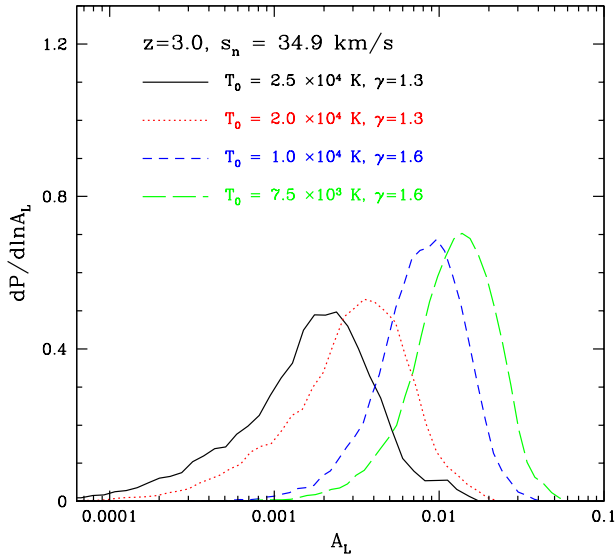}
\caption{PDF of the wavelet amplitudes for different models at $z=3$ and $s_n=34.9$ km/s. The curves show 
simulated models
for the PDF of the wavelet amplitudes, top-hat smoothed over $L = 1,000$ km/s, for several
temperature-density relations. The mean transmitted flux is fixed in this comparison. 
The black solid and red-dashed curves correspond very roughly
to temperature-density relations expected just after HeII reionization. The blue
short-dashed and green long-dashed curves, on the other hand, loosely correspond to the temperature-density
relation expected when HI and HeII are both reionized much before $z=3$.  
}
\label{fig:pdf_models_sn3}
\ec
\end{figure}

\begin{figure}
\bc
\includegraphics[width=9.2cm]{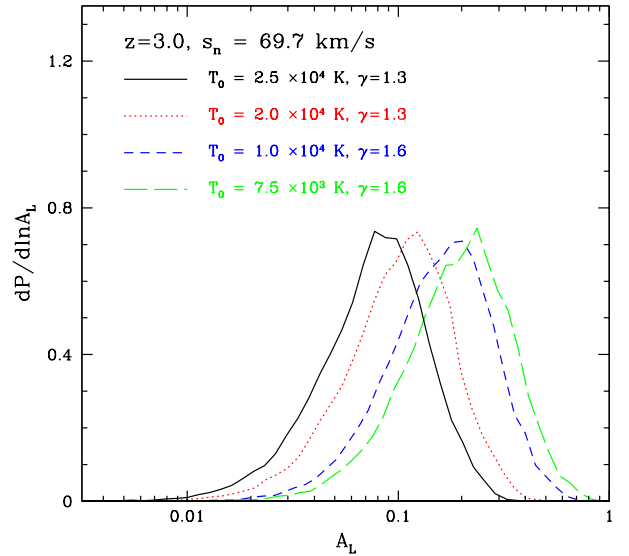}
\caption{PDF of the wavelet amplitudes for different models at $z=3$ and $s_n=69.7$ km/s. Similar to Figure \ref{fig:pdf_models_sn3}, except for $s_n=69.7$ km/s. 
}
\label{fig:pdf_models_sn4}
\ec
\end{figure}

In order to quantify this further, we calculate the probability distribution function (PDF) of the smoothed wavelet amplitudes. We do this for the two choices of small-scale smoothing adopted in this paper (see \S \ref{sec:smoothing}): $s_n=34.9$ km/s, and twice this, $s_n=69.7$ km/s.
The PDF of smoothed wavelet amplitudes will be the main statistic we consider in the present paper.  
For now, we examine models with homogeneous temperature-density relations. The models we select
for the temperature-density relation
loosely correspond respectively to what one expects right after HeII reionization ($T_0 \sim 20-25,000$ K and $\gamma=1.3$) (McQuinn et al. 2008),
and to what one expects if HI, HeI, and HeII are all ionized much 
before $z \sim 3$ ($T_0 \sim 7,500-10,000$ K and $\gamma=1.6$) (Hui \& Haiman 2003). 
The latter, cooler model, might be expected if, for example, the IGM is reionized by abundant faint quasars which have sufficiently
hard spectra to doubly ionize helium at the same time they reionize hydrogen, or if high redshift galaxies have a 
surprisingly hard spectrum and can doubly ionize helium themselves. Note that the precise $z \sim 3$ temperature in the early reionization models
is determined by residual photoheating and depends on the reprocessed spectra of the post-reionization ionizing sources (Hui \& Haiman 2003).

The PDFs in these models are shown for two choices of small-scale smoothing in Figure \ref{fig:pdf_models_sn3} 
($s_n=34.9$ km/s), and Figure \ref{fig:pdf_models_sn4} ($s_n=69.7$ km/s). A larger range of models
will be examined in \S \ref{sec:theor}. 
Considering first the smaller smoothing scale (Figure \ref{fig:pdf_models_sn3}), one
sees that the peak of the PDF in the $T_0 = 20,000$ K, $\gamma=1.3$ model is reached at a smoothed
wavelet amplitude that is roughly a factor of $2$ smaller than the peak location in 
the $T_0 = 10,0000$ K, $\gamma=1.6$ model. The PDFs in the 
hotter $T_0 \sim 25,000$ K model and the colder $T_0 = 7,500$ K model differ by even more. In the 
midst of HeII reionization,
one expects an inhomogeneous temperature field and the true temperature-density relation may be a mix
of the models shown here. 
At any rate, the wavelet PDFs differ significantly in the models with $20,000$ K and those with cooler
temperatures. This further demonstrates -- beyond the visual inspection 
of Figure \ref{fig:wavelet_illust} -- 
that the wavelet PDF is a useful statistic for constraining the thermal history and HeII reionization.
The typical wavelet amplitude in each model is significantly larger 
at $s_n=69.7$ km/s (Figure \ref{fig:pdf_models_sn4}), a consequence of the roughly exponential fall-off
in flux power towards high $k$ (Zaldarriaga et al. 2001). The PDFs still vary significantly with 
temperature-density relation at this larger smoothing scale, although the sensitivity is a 
little bit reduced.

\subsection{Smoothing Scales}
\label{sec:smoothing}

\begin{figure}
\bc
\includegraphics[width=9.2cm]{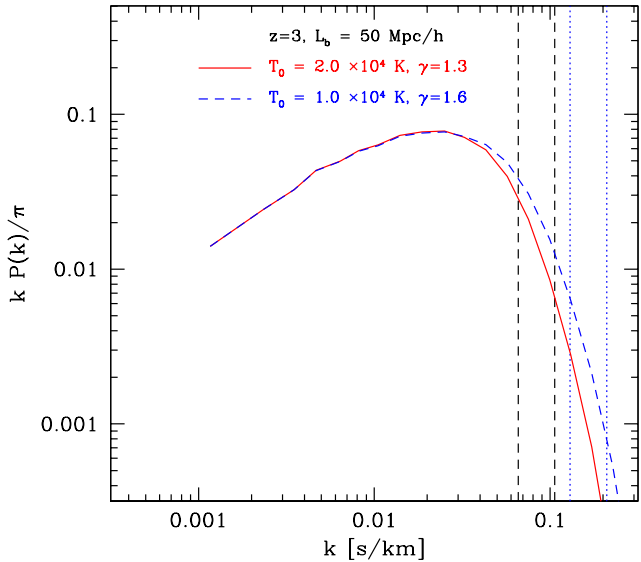}
\caption{Relation between the mean wavelet amplitude and flux power spectrum. The red solid and blue 
dashed lines show
the usual flux power spectrum for simulated models with two different temperature-density 
relations at $z=3$, with the mean transmitted flux fixed at $\avg{F}=0.680$ for each model. The black dashed
vertical lines indicate the range of scales ($\pm 1-\sigma$) extracted by the $s_n =69.7$ km/s 
wavelet filter,
while the blue dotted vertical lines indicate the same for the $s_n = 34.9$ km/s filter.  
}
\label{fig:pauto_mean_wave}
\ec
\end{figure}

Before we move on to analyze observational data, let us consider further 
the two smoothing scales, $s_n$ and $L$, in
our calculations.  We make measurements for two choices of small-scale 
smoothing: $s_n = 34.9$ km/s and $s_n = 69.7$ km/s.\footnote{The precise values are chosen because it is 
convenient for the
smoothing scale to be related to the pixelization of our data $\Delta u$ (see \S \ref{sec:data_analysis}) by $s_n = 2^n \Delta u$ for 
some choice of $n$.} 
For the former choice of smoothing scale $|\Psi_n(k)|^2$ is proportional to a Gaussian centered on $k_0 = 6/s_n = 0.17$ s/km, with width
$\sigma_k = \sqrt{2}/s_n = 0.04$ s/km. The latter choice of smoothing scale centers the 
Gaussian on $k_0 = 6/s_n = 0.086$ s/km, with
a width of $\sigma_k = \sqrt{2}/s_n = 0.02$ s/km. 
The range of scales probed by these filters is shown in comparison to simulated flux power spectra in
Figure \ref{fig:pauto_mean_wave}.
As illustrated in Figures \ref{fig:pdf_models_sn3},
\ref{fig:pdf_models_sn4}, and \ref{fig:pauto_mean_wave}, the wavelet PDFs are slightly less sensitive to 
the IGM temperature for the larger
smoothing scale filter. On the other hand, the results at the larger smoothing scale are less sensitive to metal
line contamination and other systematics. Increasing the smoothing by still another factor of two would
almost completely remove the sensitivity to temperature (see Figure \ref{fig:pauto_mean_wave}). 
Decreasing $s_n$ by an additional factor of two (to 
$s_n=17.4$ km/s) increases the fractional difference between model curves, but brings one very far out on 
the exponential
tail of the power spectrum (Figure \ref{fig:pauto_mean_wave}) and makes the results very sensitive
to metal line contamination, detector noise, and pixelization effects. The two choices of filtering scale
used here represent a compromise between discriminating power and systematic
effects. Considering both choices of filtering scale gives a consistency check on the results and helps to
protect against systematic effects.

Let us now consider the large scale smoothing, $L$. Naively, one would want to tune this filtering to precisely the scale on which 
the temperature field is inhomogeneous. Since the power spectrum of temperature fluctuations during HeII reionization has
a relatively well defined peak (McQuinn et al. 2008), one might expect the variance of the wavelet amplitudes to also 
show
a clear maximum at some characteristic smoothing scale. 
However, in practice we find that this is washed out
in Ly-$\alpha$ forest spectra, which as one dimensional skewers suffer owing to 
aliasing from high-$k$ modes transverse to the line of sight (Kaiser \& Peacock 1991). To illustrate this, consider 
the two-point function of the
wavelet amplitudes (squared),
\beqa
\xi_A(|v_1 - v_2|) = \frac{\avg{A(v_1) A(v_2)} - \avg{A}^2}{\avg{A}^2},
\label{eq:atwop_dem}
\eeqa 
and its Fourier transform, the power spectrum of wavelet-amplitudes squared, $P_A(k)$. Here $v_1$ and $v_2$ are two 
points along a quasar spectrum and
$\avg{A}$ is the globally averaged wavelet amplitude squared, and we have normalized this two-point function
by the (square of the) mean wavelet amplitude squared. The power spectrum of wavelet amplitude squared 
fluctuations encodes how much the small-scale power spectrum
fluctuates across a quasar spectrum as a function of scale.  It involves a product of four values of 
$\delta_F$ and is hence a four-point function.

\begin{figure}
\bc
\includegraphics[width=9.2cm]{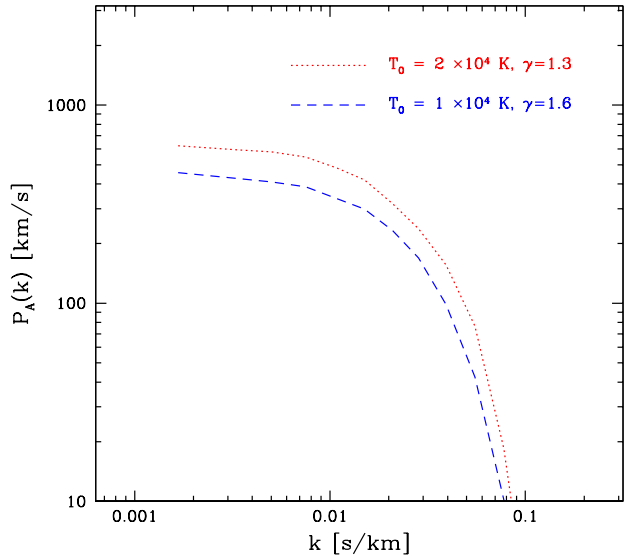}
\caption{Power spectra of the squared wavelet amplitudes. The curves show power spectra for 
two different (homogeneous temperature-density relation) models. Aside from the small-scale turn-over, which
owes to the smoothing (on scale $s_n=34.9$ km/s) from the wavelet filter, the model curves are quite flat as a function of wavenumber. 
}
\label{fig:power_ampsq}
\ec
\end{figure}

We show two simulated examples of $P_A(k)$ in Figure \ref{fig:power_ampsq} for $s_n=34.9$ km/s. One can see 
that, except for the small-scale cut-off,
the power spectra are quite flat as a function of scale.  This is somewhat unfortunate, as one would naively
hope that the scale dependence of $P_A(k)$ would directly reveal {\em the scale dependence of temperature fluctuations}, but the
flatness we find is
a direct consequence of aliasing. 
We have experimented with various inhomogeneous temperature models, including
simulated models from McQuinn et al. (2008) and find similarly flat power spectra.
One might be able to get around this by using quasar pairs to measure the power spectrum of
wavelet amplitude squared transverse to the line of sight. We defer, however, investigating this to future work. For the moment, our main
conclusion is that, owing to the flatness of $P_A(k)$, the precise smoothing scale $L$ is relatively unimportant. Hence we 
generally stick to $L=1,000$ km/s as a convenient choice. We nevertheless investigate the dependence on 
large scale smoothing from observational and simulated data in \S \ref{sec:smooth_large}.

To summarize, by applying a very simple filter to a quasar spectrum, we can measure the small-scale power spectrum of
transmission fluctuations as a function of position across each spectrum, and thereby constrain the temperature of 
the IGM.
Note that our procedure does not involve identifying absorption lines and fitting profiles to identified lines, (although
we find in \S \ref{sec:data_analysis} that it is important to identify metal absorbers in the forest which does 
involve line-fitting). It is instead within the spirit of treating the forest as a one dimensional random field and
measuring the statistics of this continuous field (e.g. Croft et al. 1998). This is more appropriate given the modern
understanding that the forest arises from fluctuations in the line of sight density field, rather than discrete absorbing
clouds (e.g. Hernquist et al. 1996, Miralda-Escud\'e et al. 1996, Katz et al. 1996).
In this way our approach is very similar to 
Theuns \& Zaroubi (2000) and Zaldarriaga (2002), and somewhat resembles Zaldarriaga et al. (2001), but is rather different
than Schaye et al. (2000), Ricotti et al. (2000) and McDonald et al. (2001). 

Additionally, recall that the widths of {\em most} of the 
absorption lines in the Ly-$\alpha$ forest
are dominated by the Hubble expansion across an absorber, and not by thermal broadening (Hernquist et al. 1996,
Weinberg et al. 1998).
In order
to determine the temperature with a line fitting method, one typically looks for a low-end cut-off in the distribution
of line widths (e.g. Schaye et al. 2000). One might worry that this throws out information as thermal broadening smooths 
the spectrum {\em everywhere}.
In practice, though, it appears that most of the signal and information in our method also arises from deep narrow lines which produce a large response after wavelet filtering.
Another possible issue is that the precise interpretation of the line 
width cut-off in the line-fitting studies is unclear 
when the temperature
field is inhomogeneous. It would certainly be interesting to compare more closely the different 
methods, but we defer this to 
future work. For now, note that our method is very simple to apply.

\section{Data Analysis} \label{sec:data_analysis}

We now move on to apply the method to observational data. The main result
will be a measurement of the PDF of the smoothed wavelet amplitudes at $z \sim 2.2-4.2$. 
Our data set consists of $40$ quasar spectra observed with UVES on the VLT, described and reduced
as in Dall'Aglio et al. (2008). We have identified metal lines in the Ly-$\alpha$ forest for
$11$ of these spectra, as described in \S \ref{sec:metals}.
The spectra have high $S/N$ ranging from 
$S/N \sim 30-130$ (quoted at the
continuum level per $0.05 \AA$ pixel), and high spectral resolution, $\rm{FWHM} \sim 6$ km/s. High spectral resolution
and $S/N$ are essential to reliably probe high-$k$ modes in the spectra and to estimate the temperature 
of absorbing
gas. A detailed list of the quasar spectra, with redshift estimates and other properties, can be
found in Dall'Aglio et al. (2008).

\subsection{Raw Measurements} \label{sec:measurements}

We aim to estimate the small-scale power in a way that minimizes sensitivity
to uncertainties in the quasar continuum. Dall'Aglio et al. (2008) carefully
continuum fit the data we use here, and used Monte Carlo simulations to check the accuracy
of their fits. We can further mitigate uncertainties by considering fluctuations
in the transmission around the mean, {\em relative to the mean}. This is helpful
because the overall normalization of the continuum divides out.
Provided that the continuum varies slowly across each spectrum in comparison with the fluctuations in the 
forest, we can additionally remove any slowly varying trend produced by the quasar continuum -- or any 
slowly-varying residuals in the case of data that has previously been continuum fitted -- 
and obtain an unbiased estimate of the small-scale structure in the forest (Hui et al. 2001).
For each spectrum, we estimate a running mean flux
by filtering the data on large scales as in Croft et al. (2002), Kim et al. (2004), and 
Lidz et al. (2006). Our
estimate of the fractional transmission is then:
\beqa
\hat{\delta}_F(\Delta v) = \frac{F(\Delta v) - F_R(\Delta v)}{F_R(\Delta v)}.
\label{eq:deltaf}
\eeqa
Here $F(\Delta v)$ is the flux at velocity
separation $\Delta v$, and $F_R(\Delta v)$ is the spectrum smoothed with a large
radius filter. We use here a Gaussian filter with radius $R = 2,500$ km/s. One may
form $\hat{\delta}_F$ using either the raw flux or a continuum-normalized flux. 
In the present work, we
use the continuum fitted data from Dall'Aglio et al. (2008) throughout.
The large scale filter removes any slowly-varying trend owing to structure in the underlying
quasar continuum from, e.g. weak emission lines, or slowly varying residuals in the case of 
continuum fitted data. It also 
means that we sacrifice measuring large
scale modes in the Ly-$\alpha$ forest, but we presently focus on small-scale structure, and 
sufficiently large scale modes are regardless dominated by structure in the quasar continuum.
We refer the reader to Croft et al. (2002) and Lidz et al. (2006) for some tests illustrating the
robustness of $\hat{\delta}_F$ to continuum-fitting uncertainties. As a double-check that the present 
results are insensitive to the precise $\hat{\delta}_F$ estimator, we also generated $\hat{\delta}_F$ with
a different choice of large scale smoothing for one of our redshift bins, $R=10,000$ km/s -- i.e., close to the 
flat mean case -- and found a nearly identical wavelet PDF.

We begin by estimating $\hat{\delta}_F$ across each spectrum, first re-binning, using linear interpolation, 
all of the data onto uniform pixels in velocity space with $\Delta u = 4.4$ km/s. We consistently use the
same binning in constructing simulated spectra. 
This avoids effects from variable pixelization,
while still preserving the scales of interest.\footnote{We estimate that rebinning reduces
the mean wavelet amplitude by $\lesssim 5\%$ for $s_n = 69.7$ km/s.} 
After forming $\hat{\delta}_F$ across each spectrum,
we break the data into several (contiguous and non-overlapping) redshift bins of full-width
$\Delta z = 0.4$, centered 
around $\bar{z} = 2.2, 2.6, 3.0, 3.4, 3.8$,
and $4.2$.  Owing to uneven redshift sampling in the data set, the redshift bin 
at $\bar{z} = 3.8$ (Dall'Aglio et al. 2008) would be almost entirely empty and so we do not consider 
it further here.
This occurs because most of the spectra in the Dall'Aglio et al. (2008) sample have emission redshift
$z_{\rm em} \lesssim 3.7$, but the sample has two high quality spectra at emission redshift above
$z_{\rm em} \gtrsim 4.6$, which contribute extended ($\gtrsim 150$ co-moving Mpc) stretches 
to our highest redshift bin at $\bar{z}=4.2$. 
We select only spectral regions that lie between rest frame wavelengths of 
$\lambda_r = 1050$ $\AA$ and $\lambda_r = 1190$ $\AA$. This conservative cut serves to remove spectral
regions that may be contaminated by either the proximity effect, by the Ly-$\beta$ forest (and other
higher Lyman series lines), or by Ly-$\beta$ and OVI emission features.   
We then form the wavelet amplitude squared field, smoothed at $L=1,000$ km/s, using 
Equations \ref{eq:filt_real} -- \ref{eq:smoothed_waveamp}. 
The resulting spectra and wavelet amplitudes
are visually inspected. Regions impacted by DLAs, or with obvious spurious stretches,
are removed from the data sample by hand.

\begin{figure}
\bc
\includegraphics[width=9.2cm]{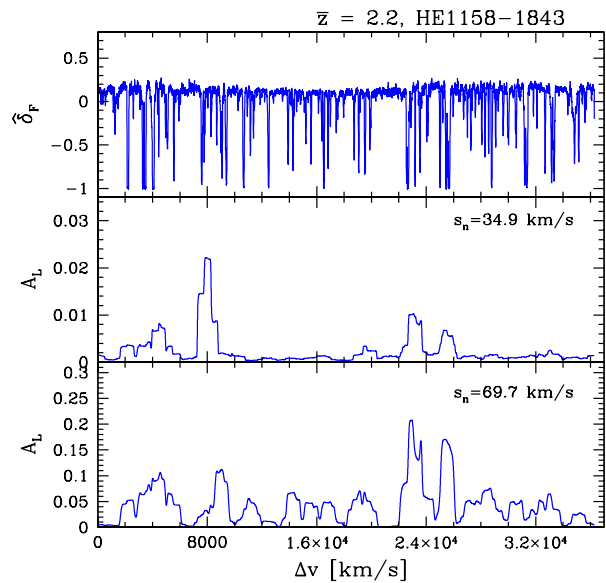}
\caption{Example spectrum and smoothed wavelet amplitudes from the $\bar{z}=2.2$ bin.
{\em Top panel:} The fractional transmission fluctuations $\hat{\delta}_F$ for the spectrum
of the quasar HE1158-1843. {\em Middle panel:} The amplitude squared of the wavelet filtered field, 
formed with a $s_n = 34.9$ km/s filter, smoothed
over $L=1,000$ km/s. {\em Bottom panel}: Similar to the middle panel, but using a Morlet wavelet with
$s_n = 69.7$ km/s. Note that the y-axis in the bottom two panels have rather different ranges. This is required because
of the strong dependence of small scale power on smoothing scale.}
\label{fig:wave_z2.2_examp}
\ec
\end{figure}

\begin{figure}
\bc
\includegraphics[width=9.2cm]{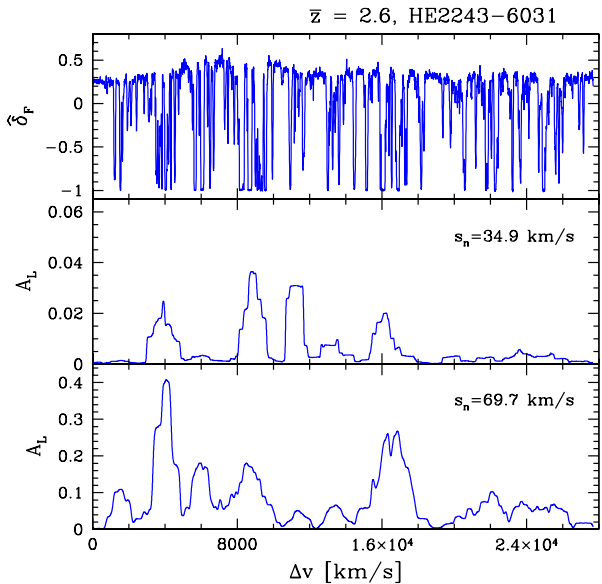}
\caption{Example spectrum and smoothed wavelet amplitudes from the $\bar{z}=2.6$ bin.
Similar to Figure \ref{fig:wave_z2.2_examp}, but for the spectrum of HE2243-6031. 
Note that the x and y
axes have different ranges than in the previous figure. The x-axis range is set by the portion of
the forest that we use from the example spectrum in a given redshift bin. We vary the y-axis range because the mean
wavelet amplitude changes strongly with redshift, owing mostly to evolution in the mean absorption, and so a varying
range is necessary for visual clarity.}
\label{fig:wave_z2.6_examp}
\ec
\end{figure}

\begin{figure}
\bc
\includegraphics[width=9.2cm]{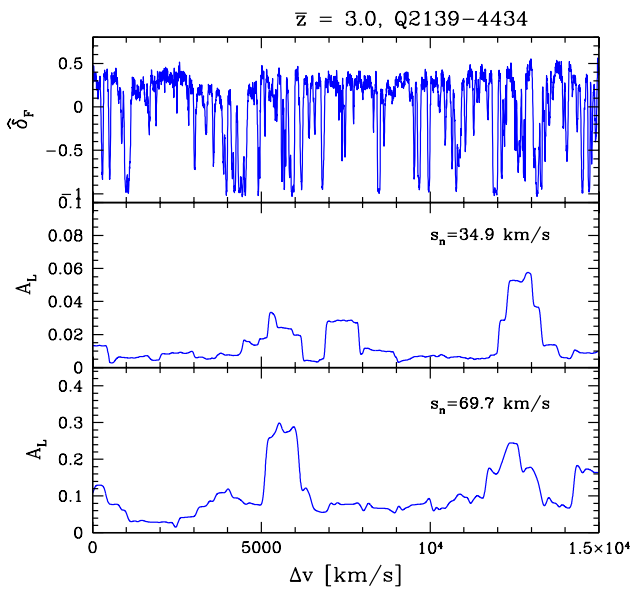}
\caption{Example spectrum and smoothed wavelet amplitudes from the $\bar{z}=3.0$ bin.
Similar to Figure \ref{fig:wave_z2.2_examp}, but for the spectrum of Q2139-4434. Note that the x and y
axes have different ranges than in the previous figures.}
\label{fig:wave_z3.0_examp}
\ec
\end{figure}

\begin{figure}
\bc
\includegraphics[width=9.2cm]{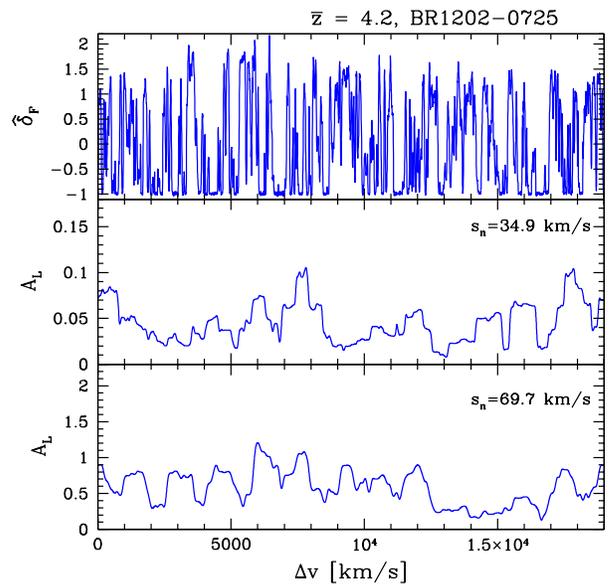}
\caption{Example spectrum and smoothed wavelet amplitudes from the $\bar{z}=4.2$ bin.
Similar to Figure \ref{fig:wave_z2.2_examp}, but for the spectrum of BR1202-0725. Note that the x and y
axes have different ranges than in the previous figures.}
\label{fig:wave_z4.2_examp}
\ec
\end{figure}

It is instructive to examine a few example spectra visually before measuring their detailed 
statistical properties. 
In Figures \ref{fig:wave_z2.2_examp} -- \ref{fig:wave_z4.2_examp} we show several spectra, along 
with
the corresponding (smoothed) wavelet amplitudes squared for a few redshift bins. 
The most conspicuous change
across the different redshift bins is the increasing average absorption with increasing redshift. Since we 
are considering
fractional fluctuations, this manifests itself as an increase in the fraction of pixels with $\hat{\delta}_F$ close to $-1$, with occasional
excursions to very large $\hat{\delta}_F$. The next impression provided by the spectra appears at first 
tantalizing: most regions have low $A_L$, but
there are occasional upward excursions over portions of the spectrum. This behavior is especially 
apparent for the
smaller of the two filtering scales, and is less apparent in the highest redshift 
case (Figure \ref{fig:wave_z4.2_examp}). 

Consider for example the spectrum Q2139-44, in the $\bar{z}=3.0$ bin, convolved with a $s_n = 34.9$ km/s Morlet filter,
as shown in Figure \ref{fig:wave_z3.0_examp}. In this spectrum the
regions near $\Delta v = 5,000$, $7,500$, and $12,500$ km/s all have relatively high wavelet amplitudes, $A_L \gtrsim 0.02$, while
the rest of the spectral regions have low amplitude. Inspecting the simulated PDF of Figure \ref{fig:pdf_models_sn3}, the low amplitude 
floor with
$A_L \sim 0.005$ seems to indicate hot $T_0 \sim 20,000$ K gas, while the regions with $A_L \gtrsim 0.02$ seem to require cooler gas
$T_0 \lesssim 7,500$ K gas.

\begin{figure}
\bc
\includegraphics[width=9.2cm]{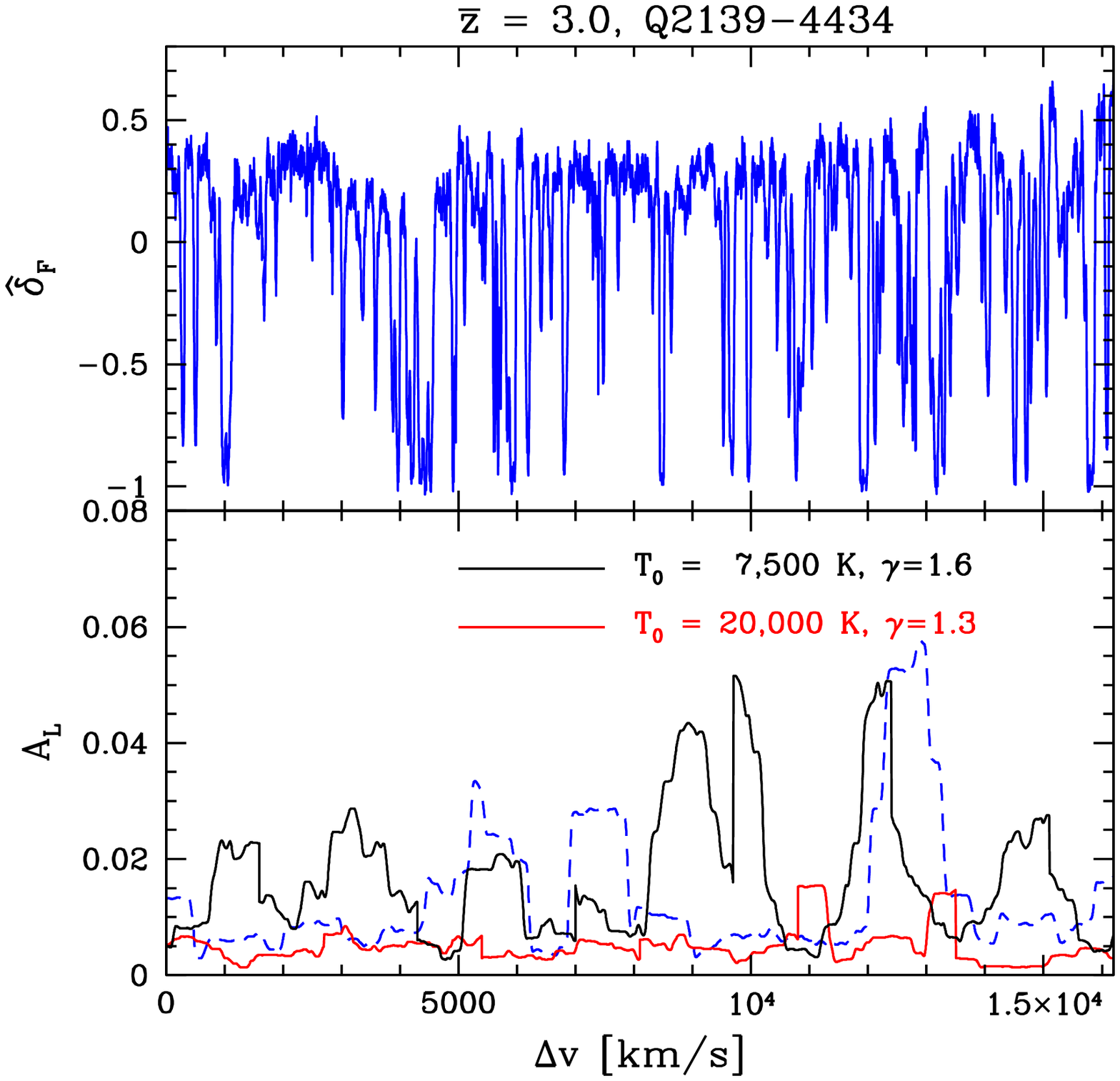}
\caption{Example wavelet amplitude field compared with models. The smoothing scale is
$s_n = 34.9$ km/s here.
The blue lines are the same as in 
Figure \ref{fig:wave_z3.0_examp}. The observed wavelet amplitudes are shown by a dashed line to avoid confusion
with the model curves.
The red and black lines in the bottom panel are
simulated sightlines for a hot IGM model (red), and a cold IGM model (black). Random noise has been added to the
simulated spectra (see \S \ref{sec:smooth_small}).
The wavelet amplitudes in 
most spectral regions are roughly consistent with the hot IGM model, but the high wavelet amplitude excursions
(near $\Delta v = 5,000$, $7,500$, and $12,500$ km/s)
look naively like cold gas. In \S \ref{sec:metals}, we show that these apparent cold regions are spurious and
are instead consistent with being hotter gas contaminated by metal lines.}
\label{fig:wave_z3_vs_models}
\ec
\end{figure}

At first glance, these upward wavelet amplitude excursions seem to be cold regions embedded in an 
otherwise hot IGM. This is 
what one naively
expects in the midst of HeII reionization: cool regions where HI and HeI reionized long ago, and hotter regions where helium
is doubly ionized. Before we dispel this fantasy -- these regions are contaminated by metal absorbers (see \S \ref{sec:metals}) --
let us add some sightlines from simulated models to further illustrate 
this (Figure \ref{fig:wave_z3_vs_models}).\footnote{The mock spectra are described in \S \ref{sec:sims}. These examples
are longer than the side-length of the simulation box, and are produced by splicing together the wavelet amplitudes from shorter
mock spectra.} 
The sightlines show that the low wavelet amplitude floor in the observed spectrum roughly matches the hot IGM 
sightline. This implies that there are indeed significant quantities of hot $\sim 20,000$ K gas in the IGM at 
$z=3$. However, the hot model fails to produce the high wavelet amplitude excursions seen in the data. Matching these seems, at first glance, to require
a cooler model -- one with roughly $T_0 \sim 7,500$ K, $\gamma=1.6$, for example. (To be clear, note that the simulation and 
observational
data are drawn from different realizations, so one does not expect the simulated case to match the observations region-by-region or
feature-by-feature. The meaningful comparison is the overall number of regions with high or low wavelet amplitude.)
It is at first tempting to conclude that we are detecting temperature inhomogeneities from incomplete HeII reionization.

\subsection{Metal Line Contamination} \label{sec:metals}

We need, however, to consider a very important systematic. A hot region that lands at the same wavelengths
as a `clump' of prominent narrow metal lines may look
to us like a cold region. The wavelet filter just tells us the total level of small-scale power from place 
to place, and
does not distinguish whether absorption arises from HI or some other element. To make a robust estimate of the IGM temperature,
we need to identify metal line absorbers within the Ly-$\alpha$ forest.\footnote{An alternate approach is to remove
metal contamination statistically. This can be done by using a set of lower redshift quasars where the metal absorbing gas of interest
lies redward of Ly-$\alpha$ (McDonald et al. 2006). This procedure only works for lines with rest frame wavelength longer than that of Ly-$\alpha$,
however. Presently, we do not have the data sample to explore the impact of metals on the small-scale wavelet amplitudes in
this way, but it might be interesting to investigate this in future work.} We expect metal line contamination to be most severe in the low redshift bins, where the fractional contribution of
metals to the overall opacity in the forest is highest (e.g. Faucher-Gigu\`ere et al. 2008b), and on 
the smaller of our two filtering scales (see Appendix B). 

Naturally, distinguishing metal absorption lines and Ly-$\alpha$ lines within the Ly-$\alpha$ forest is a challenging
and imperfect process. We do, however, have a few separate handles on distinguishing metal lines from Ly-$\alpha$ lines within the 
forest.
First, we identify all of the metal absorbers redward of Ly-$\alpha$ and look for `partner' transitions. The partner
transitions are additional transitions that lie at the same redshift as an identified red-side line, yet which 
land within the Ly-$\alpha$ forest. 
Next, we search for doublets within the Ly-$\alpha$ forest, which can be identified by their 
distinctive optical
depth ratios and by the characteristic separation between a doublet's two components.
For instance, CIV is a doublet with a strong component at $\lambda_r = 1548.2 \AA $, and a weaker
component at $\lambda_r = 1550.8 \AA$, and the ratio of the absorption cross sections of the two 
components is $2$. So CIV should
stand out as a doublet with the two components separated by $\sim 640$ km/s, with the lower wavelength line a 
factor of two 
stronger than its partner component. MgII is another prominent doublet. After identifying a doublet, one can use the
estimated redshift of an identified doublet to search for additional transitions at the same 
redshift: we look for CII/III/IV, NII/III/V, OI/VI, MgI/II,
AlII/III, SiII/III/IV, SVI, and FeII, and consider further transitions for DLAs.  This approach already identifies a host of metal lines
within the forest, but there are inevitably some remaining metal lines left within the forest. For example, there are sometimes
absorbers where the doublet features are undetectable owing to line blending. To further mitigate metal line
contamination, our final step is to mark extremely narrow lines (with b-parameter $b \lesssim 7$ km/s) as metals. 
This final
cut amounts to only $25\%$ of the identified lines. Clearly, one needs to be careful about making cuts based
on line width: doing so could bias us against detecting cold regions. 
However, for an HI line to have a linewidth of $b \lesssim 7$ km/s it needs to have an implausibly low temperature
of $T \lesssim 3,000$ K. Hence, we are confident that this final cut does not bias our results, yet it helps
protect against remaining unidentified metal lines within the forest.
We will subsequently present tests to check how much the results depend on 
the precise way in which we excise metal line contaminated regions. 
 
In this paper, we have identified metal lines for $11$ of the $40$ spectra in our data sample.
The identified metals come entirely from portions
of spectrum absorbing at $z \lesssim 3$ -- where we expect the metal line contamination to be strongest -- and not in 
the higher redshift bins. That is, we do not presently have
estimates of metal line contamination in the redshift bins centered around $\bar{z}=3.4$ and $\bar{z}=4.2$. In these redshift
bins, we will focus entirely on the larger ($s_n = 69.7$ km/s) filtering scale where the metal line contamination is less
of an issue (Appendix B).  

In order to check the influence of metal line contamination, we calculate the wavelet amplitudes as before, and excise regions
impacted by metal line contamination. 
An important assumption here is that gas absorbing in a metal line transition at a given wavelength is spatially uncorrelated
with gas absorbing in Ly-$\alpha$ at nearby wavelengths. If this assumption were violated, we could bias ourselves by preferentially
removing regions of above average {\em hydrogen} absorption when excising metal contaminated regions. Fortunately, 
most of
the metal line transitions have rest-frame wavelengths that are very different than that of Ly-$\alpha$ and so the gas 
absorbing in a metal transition at a given
wavelength is very widely separated (in physical space) from that absorbing in Ly-$\alpha$. Hence the metal and Ly-$\alpha$ 
absorption are 
uncorrelated. This justifies our 
approach.\footnote{There are exceptions to this. For example, SiIII absorbs at $\lambda_r = 1206.5 \AA$ and is only
separated from Ly-$\alpha$ by $\sim 2300$ km/s, which leads to a distinctive yet small feature in the Ly-$\alpha$ transmission
power spectrum (McDonald et al. 2006). Fortunately, these HI-correlated transitions only produce a small fraction 
of the total 
metal opacity, and should not bias us significantly.}
Since the wavelet filter is not completely local, pixels with metal line absorption will
contaminate neighboring pixels after filtering. Furthermore, we generally smooth the wavelet squared field over $L=1,000$ km/s.
As a simple and conservative cut, we examine the fraction of contaminated pixels within a smoothing length $L$ around each pixel, and
discard a pixel if less than $f_m = 95\%$ of its neighbors (within a smoothing length) are metal free.

\begin{figure}
\bc
\includegraphics[width=9.2cm]{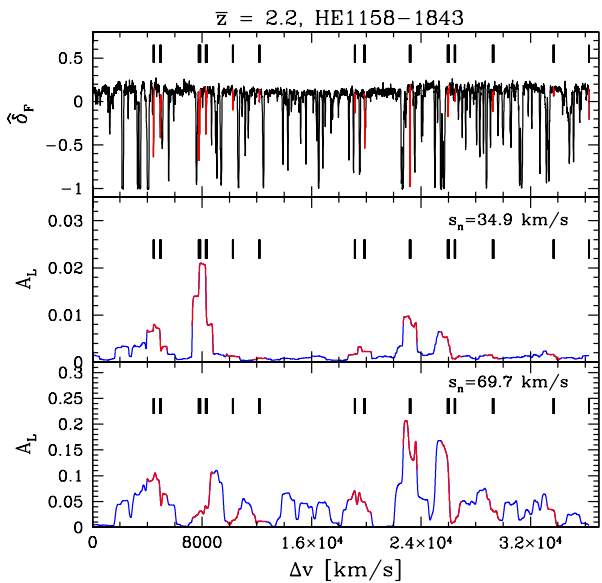}
\caption{Example of the impact of metal line contamination from the $\bar{z}=2.2$ bin. Identical to Figure \ref{fig:wave_z2.2_examp}, 
except illustrating the impact of metal line contamination. {\em Top panel}: Red lines, and short black dashed
lines above the spectrum, indicate identified 
metal
lines within the Ly-$\alpha$ forest. {\em Middle panel}: The short black lines identify the centers of pixels 
with identified metal
lines. The red lines indicate the approximate regions where metal lines impact the wavelet amplitudes (for $f_m = 0.95$,
see text). Most of the wavelet amplitude peaks correspond to metal line contaminated regions for this filtering scale
($s_n = 34.9$ km/s). {\em Bottom panel}: Similar to the middle panel, for $s_n = 69.7$ km/s. 
}
\label{fig:wave_z2.2_examp_metals}
\ec
\end{figure}

\begin{figure}
\bc
\includegraphics[width=9.2cm]{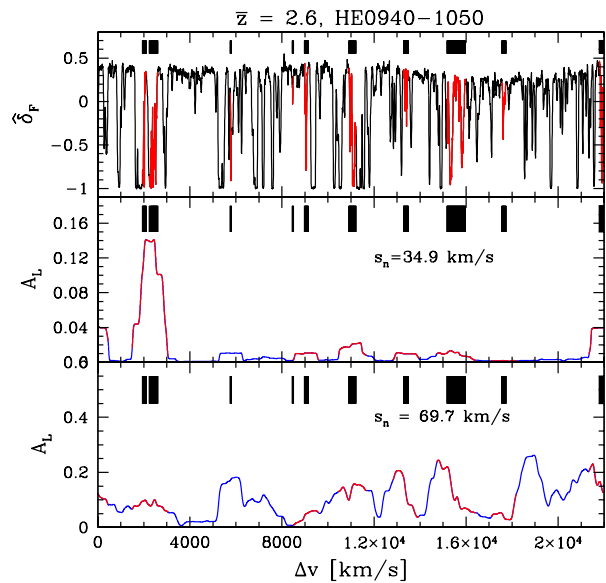}
\caption{Example of the impact of metal line contamination from the $\bar{z} = 2.6$ bin.
Similar to Figure \ref{fig:wave_z2.2_examp_metals} but for the spectrum HE0940-1050 in the
$\bar{z} = 2.6$ bin. Notice in particular that the very large wavelet amplitudes near $\Delta v = 2,000$ km/s for
$s_n = 34.9$ km/s
correspond closely to several strong metal lines. Again the wavelet peaks at this filtering scale trace 
mostly metal line contaminated
regions. The lower wavelet amplitude regions, and not these high amplitude portions, indicate the IGM temperature.
Note that the metal line contamination is less severe for the larger smoothing scale filter in the bottom panel.  
}
\label{fig:wave_z2.6_examp_metals}
\ec
\end{figure}

\begin{figure}
\bc
\includegraphics[width=9.2cm]{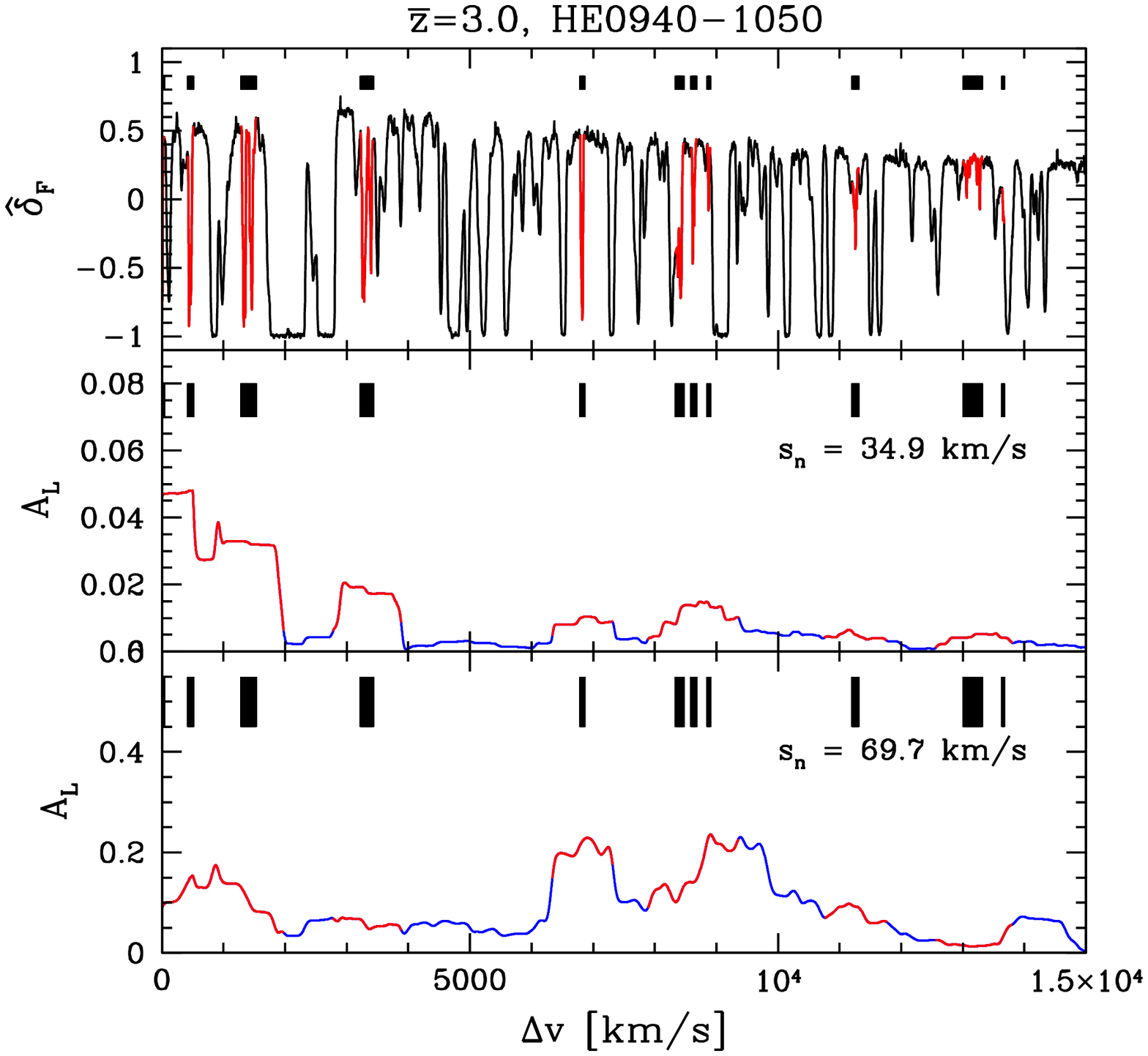}
\caption{Example of the impact of metal line contamination from the $\bar{z} = 3.0$ bin.
Similar to Figure \ref{fig:wave_z2.2_examp_metals} and Figure \ref{fig:wave_z2.6_examp_metals}, but for the 
portion of HE0940-1050 in the $\bar{z} = 3.0$ bin. Once again the large wavelet amplitude regions at filtering
scale $s_n=34.9$ km/s
are metal contaminated.
}
\label{fig:wave_z3.0_examp_metals}
\ec
\end{figure}

We find that metal line contamination can have a significant impact, especially for $s_n=34.9$ km/s, and
at $z \lesssim 3$.
We show a few example sightlines in Figures \ref{fig:wave_z2.2_examp_metals} -- \ref{fig:wave_z3.0_examp_metals}. 
It is striking that the most prominent peaks in the wavelet filtered field at $s_n = 34.9$ km/s, shown in 
the figures, correspond
very closely to metal lines. Essentially, our filter was designed to look for temperature 
inhomogeneities, but
it appears most effective at identifying metal-line contaminated regions! In fact, wavelet filtering may be
a good way to quickly identify prominent metals in the forest. The metal line contamination is less 
severe for
spectra passed through the larger wavelet 
filter ($s_n = 69.7$ km/s). The amplitude of fluctuations in the forest is much greater on this
smoothing scale. The metals also generally contribute more power on the larger smoothing scale, but the 
amplitude of fluctuations from HI increases more strongly with smoothing scale, and so the metals are fractionally
less important on larger scales.  
This is perhaps seen most easily in the example 
of Figure \ref{fig:wave_z2.6_examp_metals}. In the $s_n = 34.9$ km/s panel of this figure, all of the prominent 
peaks are 
metal contaminated regions. In the larger smoothing scale panel, there are some peaks from HI and some from metals,
and the heights of the various peaks are comparable. The more significant contamination of the metals on the
smaller smoothing scale likely results because the metal lines tend to be narrower than the HI lines. 
In Appendix B we find qualitatively similar results by adding metal line absorbers, with empirically
derived properties, to mock Ly-$\alpha$ forest spectra.

Since we can attribute many of the peaks observed in the wavelet amplitudes to metal lines, this does imply, however,
that the temperature inhomogeneities cannot be too large.
If temperature inhomogeneities
were present and large, we would expect to see more high wavelet amplitude regions left over after excising the metals.
In particular, consider Figure \ref{fig:wave_z3_vs_models}. In this example, we found that the low wavelet amplitude
regions of the spectrum are consistent with hot $20,000$ K gas. While we have not identified metal lines
for this particular spectrum, our results from other lines of sight 
clearly suggest that the high wavelet amplitude regions are metal-contaminated
rather than genuine cold regions with $T_0 \sim 7,500-10,000$ K. The lack of high wavelet amplitude 
regions after metal
excision implies there are few such cold regions left, and 
that {\em most of the volume of the IGM at $z \sim 3$ is hot with
$T_0 \sim 20,000$ K} (although see \S \ref{sec:sims} for a discussion regarding the dependence of 
our results on $\gamma$). 

It is clear, however, that metal line contamination is a very important systematic for these measurements, although 
the contamination is less of an issue on the larger smoothing scale and for the high redshift measurements.
This issue is not unique
to our method, although the detailed impact of metal lines will depend on the precise algorithm for constraining the IGM temperature.
For instance, measurements based on fitting the minimum width of absorption lines in 
the Ly-$\alpha$ forest need
to carefully avoid including metal lines in the sample of lines used to estimate the temperature. Power spectrum based 
temperature estimates need
to account for the small-scale power contributed by metal absorbers or mask the metal 
absorbers before
estimating the power spectrum.

\subsection{The Wavelet Amplitude Squared PDF} \label{sec:pdf}

Let us now move past mere visual inspection and measure statistical properties from the observed spectra.
We focus mostly on the PDF of $A_L$ for our fiducial choices of $s_n = 34.9$ km/s, $s_n = 69.7$ km/s, 
$L = 1,000$ km/s, and
$f_m = 0.95$. In each redshift bin, we find the minimum and maximum $A_L$ and then choose $10$ evenly-spaced 
logarithmic
bins in $A_L$ for the PDF measurement. We tabulate the average $A_L$ and the differential PDF 
in each $A_L$ bin for each redshift bin. The average redshift of pixels in a redshift bin is typically close
to the redshift at bin center, and the error bars are still fairly large, so we ignore any issues 
associated with redshift evolution across each bin and quote all results at the bin center.

We use a jackknife resampling technique to calculate error bars for the PDF measurements.
We first estimate the PDF from the entire data sample within a given redshift bin, $\hat{P}(A_i)$. Here $\hat{P}(A_i)$ is the
PDF estimate for the ith $A_L$ bin, and $A_i$ is the average wavelet amplitude squared and smoothed within the bin.
Next we divide the data set into $n_g = 10$ subgroups, and estimate the PDF of the data sample {\em omitting each subgroup}.
Let $\tilde{P}_k(A_i)$ represent the PDF estimate omitting the pixels in the kth subgroup.
Then our estimate of the jackknife covariance between bins $i$ and $j$, $C(i,j)$, is:
\beqa
C(i,j) =
\sum_{k=1}^{n_g} \left[\hat{P}(A_i) - \tilde{P}_k(A_i)\right]\left[\hat{P}(A_j) - \tilde{P}_k(A_j)\right].\nonumber \\
\label{eq:coveq}
\eeqa

In practice our estimates of the off-diagonal elements of the covariance matrix are very noisy. Consequently,
we will be forced to ignore the off-diagonal elements of $C(i,j)$. We have tested the jackknife error estimator 
with lognormal mocks (see McDonald et al. 2006, Lidz et al. 2006)
that approximately mimic the properties of the current data set. We generate $10,000$ mock realizations of a $z=3$ 
data set
and compare error bars estimated from the dispersion across the mock realizations with the jackknife error estimates. 
In the mock data, we find that neglecting the off-diagonal elements in the covariance matrix increases the average 
value of 
$\chi^2$ by $\sim 1$ for $14$ degrees of freedom (the mock PDFs had $15$ rather than $10$ $A_L$ bins), and so ignoring the 
off-diagonal elements is likely a good 
approximation.
The jackknife estimates of the diagonal elements of the covariance matrix agree with direct estimates of the 
dispersion across
the mock data to better than $20\%$ on average, although the jackknife estimator sometimes under-predicts the 
errors in the tails of the PDF more severely. We provide tables of the wavelet PDF measurements in 
Tables \ref{table:pdfz4.2}--\ref{table:pdfz2.2}.

\subsection{Shot Noise} \label{sec:shot}

We plot the measured wavelet PDF in the next section, but pause to consider first the impact of 
shot-noise. The observed Ly-$\alpha$ forest spectra are contaminated by random noise owing to Poisson fluctuations
in the discrete photon count and around the mean night sky background count, as well as by random read-out noise
in the CCD detector (see e.g. Hui et al. 2001 for discussion). We need to consider how this noise impacts the wavelet PDF
measurements.  

In Appendix A, we derive estimates of the noise bias for measurements of the first two moments of the wavelet amplitude
PDF. We apply these formulae here to estimate the impact of noise on the present measurements. On the larger
smoothing scale, $s_n=69.7$ km/s, we find that shot-noise bias is unimportant for our present data set. 
For example, at $z=3$, applying the formulae of Appendix A, we find that the noise contamination to the mean wavelet amplitude is less
than one-third of the 1-$\sigma$ statistical error on this quantity for our present 
data sample. Similarly, in this redshift bin and for this smoothing scale,
we find that the wavelet amplitude variance is biased by random noise only at the $\sim 1\%$ level.
However, the shot-noise bias is not negligible on the smaller smoothing scale, $s_n= 34.9$ km/s. For instance, a quasar
spectrum with $S/N \sim 50$ at the continuum contributes a mean wavelet amplitude owing to noise of $\avg{|a_n^{\rm noise}|^2} \sim
(N/S)^2/\avg{F} \sim 6 \times 10^{-4}$ at $z \sim 3$ (Appendix A, Hui et al. 2001). This is comparable to the wavelet amplitude
signal in the tail of the PDF in the favored hot IGM models (see Figure \ref{fig:pdf_models_sn3}). The more significant
noise contamination on the smaller smoothing scale owes to the rapid decline in signal power towards small scales. 
To guard against noise 
bias at the smaller smoothing scale, we cut spectra with $S/N \leq 50$ redward
of Ly-$\alpha$ from the sample used in the smaller smoothing scale measurement. We cut based on the red side noise, rather than using 
a noise estimate in the forest, 
to avoid introducing any possible selection bias. Further, we add noise to the mock spectra when comparing with 
the measurement on the smaller smoothing scale (\S \ref{sec:smooth_small}).

\section{Theoretical Interpretation} \label{sec:theor}

In this section, we compare the wavelet PDF measurements with cosmological simulations in order to estimate the
implied IGM temperature. A particular goal here is to determine whether the IGM is closer to the thermal state 
expected in the midst
of HeII reionization ($T_0 \sim 20-25,000$ K, $\gamma=1.3$) or whether it more closely resembles the state much
after a reionization event ($T_0 \sim 7,500-10,000$ K, $\gamma=1.6$).   
Furthermore, we aim to check whether the 
data indicate large temperature inhomogeneities. 
We perform this comparison over the full redshift 
range of our data set, $\bar{z}=2.2-4.2$.

\subsection{Cosmological Simulations} \label{sec:sims}

For the purpose of this project and related Ly-$\alpha$ forest work, we have run a new suite of cosmological smoothed 
particle hydrodynamics (SPH) simulations
using the simulation code Gadget-2 (Springel 2005). The simulations adopt a LCDM cosmology parameterized
by: $n_s=1$, $\sigma_8 = 0.82$, $\Omega_m = 0.28$, $\Omega_\Lambda=0.72$, $\Omega_b = 0.046$, 
and $h=0.7$ (all symbols have their usual meanings), consistent with the WMAP constraints
from Komatsu et al. (2009). Each simulation was started from $z=299$, with the 
initial conditions generated using the Eisenstein \& Hu (1999)
transfer function. We ran several simulations to test the convergence of our results with boxsize,
as well as mass and spatial resolution (see \S \ref{sec:box_res}). From these tests, we determined
that the best choice simulation for the present project has a  
boxsize of $L_{\rm box} = 25$ Mpc/$h$ and $N_p = 2 \times 1024^3$ particles, and this run is the fiducial
simulation in what follows. This simulation represents
a fairly significant improvement in boxsize and resolution compared to most previous work (see \S \ref{sec:compare_previous} for details). It
has approximately the gas mass recommended for resolution convergence in a recent study 
by Bolton \& Becker (2009), and tracks over an order of magnitude more particles than the simulations
of these authors. 
In each run, the softening length was taken to be $1/20$th of the
mean inter-particle spacing. In order to speed up the calculation, we chose an option in Gadget-2 that
aggressively turns all gas at density greater 
than $1000$ times the cosmic mean density into stars (e.g. Viel et al. 2004).
Since the forest is insensitive to gas at such high densities, this is a very good approximation. 

The simulations were run using the Faucher-Gigu\`ere et al. (2009) photoionizing background, which
is an update of the Haardt \& Madau (1996) 
model (see, also Katz et al. 1996a, Springel \& Hernquist 2003)\footnote{Tables are 
available electronically
at http://www.cfa.harvard.edu/$\sim$cgiguere/uvbkg.html}. The ionizing background was turned on at
$z=7$ in the simulations.
This model for the ionizing background determines the photoheating and gas temperature in the 
simulation.
We would like, however, to explore a wide range of thermal histories. In order to do this, we make an 
approximation. The approximation is to fix the fiducial ionization history to the
Faucher-Gigu\`ere et al. (2009) model 
for the purpose
of running the simulation and accounting for gas pressure smoothing, but to vary the temperature-density
relation (Equation \ref{eq:tdelta}) when constructing simulated spectra. This `post-processed spectra' approximation neglects
the dependence of Jeans smoothing on the detailed thermal history of the IGM, but correctly incorporates thermal broadening
for a given temperature-density relation model, parametrized by $T_0$ and $\gamma$. It also neglects the 
inhomogeneities in $T_0$ and $\gamma$ expected during HeII reionization. Finally, by assuming a perfect 
temperature-density relation in constructing 
mock absorption spectra, we also neglect the impact of shock heating -- which adds scatter to the temperature
density relation (Hui \& Gnedin 1997) -- on the amount of thermal broadening. 
We caution against taking the results of these first pass, homogeneous
temperature-density relation calculations too literally: if the IGM temperature is significantly inhomogeneous, these
calculations provide only a crude approximation. The calculations are meant only to get a sense for whether the IGM is
mostly at $T_0 \sim 20,000$ K, or instead at $T_0 \sim 10,000$ K, and to check whether large temperature inhomogeneities are present. We 
intend to make more detailed theoretical calculations in future work. 

Although our measurements of the wavelet amplitude PDF are robust to uncertainties in fitting the quasar continuum, our
{\em interpretation} of the measurements still relies on estimates of the mean transmitted flux. 
Specifically, we follow the normal procedure of adjusting the intensity of the simulated ionizing background in a post-processing
step, so that the simulated mean transmitted flux (averaged over all sightlines) matches the observed mean flux.
We assume here
the best-fit measurements of Faucher-Gigu\`ere et al. (2008b), and subsequently explore the impact of uncertainties
in the mean flux (\S \ref{sec:uncert}). We adopt their estimates in $\Delta z = 0.2$ bins, and use their 
measurements that include 
a correction for metal line opacity
based on Schaye et al. (2003), and a continuum-fitting correction (which accounts for the rarity of regions with nearly 
complete transmission, $F = 1$, 
at high redshift). The 
corresponding Faucher-Gigu\`ere et al. (2008b) measurements in our
redshift bins are: $(\bar{z}, \avg{F}) = (2.2, 0.849); (2.6, 0.778); (3.0, 0.680); (3.4, 0.566);$
and $(4.2, 0.346)$. 
We output simulation data at every $\Delta_z =0.5$ between $z=4.5$ and $z=2$. 
In order
to generate model wavelet PDFs at redshifts in between two stored snapshots, we measure the simulated
wavelet PDF from each stored snapshot and linearly interpolate to find the PDF at the precise desired redshift.

\subsection{Comparison with Measurements} \label{sec:compare_th_data}

\begin{figure}
\bc
\includegraphics[width=9.2cm]{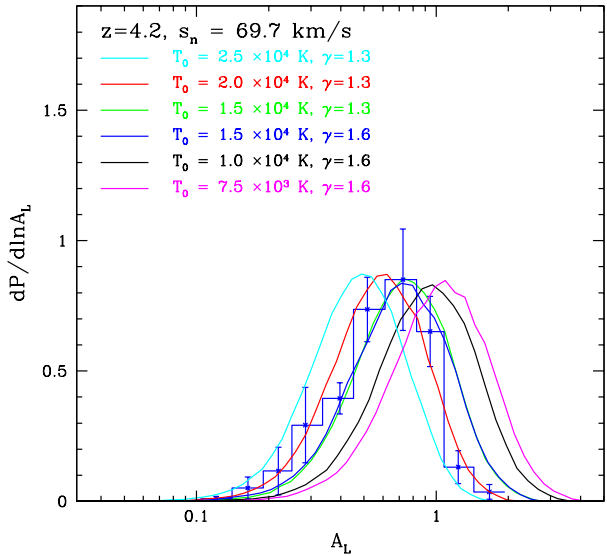}
\caption{Comparison between the measured wavelet PDF in the $\bar{z}=4.2$ bin with $s_n = 69.7$ km/s, $L=1,000$ km/s and simulated
models. The blue histogram with points and ($1-\sigma$) error bars is the measured PDF, uncorrected for metal line
contamination.}
\label{fig:pdf_wave_z4.2_sm_v_models}
\ec 
\end{figure}

\begin{figure}
\bc
\includegraphics[width=9.2cm]{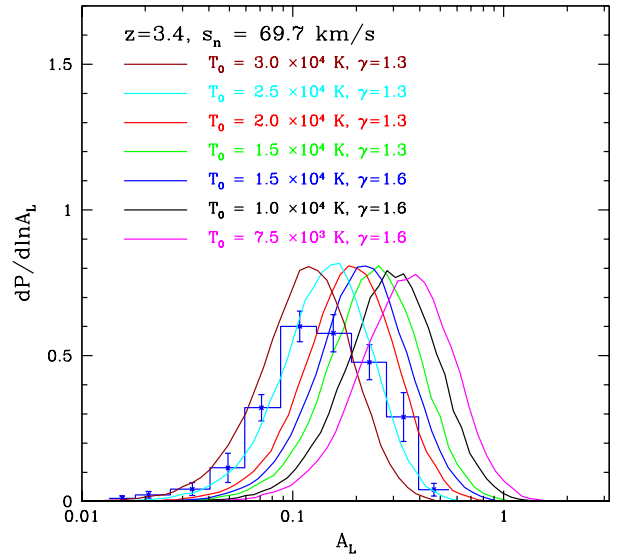}
\caption{Comparison between the measured wavelet PDF in the $\bar{z}=3.4$ bin with $s_m = 69.7$ km/s and simulated
models. Similar to Figure \ref{fig:pdf_wave_z4.2_sm_v_models}, but at $\bar{z}=3.4$.}
\label{fig:pdf_wave_z3.4_sm_v_models}
\ec 
\end{figure}

\begin{figure}
\bc
\includegraphics[width=9.2cm]{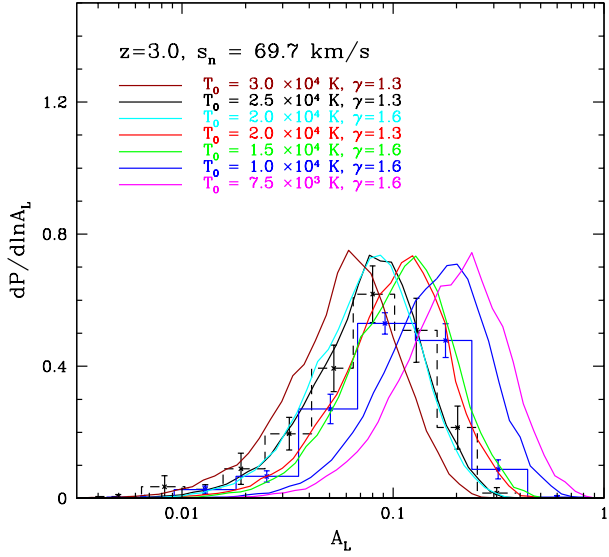}
\caption{Comparison between the measured wavelet PDF in the $\bar{z}=3.0$ bin with $s_m = 69.7$ km/s and simulated
models. Similar to Figure \ref{fig:pdf_wave_z4.2_sm_v_models}--Figure \ref{fig:pdf_wave_z3.4_sm_v_models}, but 
at $\bar{z}=3.0$. The blue histogram shows the PDF estimated from all spectral regions, while the
black dashed histogram removes regions with metal line contamination. The histogram with metal contaminated
regions removed comes from the subset of the data in this redshift bin for which we have identified metals.}
\label{fig:pdf_wave_z3.0_sm_v_models}
\ec 
\end{figure}

\begin{figure}
\bc
\includegraphics[width=9.2cm]{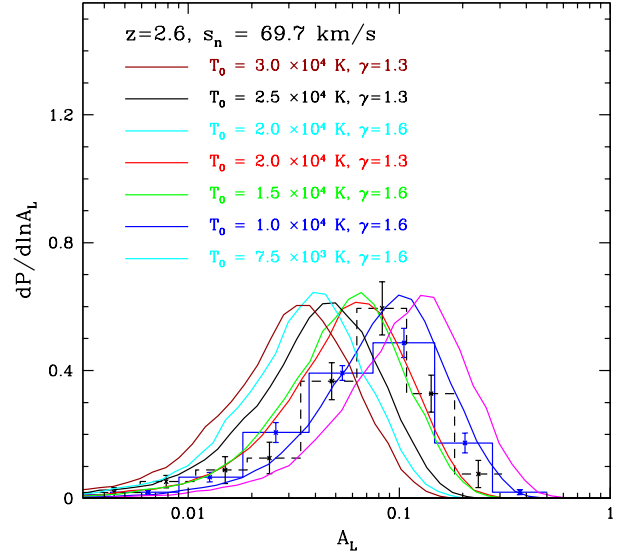}
\caption{Comparison between the measured wavelet PDF in the $\bar{z}=2.6$ bin with $s_m = 69.7$ km/s and simulated
models. Similar to Figure \ref{fig:pdf_wave_z4.2_sm_v_models}--Figure \ref{fig:pdf_wave_z3.0_sm_v_models}, but 
at $\bar{z}=2.6$.}
\label{fig:pdf_wave_z2.6_sm_v_models}
\ec 
\end{figure}

\begin{figure}
\bc
\includegraphics[width=9.2cm]{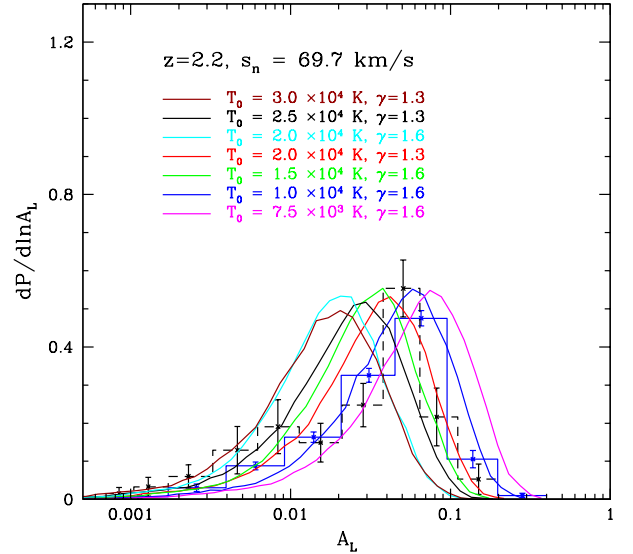}
\caption{Comparison between the measured wavelet PDF in the $\bar{z}=2.2$ bin with $s_m = 69.7$ km/s and simulated
models. Similar to Figure \ref{fig:pdf_wave_z4.2_sm_v_models}--Figure \ref{fig:pdf_wave_z3.0_sm_v_models}, but 
at $\bar{z}=2.2$.}
\label{fig:pdf_wave_z2.2_sm_v_models}
\ec 
\end{figure}

Let us first compare the measured PDF in the different redshift bins for $s_n = 69.7$ km/s. The results 
of these
calculations are shown in Figures \ref{fig:pdf_wave_z4.2_sm_v_models} -- \ref{fig:pdf_wave_z2.2_sm_v_models}.
We start with a qualitative `chi-by-eye' assessment, and provide more quantitative constraints in \S \ref{sec:therm_constrain}.

The blue histogram with error bars in Figure \ref{fig:pdf_wave_z4.2_sm_v_models} shows the
measured PDF at $\bar{z}=4.2$, uncorrected for metal line contamination. We have not identified
metal lines in the high redshift spectra contributing to this redshift bin, but we expect metal 
line contamination to have
only a small effect on the wavelet PDF at this redshift and smoothing scale (see Appendix B).
The model curves with $T_0 \sim 7,5000-10,000$ K and $\gamma=1.6$ correspond roughly to models in which
HI is reionized early, and HeII is not yet ionized. One expects a similarly low temperature in models
in which each of HI, HeI and HeII are all ionized early. 
Interestingly, these models produce
too many large wavelet amplitude regions and too few small wavelet amplitude regions compared to the data. 
The model curves with $T_0 = 15,000$ K, and each of $\gamma=1.3$ and $\gamma=1.6$ are fairly close to the 
measurements,
but overproduce slightly the high amplitude tail. These two curves are almost completely degenerate because
the wavelet amplitude PDF is sensitive to the temperature over only a limited range in overdensity. At this redshift the measurements
appear most sensitive to the temperature at densities near the cosmic mean, and so the models depend
sensitively on $T_0$ but not on $\gamma$. The model with $T_0 = 2 \times 10^4$ K, $\gamma=1.3$ is the
best overall match to the data of the models shown, although it over-predicts the point near $A_L \sim 0.4$ by more
than $2.5-\sigma$.
Finally, the model with $T_0 = 2.5 \times 10^4$ K seems to produce too many low
wavelet amplitude regions, and too few high amplitude pixels. It is also interesting that the measured PDF is not much
wider than the model PDFs. Taken at face
value, this argues against the temperature field being very inhomogeneous at this redshift. 

The results are tantalizing because they suggest the IGM is fairly hot 
with $T_0 \sim 15-20,000$ K at 
$z=4.2$. This requires some amount of early HeII photoheating and/or HI reionization to end late and 
heat the
IGM to a high temperature. If metal line contamination is in fact significant, 
this only strengthens the argument for a high temperature at $\bar{z}=4.2$:
metal lines can only add power and increase the number of high wavelet amplitude regions. 
Similarly, the finite resolution of our numerical simulations causes us to {\em underestimate} the IGM 
temperature (see \S \ref{sec:box_res}). While we show that our results are mostly converged with respect to
simulation resolution in \S \ref{sec:box_res}, convergence is most challenging at high redshift and this
may lead to a small systematic underestimate in this redshift bin.
On the other hand,
we show in \S \ref{sec:uncert} that a cooler IGM model ($T_0 \sim 10,000$ K) can match the PDF measurement at this
redshift if the true mean transmitted flux is $2-\sigma$ higher than the best fit value estimated by 
Faucher-Gigu\`ere et al. (2008b).

The measurements in our next redshift bin ($\bar{z}=3.4$) suggest the presence of even hotter gas in
the IGM (Figure \ref{fig:pdf_wave_z3.4_sm_v_models}). At this redshift the best overall match is the
model with $T_0 = 2.5 \times 10^4$ K, $\gamma=1.3$. Even a fairly hot model with 
$T_0 \sim 2 \times 10^4$ K, $\gamma=1.3$ produces too few low wavelet amplitude regions, and too many
high amplitude ones. Models with lower temperatures are clearly quite discrepant. At this redshift,
the measured PDF is a bit wider than the simulated ones. This might owe to temperature inhomogeneities,
or it may indicate some metal line contamination since, as with the $\bar{z}=4.2$ data, we have not
identified and excised metal lines in this redshift bin.  
In either of these cases, the measurements may allow for
some even hotter gas at $T_0 \sim 3 \times 10^4$ K.

The measurements at $\bar{z}=3.0$ indicate similarly hot gas (Figure \ref{fig:pdf_wave_z3.0_sm_v_models}). 
By this redshift, the average absorption in the forest is increased and the wavelet PDF is most sensitive
to gas a little more dense than the cosmic mean, at roughly $1 + \delta = \Delta \sim 2$ for our method. This
means that models that have a lower temperature at mean density ($T_0$), yet a steeper temperature-density
relation ($\gamma$) give similar wavelet PDFs to models with higher $T_0$ and flatter $\gamma$ at this
redshift. This explains why the model curves with $T_0 = 2.5 \times 10^4$ K, $\gamma=1.3$ and
$T_0 = 2 \times 10^4$ K, $\gamma=1.6$ are nearly identical to each other, as are the models with
$T_0 = 2 \times 10^4$ K, $\gamma=1.3$, and $T_0 = 1.5 \times 10^4$ K, $\gamma=1.6$. At this redshift,
the metal line correction appears fairly important: it shifts the peak of the PDF to lower amplitude
and narrows the histogram somewhat (as seen by comparing the black dashed histogram and the blue solid
histogram in the figure). The error bars are significantly larger for the metal-cleaned measurement than
for the full measurement. This is
mostly because we only have metal line identifications for some of the spectra in this bin and the
metal-cleaned measurement hence comes from a smaller number of spectra, and also because we use a smaller portion
of each spectrum after metal cleaning. The mean wavelet amplitude changes by less than the $1-\sigma$ error bar
as we vary $f_m$ between $f_m=0.8$ and $f_m=1$, and so $f_m=0.95$ is a conservative choice, and we hence stick to this choice
throughout.
After accounting for metal contamination, the model curves with
$T_0 = 2 \times 10^4$ K, $\gamma=1.3$, and $T_0 = 1.5 \times 10^4$ K, $\gamma=1.6$ are somewhat disfavored.
Again, the cooler models with $T_0 = 7,500-10,000$ K differ strongly with the measurement, regardless of
the metal correction. The hottest model shown with $T_0= 3.0 \times 10^4$ K, $\gamma=1.3$ produces too many low-amplitude,
and two few high amplitude regions. The models with $T_0 = 2.5 \times 10^4$ K, $\gamma=1.3$
and $T_0 = 2.0 \times 10^4$ K, $\gamma=1.6$ are strongly degenerate and each roughly match the measured PDF.

Proceeding to lower redshift, the data at $\bar{z}=2.6$ disfavor some of the hotter IGM 
models (Figure \ref{fig:pdf_wave_z2.6_sm_v_models}). 
At this redshift, the models shown with $T_0 = 3.0 \times 10^4$ K, $\gamma=1.3$; $T_0 = 2.5 \times 10^4$ K,
$\gamma=1.3$; $T_0 = 2.0 \times 10^4$ K, $\gamma=1.6$ all produce too many low amplitude regions, and too
few high amplitude ones. The other models shown with $T_0 = 2.0 \times 10^4$ K, $\gamma=1.3$; 
$T_0 = 1.5 \times 10^4$ K, $\gamma=1.6$, and $T_0 = 1.0 \times 10^4$ K, $\gamma=1.6$ are closer to
the measurements, although none of the models are a great fit. 
The cooler model with $T_0 \sim 7,500$ K is again a poor 
match to the measurement. The preference
for somewhat more moderate temperatures at this redshift may result from cooling after HeII reionization completes
at higher redshift.

Finally, the measurement in the $\bar{z}=2.2$ bin is shown in Figure \ref{fig:pdf_wave_z2.2_sm_v_models}. The general
features are similar to the results at $\bar{z}=2.6$: the hotter models are clearly a poor match to the data,
and there is some preference for cooler temperatures, although none of the models are a great match to the data.
The models with $(T_0, \gamma) = (2.0 \times 10^4 K, 1.3), (1.5 \times 10^4 K, 1.6)$ and $(1.0 \times 10^4 K, \gamma=1.6)$ are the closest 
matches of the models shown. 
At this redshift, the mean transmission is high ($\avg{F}=0.849$), and the method is sensitive to somewhat
overdense gas as a result. The similarity between the models with $T_0 = 3.0 \times 10^4$ K, $\gamma=1.3$ and
$T_0 = 2.0 \times 10^4$ K, $\gamma=1.6$ suggests that the PDFs are most sensitive to densities
around $\Delta = 3.9$ at this redshift. We expect scatter in the temperature density relation from shock-heating to
be most important at this low redshift, especially since the wavelet PDF is becoming sensitive to the temperature
of moderately overdense gas. This may be part of the reason for the poorer overall match between simulations, where the
effects of shocks on $T$ are ignored in post-processing,
and observations at this redshift. We will investigate this in more detail in the future. 

In summary, our measurements appear to support a picture where the IGM is being heated in the middle of the
redshift range probed by our data sample, with the 
temperature likely peaking between $z=3.0-3.8$, before cooling down towards lower redshifts.
The favored peak temperature appears to be around $T_0 \sim 25-30,000$ K, somewhat hotter than found by most previous
authors (see \S \ref{sec:compare_previous}), although consistent with theoretical expectations from photoheating during
HeII reionization, especially if the quasar ionizing spectrum is on the hard side of the models considered by
McQuinn et al. (2008) (see their Figure 12).

\subsection{Uncertainties in the Underlying Cosmology and Mean Transmitted Flux}
\label{sec:uncert}

In the previous section we showed model wavelet PDFs for varying temperature-density relations, but left
the underlying cosmology and mean transmitted flux fixed. Here we consider how much
the wavelet PDF varies with changes in these quantities. As far as the underlying cosmology is concerned, we 
restrict our discussion to uncertainties in the amplitude of density fluctuations. Note that there is 
some ($2-\sigma$ level)
tension between the amplitude of density fluctuations determined from the Ly-$\alpha$ forest and WMAP
constraints (Seljak et al. 2006).

\begin{figure}
\bc
\includegraphics[width=9.2cm]{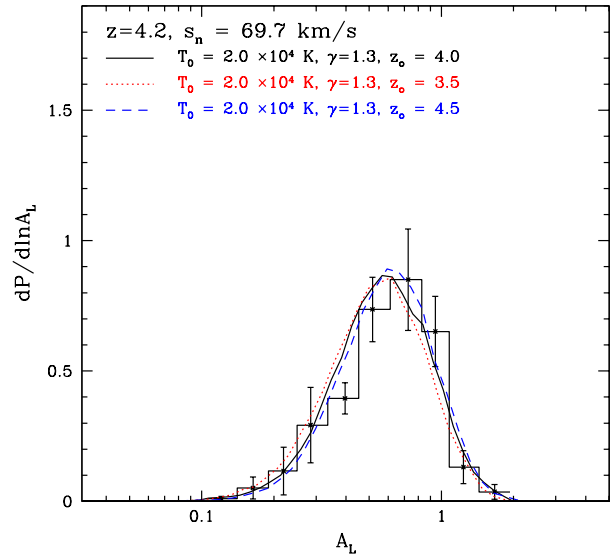}
\caption{Sensitivity of the wavelet PDF to the amplitude of underlying density fluctuations. The model
curves show the wavelet PDF for mock spectra generated using simulation snapshots at a range of redshifts for
an otherwise identical model. Snapshots at lower redshift approximate models in which the amplitude of
underlying density fluctuations is higher than our fiducial value, while the curve generated from the
$z=4.5$ model (blue dashed line) approximates a model with a lower amplitude of density fluctuations.
}
\label{fig:pdf_z4.2_v_sig8}
\ec
\end{figure}

In order to gauge the impact of uncertainties in the amplitude of density fluctuations, we generate mock spectra 
for a given model using simulation outputs of varying redshift. In particular, we consider a model at 
$\bar{z}=4.2$ with $\langle F \rangle = 0.346$, $T_0 = 2 \times 10^4$ K, and $\gamma=1.3$, which roughly
matches the measured PDF. We generate mock
spectra in this model from outputs at $z_o = 3.5, 4.0$, and $4.5$. For the prediction in our fiducial cosmology,
we linearly interpolate between wavelet PDFs generated from the $z=4.0$ and $z=4.5$ outputs. Using instead the 
model PDFs at $z_o = 3.5$ or $4.0$ (with the mean transmitted flux fixed at $\langle F \rangle = 0.346$) -- in
which structure formation is more advanced -- should mimic a model with a higher amplitude of density fluctuations,
while using the $z=4.5$ snapshot should correspond to a model with smaller density fluctuations. Our fiducial model
has $\sigma_8(z=0)=0.82$, roughly in between the preferred values inferred from the forest alone and that from WMAP-3
alone (Seljak et al. 2006). Using the outputs at $z_o = 3.5, 4.0$, and $4.5$ for the $\bar{z} = 4.2$ mock spectra 
should roughly correspond to models with $\sigma_8(z=0)=0.95,0.85$ and $0.78$ respectively. The results of these
calculations, shown in Figure \ref{fig:pdf_z4.2_v_sig8}, illustrate that the wavelet PDF is only weakly sensitive
to the underlying amplitude of density fluctuations. 
The mean small-scale power is exponentially sensitive to the temperature, which is uncertain at the factor of $\sim 2$
level, and so it is unsurprising that $\sim 10\%$ level changes in the amplitude of density fluctuations have relatively little impact.
The small effect visible in the plot is that the wavelet PDF
shifts to smaller amplitudes for the outputs in which structure formation is more advanced. This likely owes to
the enhanced peculiar velocities in models with larger density fluctuations, which suppress the small-scale
fluctuations in the forest via a finger-of-god effect (e.g. McDonald et al. 2006). The impact of uncertainties
in the amplitude of density fluctuations on the wavelet PDF are similarly small at other redshifts, and so we
do not discuss this further here. 

The amplitude of fluctuations in the forest, and the wavelet PDF, are sensitive to the mean transmitted flux and
uncertainties in this quantity impact constraints on the temperature from the PDF measurements.  The mean transmitted
flux partly determines the `bias' between fluctuations in the transmission and the underlying density fluctuations,
with the bias increasing as the mean transmitted flux decreases. This impacts the small-scale transmission power
spectrum, and the wavelet PDF, as well as fluctuations on larger scales. When the gas density is sufficiently
high, and/or the ionizing background sufficiently low -- i.e., when the mean transmitted flux is small -- even slight
density inhomogeneities produce absorption features, yielding large transmission fluctuations on small-scales.  

In the previous section, we adopted the best fit values of the mean transmitted flux 
from Faucher-Gigu\`ere et al. (2008b),
but now consider variations around these values. These authors provide estimates of the statistical and systematic
errors on their mean transmitted flux measurements. Their $1-\sigma$ errors at our bin centers are:
$(z, \avg{F} \pm 1-\sigma) = (2.2, 0.849 \pm 0.017), (2.6, 0.778 \pm 0.017), (3.0, 0.680 \pm 0.02), 
(3.4, 0.566 \pm 0.022), (4.2, 0.346 \pm 0.042)$. 
Their systematic error budget accounts for uncertainties in estimating metal line contamination, 
and for uncertainties in corrections related to the rarity of true unabsorbed regions at high redshift, among
other issues. Nonetheless, there is some tension between the measurements of different groups. We
refer the reader to Faucher-Gigu\`ere et al. (2008b) for a discussion. 

\begin{figure}
\bc
\includegraphics[width=9.2cm]{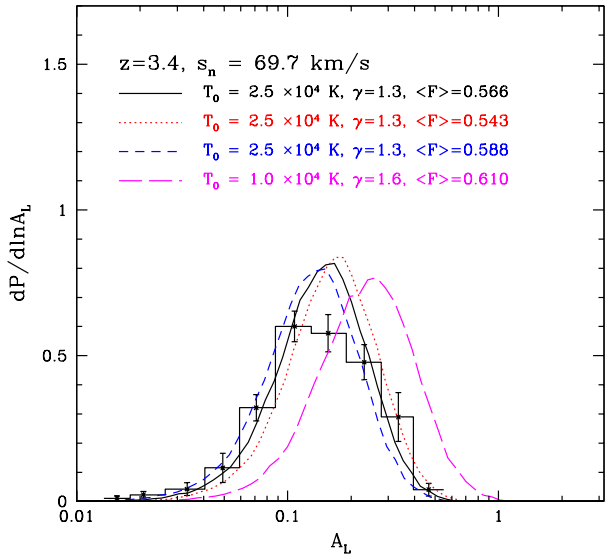}
\caption{Impact of uncertainties in the mean transmitted flux at $\bar{z}=3.4$. The black solid line shows the 
wavelet PDF
in a model with the best fit mean transmitted flux from Faucher-Gigu\`ere et al. (2008b). The red dotted line
shows the same model, except adopting a mean transmitted flux that is $1-\sigma$ less than the best fit value.
The blue dashed line shows the same, except for a mean transmitted flux $1-\sigma$ larger than the best fit.
The magenta line shows a cooler IGM model, where the mean transmitted flux is $2-\sigma$ higher than the best fit.}
\label{fig:pdf_z3.4_v_amf}
\ec
\end{figure}

Below $z \lesssim 4$ uncertainties in the mean transmitted flux have a noticeable yet fairly small impact on 
our constraints.
A typical example, in the $\bar{z} = 3.4$ redshift bin, is shown in Figure \ref{fig:pdf_z3.4_v_amf}. The solid black
line in the figure shows a model with $T_0 = 2.5 \times 10^4$ K, $\gamma=1.3$ that adopts the best fit value for the
mean transmission, $\avg{F}=0.566$. The blue dashed line is the same model, but with the mean transmitted flux
shifted up from the central value by $1-\sigma$. This reduces the amplitude of transmission fluctuations in the 
model, and shifts the wavelet PDF towards slightly lower amplitudes. Reducing the transmission by $1-\sigma$ has
the opposite effect of boosting the typical wavelet amplitudes slightly, as illustrated by the red dotted line
in the figure. While the uncertainty in the mean
transmitted flux can shift the preferred temperature around slightly, the effect at this redshift is relatively small
and has little impact on our main conclusions. For example, a cooler IGM model with $T_0 = 1.0 \times 10^4$ K 
and $\gamma=1.6$ still differs greatly from the PDF measurement, even after assuming a mean transmitted flux that
is $2-\sigma$ higher than the central value. This is demonstrated by the magenta line 
in Figure \ref{fig:pdf_z3.4_v_amf}.

\begin{figure}
\bc
\includegraphics[width=9.2cm]{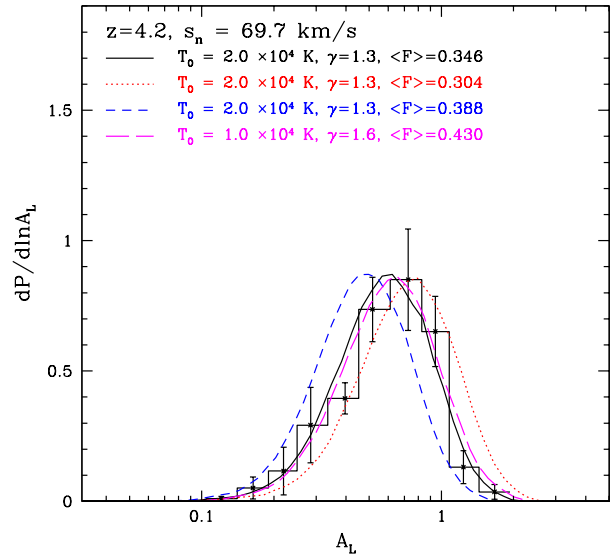}
\caption{Impact of uncertainties in the mean transmitted flux at $\bar{z}=4.2$. Similar to 
Figure \ref{fig:pdf_z4.2_v_amf}, except at $\bar{z}=4.2$.}
\label{fig:pdf_z4.2_v_amf}
\ec
\end{figure}

The impact of uncertainties in the mean transmitted flux is more important in our highest redshift bin,
at $\bar{z}=4.2$. The impact is larger at this redshift both because data samples are smaller and 
the fractional error on the mean transmitted
flux is larger at this redshift, and because the wavelet amplitudes are more sensitive to the mean transmission
once the transmission is sufficiently small. We repeat the exercise of the previous figure at $\bar{z}=4.2$ and
present the results in Figure \ref{fig:pdf_z4.2_v_amf}. In this case, the model that roughly goes through the
PDF measurement with our best fit mean transmitted flux has $T_0 = 2.0 \times 10^4$ K, and $\gamma=1.3$. After shifting
the mean transmitted flux up in this model by $1-\sigma$ it produces too many low wavelet amplitude regions, and
too few high amplitude ones, in comparison to the measurement. Indeed, at this redshift, even the cooler IGM model with
$T_0 = 1.0 \times 10^4$ K, $\gamma=1.6$ will pass through the measurement after a $2-\sigma$ upwards shift in the
mean transmitted flux. In other words, accounting for uncertainties in the mean transmitted flux, the cool IGM model
with $T_0 = 1.0 \times 10^4$ K, $\gamma=1.6$ can only be excluded at roughly the $2-\sigma$ level. 

Furthermore, systematic concerns with direct continuum-fitting are most severe at high redshift, and the agreement
between different measurements, while generally good at lower redshifts, is marginal 
above $z \sim 4$ or so (Faucher-Gigu\`ere et al. 2008b). Direct continuum 
measurements must correct for the fact that there are few genuinely unabsorbed regions at high redshift, which 
can cause
one to systematically underestimate the mean transmitted flux. Part of the disagreement can be 
traced to the fact that some of the measurements in the literature do not make this important correction.
Since Faucher-Gigu\`ere et al. (2008b) make a correction using cosmological simulations, we consider their
measurement to be more reliable than many of the other previous measurements.
However, McDonald et al. (2006) constrain the mean transmitted flux based
on a multi-parameter fit to their SDSS power spectrum measurements, which should be immune to this concern.
Their best fit to the redshift evolution of the mean transmitted flux 
gives $\avg{F} = 0.41$ at $z=4.2$. This disagrees with the Faucher-Gigu\`ere et al. (2008b) measurement at this 
redshift by $1.6-\sigma$. The overall disagreement is in fact more severe than this, because there is a similar
level of disagreement in neighboring redshift bins.
Dall'Aglio et al. (2008) also perform a direct continuum-fit, correct for the
rarity of unabsorbed regions at high redshift with a different methodology, and find a best fit to the redshift evolution of the opacity
of $\avg{F}=0.40$ at $z=4.2$. Again, this measurement is in tension with the measurement we adopt. Adopting 
either of these
measurements for the best fit mean transmitted flux would favor a cooler IGM temperature.

\subsection{Dependence on Large-Scale Smoothing}
\label{sec:smooth_large}

The measured PDF in the $\bar{z}=3.4$ redshift bin requires hot ($T_0 \gtrsim 20,000$ K) gas. Interestingly, the PDF in this
redshift bin is somewhat broader than the theoretical model curves, which assume a homogeneous temperature-density relation. 
This may be the result of uncleaned metal line contamination, but a more interesting possibility is that the wide measured
PDF indicates temperature inhomogeneities from ongoing HeII reionization. We argued in \S \ref{sec:smoothing} that
the precise choice of large scale smoothing, $L$, should be relatively unimportant. Nevertheless, to further explore
the exciting possibility that the data indicate temperature inhomogeneities in this redshift bin, we measure the PDF
for a few additional choices of $L$ and compare with theoretical models.

\begin{figure}
\bc
\includegraphics[width=9.2cm]{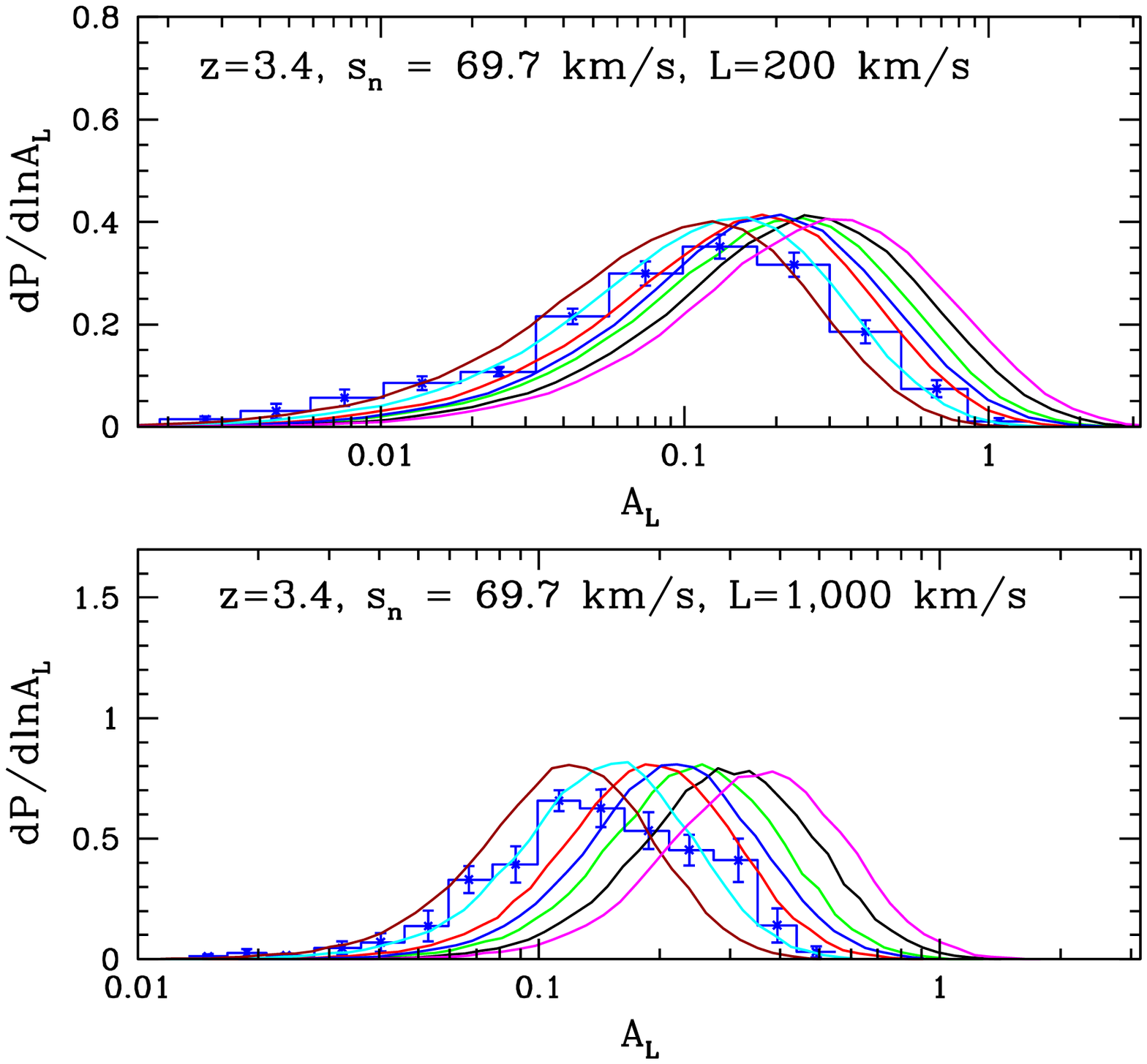}
\includegraphics[width=9.2cm]{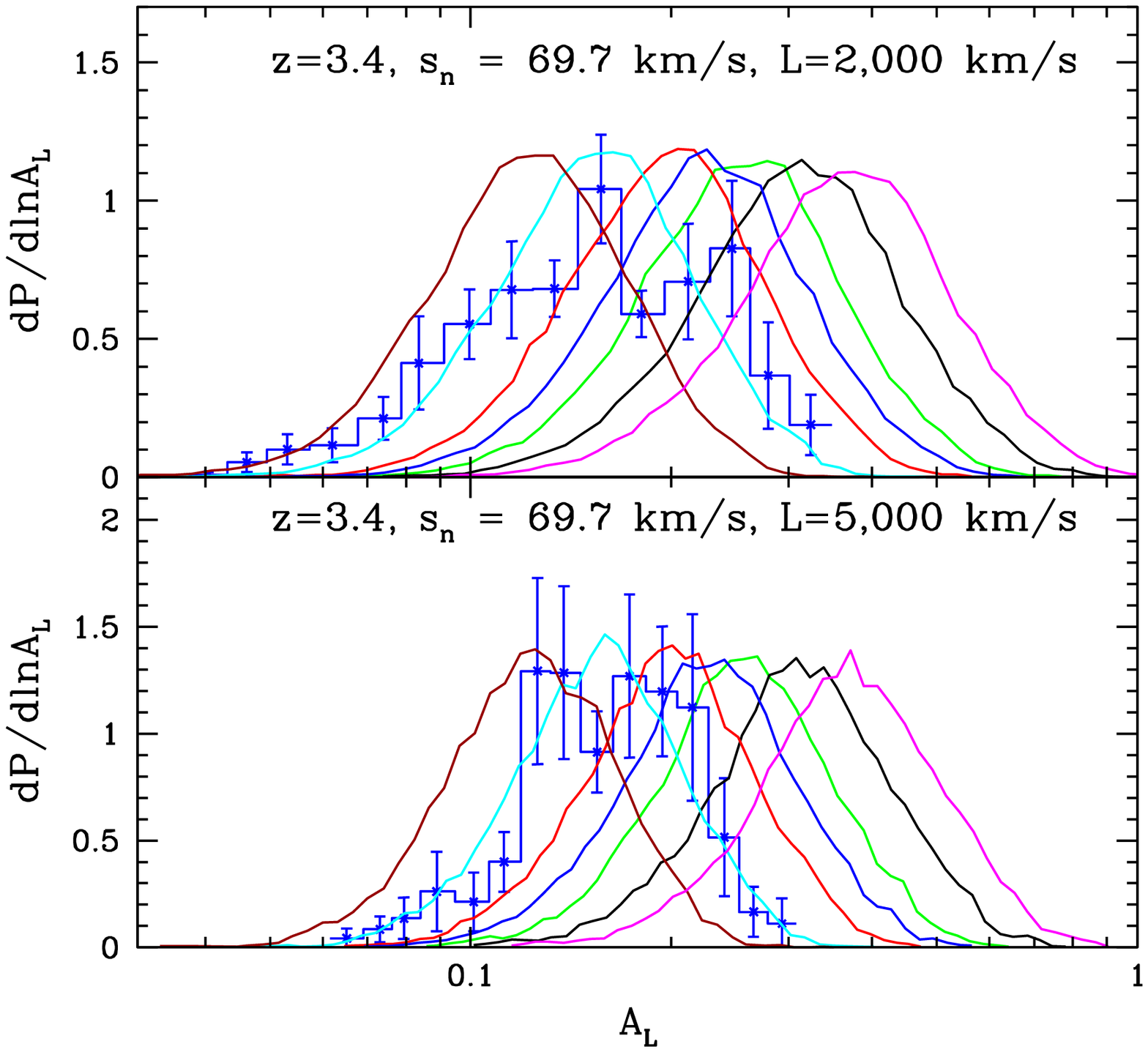}
\caption{Wavelet PDF at $\bar{z}=3.4$ as a function of large scale smoothing, $L$. The blue histogram in the
panels is the wavelet PDF for a large-scale smoothing $L$ of {\em Top}--{\em Bottom}: $200; 1,000; 2,000$; and $5,000$ km/s.
The color code for the different temperature-density relation models is identical to that in Figure \ref{fig:pdf_wave_z3.4_sm_v_models}.}
\label{fig:pdf_z3.4_lsmooth}
\ec
\end{figure}

The results of these calculations are shown in Figure \ref{fig:pdf_z3.4_lsmooth}. In addition to our usual large scale smoothing of $L=1,000$ km/s,
we also compare simulated and observational wavelet PDFs for $L=200; 2,000;$ and $5,000$ km/s. Here we use $15$ logarithmically 
spaced $A_L$ bins
for the PDF measurement, rather than $10$ as in the previous sections, to increase our sensitivity to any bi-modality in the PDF. 
The mean of the model curves with different smoothing scales is of course fixed, while the width of the PDF
increases with decreasing smoothing scale (see \S \ref{sec:smoothing}, Figure \ref{fig:power_ampsq}).
At all smoothing scales, the simulated
model with $T_0 = 25,000$ K and $\gamma=1.3$ is the best overall match to the data. The fit is poorest
at $L=2,000$ km/s, but it is not clear precisely how to interpret this since the model is a formally poor
fit at each smoothing scale.
There does appear to be a slight, yet tantalizing, hint that the PDF is bimodal on large smoothing 
scales: this trend is most apparent at $L=2,000$ km/s and $L=5,000$ km/s. 
This may be
a first indication of temperature inhomogeneities from ongoing HeII reionization, or it may be the result of uncleaned metal line contamination, as the abundance of metals can vary significantly on large smoothing scales. 
It will
be interesting to revisit this measurement with larger data samples in the future.

\subsection{Dependence on Small-Scale Smoothing}
\label{sec:smooth_small}

We found in the previous sections that our results at $s_n=34.9$ km/s are quite susceptible to metal-line contamination and
somewhat to shot-noise bias. Because of this, we will not presently use the results at this smoothing scale 
in constraining the thermal history of the IGM. Nevertheless,
as a consistency check we compare here the measured wavelet PDF at this smoothing scale with simulated models. 

\begin{figure}
\bc
\includegraphics[width=9.2cm]{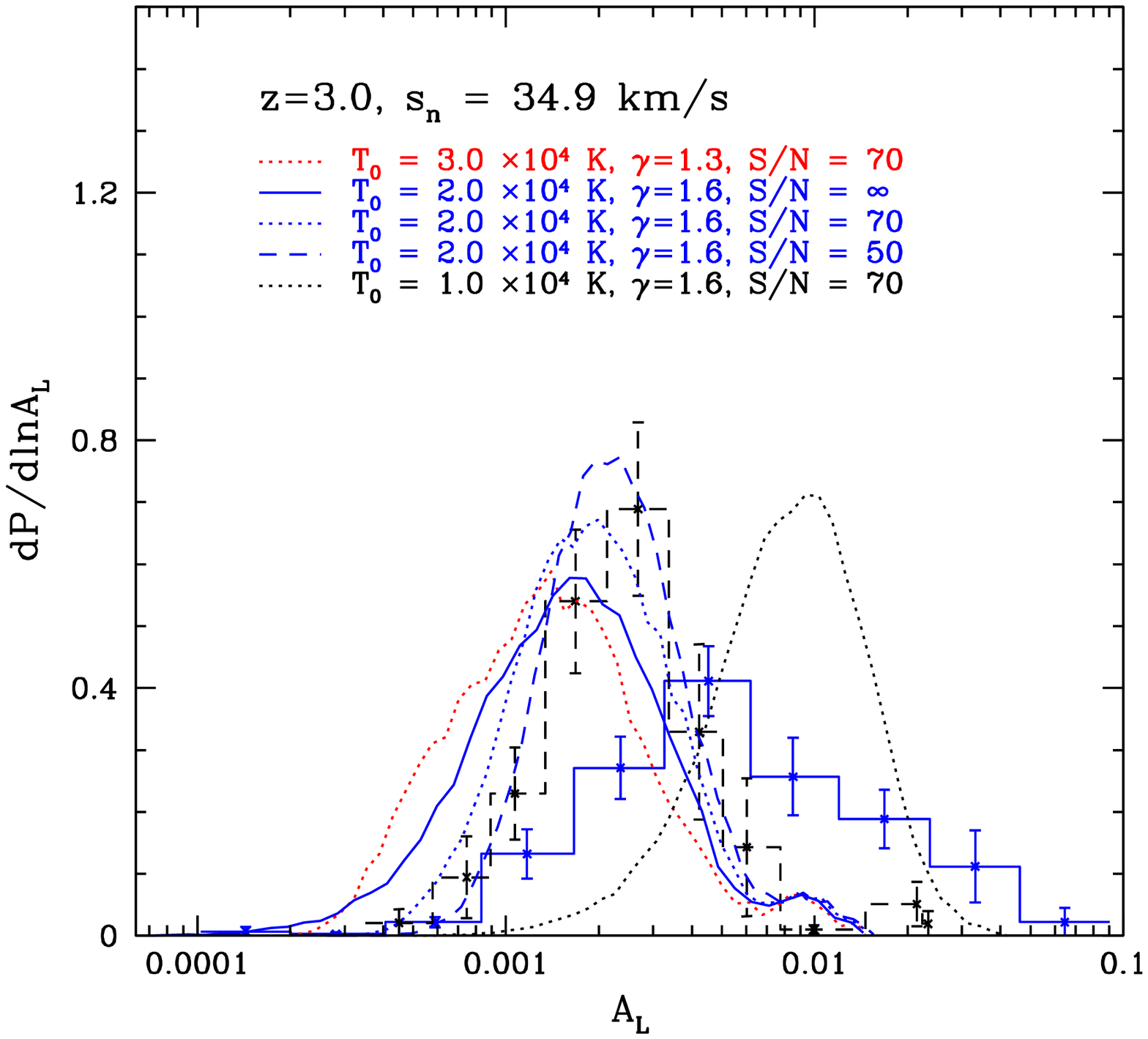}
\includegraphics[width=9.2cm]{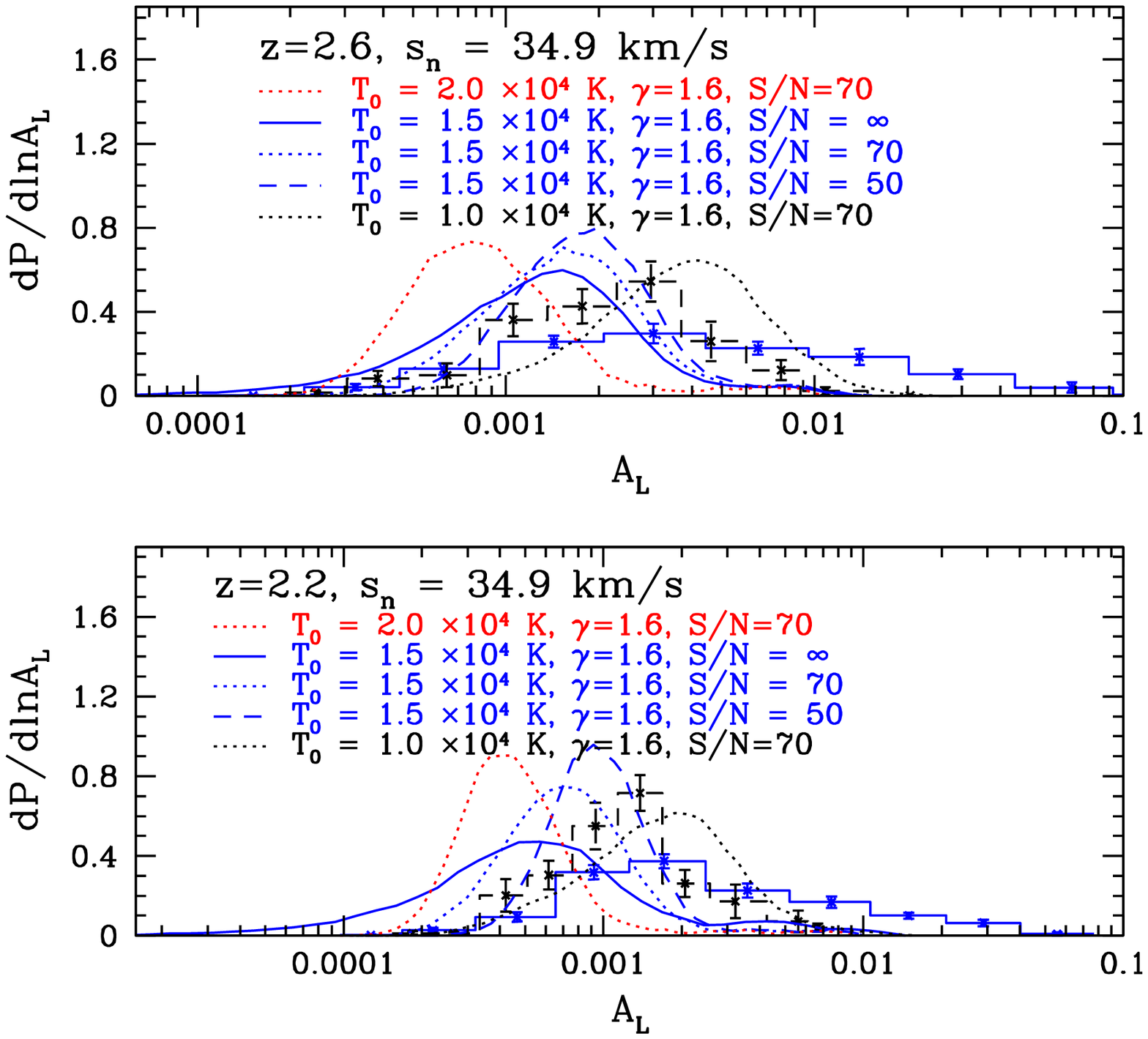}
\caption{Wavelet PDFs for the $s_n=34.9$ km/s filter at $\bar{z}=3.0, 2.6$ and $\bar{z}=2.2$ (from {\em Top} -- {\em Bottom}).
Similar to previous plots at $s_n=69.7$ km/s, the black dashed histograms with error bars show the measured wavelet PDFs, corrected
for metal line contamination. The blue solid histogram is the same, without masking metal lines. A few example model curves are
shown at each redshift, with random noise added to the mock spectra. The models that match the measurements at this smoothing are
similar to the ones at the larger smoothing scale.}
\label{fig:pdf_small_vz}
\ec
\end{figure}

As mentioned previously, to guard against shot-noise bias, we cut spectra with a (red-side) $S/N \leq 50$ and add random noise
to the mock spectra. Provided we cut out the low $S/N$ data, the random noise mainly impacts only the low wavelet amplitude tail by
decreasing the number of very low amplitude wavelet regions. We add Poisson distributed noise to the mock spectra, assuming that
the noise is dominated by Poisson fluctuations in the photon counts from the quasar itself. We have 
experimented with incorporating Poisson
distributed sky noise, and Gaussian random read-noise, and find qualitatively similar results at fixed noise level.
We estimate the average wavelet amplitude in the forest contributed by noise (after our $S/N$ cut) as described in Appendix
A, and find that it corresponds to $S/N \sim 70$, per $4.4$ km/s pixel at the continuum for the $\bar{z}=3.0$ bin.   In Figure
\ref{fig:pdf_small_vz} we compare some example model PDFs with the measurements, and find results gratifyingly close to
those at larger smoothing scale. In particular, the model with $T_0 = 2.0 \times 10^4$ K, and $\gamma=1.6$ at $\bar{z}=3.0$ that
roughly matched the measurement on larger scales, matches the PDF on this smaller scale as well. For contrast, we show a hotter
and a colder IGM model which are again a poor match. At $\bar{z}=2.2$ and $\bar{z}=2.6$ the results are similar to the
previous ones, suggesting a cooler IGM at these redshifts.
Comparing the blue and black dashed histograms, it is clear that the metal contamination
correction is quite important at this scale and we do not use these results in what follows.

We have also compared the $s_n=34.9$ km/s wavelet PDF in the two highest redshift bins -- where we have not identified metal lines --
with model PDFs. The measured PDF at $z=3.4$ looks similar to the $T_0 = 2.5 \times 10^4$ K, $\gamma=1.3$ model that we previously 
identified
as the best general match of our example models at $s_n=69.7$ km/s, except with a fairly prominent tail towards high 
wavelet amplitudes. We expect more significant metal contamination at this smoothing scale (Appendix B), and so this is in line
with our expectations. Indeed, the tail towards high wavelet amplitude 
looks similar to the one in the top panel of Figure \ref{fig:mock_metals_pdf}.
Similar conclusions hold at $z=4.2$, except the agreement without excising metals is better, likely owing to the smaller impact
of metals at this redshift (Appendix B).

\subsection{Approximate Constraints on the Thermal History of the IGM}
\label{sec:therm_constrain}

In this section, we perform a preliminary likelihood analysis, in order to provide a more quantitative constraint on the thermal history
of the IGM from the wavelet measurements. 
We confine our analysis to a
three-dimensional parameter space, spanning a range of values for $T_0, \gamma$, and $\avg{F}$. The results of the previous section 
suggest that
CDM models close to a WMAP-5 cosmology should all give similar wavelet PDFs, and so it should be unnecessary to vary the cosmological 
parameters in this analysis. 
In order to facilitate this calculation, we adopt here an approximate approach to cover the
relevant parameter space. We generate the wavelet PDF for a range of models by expanding around 
a fiducial
model in a first order Taylor series (see Viel \& Haehnelt 2006 for a similar approach applied to SDSS
flux power spectrum data).  In particular let $\bbox{p}$ denote a vector in the three-dimensional parameter space. Then
we calculate the wavelet PDF at a point in parameter space assuming that:
\beqa
P(A_L,\p) = && P(A_L,\pz) + \nonumber \\ && \sum_{i=1}^3 \frac{\partial P(A_L, p_i)}{\partial p_i} 
\bigg|_{\p=\pz} (p_i - p^0_i).
\label{eq:taylor}
\eeqa

Although inexact, this approach suffices to determine degeneracy directions, approximate confidence intervals, and the main trends with
redshift.
We use the results of the previous section to choose the fiducial model to expand around: at each 
redshift
we choose the best match of the example models in the previous section
as the fiducial model. Using the Taylor expansion approximation of Equation \ref{eq:taylor}, we then estimate the wavelet PDF for a large range of 
models, spanning $T_0 = 5,000-35,000$ K, $\gamma=1.0-1.6$, and $\avg{F} = F_c \pm 3 \sigma_F$ (subject to a $\avg{F}$ prior). 
Here $F_c$ denotes the central value
from Faucher-Gigu\`ere et al. (2008b), and $\sigma_F$ denotes their estimate of the $1-\sigma$ uncertainty on the mean transmitted flux. For each
model PDF in the parameter space, we first compute $\chi^2$ between the model and the wavelet PDF data, ignoring 
off-diagonal terms in the co-variance matrix. We then add to this $\chi^2$ an additional term to account
for the difference between the model mean transmitted flux and the best fit value of Faucher-Gigu\`ere et al. (2008b). Finally, we marginalize over $\avg{F}$ (subject to the above prior based on the results of Faucher-Gigu\`ere et al. (2008b))
to compute two-dimensional likelihood surfaces in the $T_0 - \gamma$ plane
at each redshift, and marginalize over $\gamma$ to obtain reduced, one-dimensional 
likelihoods
for $T_0$. We assume Gaussian statistics, so that $1-\sigma$ ($2-\sigma$) two-dimensional likelihood regions
correspond to $\Delta \chi^2 = 2.30 (6.17)$, while one-dimensional constraints correspond to 
$\Delta \chi^2 = 1 (4)$. 

The best fit models at $z = 4.2, 3.4, 3.0, 2.6,$ and $2.2$ have $\chi^2 = 9.5, 19.8, 5.7, 8.0,$ and $23.1$ respectively 
for $7$ degrees of freedom (10 $A_L$ bins minus
$1$ constraint since the PDF normalizes to unity, minus two free parameters). The fits at $z=4.2$, $3.0$, and $2.6$ are 
acceptable, while the $\chi^2$ values in the $z=3.4$ and $z=2.2$ bins are high
(p-values of $6 \times 10^{-3}$ and $2 \times 10^{-3}$ respectively). The poor $\chi^2$ in these redshift 
bins results because the measured
PDFs are broader than the theoretical models in these bins, as discussed previously. We will nevertheless consider how $\chi^2$ changes
around the best fit models in these redshift bins, although we caution against taking the results too literally.

\begin{figure}
\bc
\includegraphics[width=9.2cm]{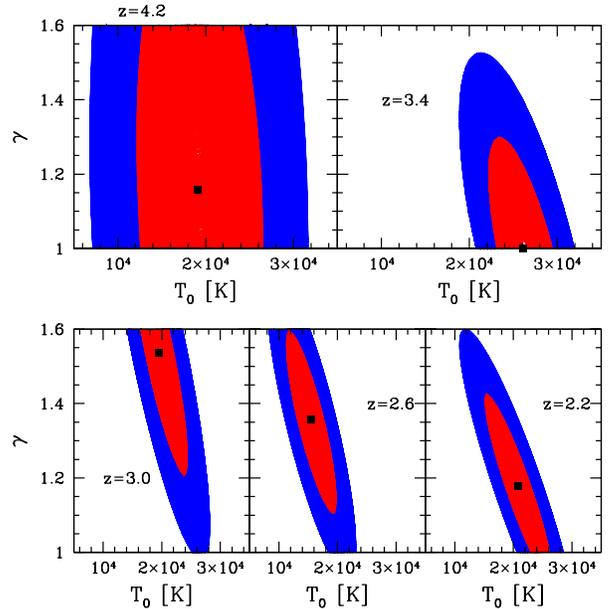}
\caption{Approximate constraints in the $T_0 -\gamma$ plane. The panels show $1-\sigma$ (red) and $2-\sigma$ (blue) constraints
in the $T_0-\gamma$ plane at different redshifts, marginalized over the mean transmitted flux.}
\label{fig:tzero_gamma_plane}
\ec
\end{figure}

\begin{figure}
\bc
\includegraphics[width=9.2cm]{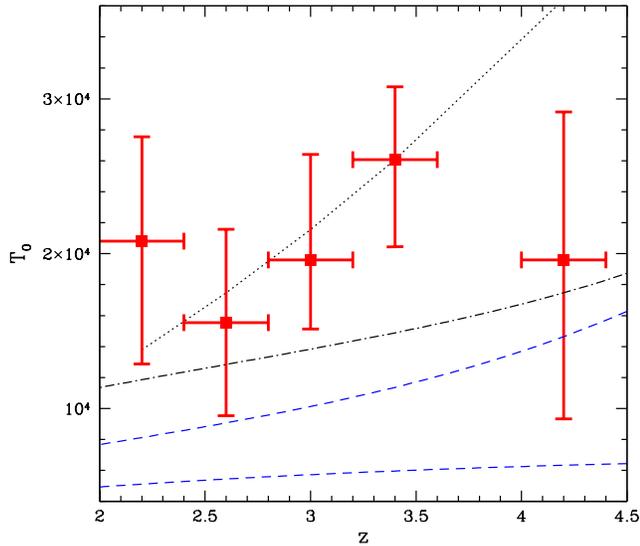}
\caption{Approximate constraints on $T_0$ as a function of redshift. The red points and error bars show $2-\sigma$ constraints on the 
temperature
at mean density in each redshift bin, after marginalizing over $\avg{F}$ and $\gamma$ at each redshift.  
The dotted, dashed and dot-dashed lines are for comparison.
The black dotted line varies as $(1+z)^2$, after passing through the highest temperature
point at $z=3.4$. The upper blue dashed line shows a model in which
HI/HeI reionization completes late, the IGM is reionized to a high temperature, and HeII is not yet reionized. The lower blue dashed line
is similar, except in this case HI/HeI reionize early. The black dot-dashed line is for a model in which HI/HeI/HeII are all reionized
together at $z=6$ by sources with a quasar like spectrum.  This curve is roughly an {\em upper limit} to the temperature without
late time HeII reionization. 
A flat $T_0 \sim 20,000$ K thermal history is consistent within the errors, but an implausibly hard ionizing spectrum is required to achieve
such a high temperature from residual photoheating after reionization.
This comparison suggests late time HeII reionization, perhaps completing 
near $z \sim 3.4$.}
\label{fig:tzero_v_z}
\ec
\end{figure}

The constraints from these calculations are shown in Figure \ref{fig:tzero_gamma_plane} and Figure \ref{fig:tzero_v_z}. They are 
qualitatively consistent with the example models shown in the previous section. The degeneracy
direction of the constraint ellipses results because the $z=4.2$ measurements are sensitive
only to the temperature close to the cosmic mean density, while the lower redshift measurements 
start to
constrain only the temperature of more overdense gas. The best fit model at $z=4.2$ has $T_0 \sim 20,000$ K, but
uncertainties in the mean transmitted flux allow cooler models with $T_0 \sim 10,000$ K at $\sim 2-\sigma$, as discussed
previously. The
$z=3.4$ measurements indicate the largest temperatures, and require that $T_0 \gtrsim 20,000$ K at $2-\sigma$ 
confidence. The lower redshift measurements, particularly that at $z=2.6$, generally favor cooler temperatures although at only
moderate statistical significance.

Figure \ref{fig:tzero_v_z} shows ($2-\sigma$) constraints on the temperature at mean density after
marginalizing over $\gamma$ and $\avg{F}$. 
We conservatively allow $\gamma$ to vary over $\gamma=1.0-1.6$, even though $\gamma \gtrsim 1.2$ is 
expected theoretically (McQuinn et al. 2008). If we enforced a prior that $\gamma$ be steeper than $1.2$,
then the results at $z \lesssim 3.4$ would disfavor some of the higher $T_0$ models.
The $T_0$ results are consistent with the IGM temperature
falling off as $T_0 \propto (1+z)^2$ below $z=3.4$, i.e., below this redshift the temperature 
evolution appears
consistent with simple adiabatic cooling owing to the expansion of the Universe. Theoretically, we expect
the temperature fall-off to be similar, but slightly slower, than the adiabatic case just after reionization with
the temperature evolution eventually slowing owing to residual photoionization heating (Hui \& Gnedin 1997).
The statistical errors are however still large, and a flat temperature evolution is also consistent
with the $T_0$ constraints, although this case is disfavored theoretically (see below). 
Note also that enforcing a $\gamma \geq 1.2$ prior would disfavor the high $T_0$ models
that are otherwise allowed at $z=2.2$ and $z=2.6$, strengthening the case for cooling below $z \sim 3.4$.

Moreover, the high temperatures at $z=3.0$ and $z=3.4$ suggest recent HeII photoheating. To illustrate this point, we show
several example thermal history models in Figure \ref{fig:tzero_v_z}, considering both cases without any HeII photoheating, and
ones in which HI/HeI/HeII are all reionized together at high redshift ($z \geq 6$).
The upper blue dashed
line is a late HI reionization model ($z_r=6$), with a high temperature at reionization ($T_r = 3 \times 10^4 K$),  and a hard
spectrum near the HI/HeI ionization thresholds (with a specific intensity near threshold of 
$J_\nu \propto \nu^{-\alpha}$ and $\alpha=0$). This case 
should roughly indicate the highest possible temperature without HeII photoheating over the redshift range probed. Note that this is a
rather extreme situation, since even if reionization completes as late as $z=6$, much of the volume will be reionized significantly earlier
(e.g. Lidz et al. 2007).
The lower blue
dashed line is an early reionization model ($z_r=12$ and $\alpha=2$) that approximately indicates the lowest plausible temperature
without HeII photoheating. 

Finally, perhaps the most interesting case is the black dot-dashed line which shows a model in which
HI/HeI/HEII are all reionized together at $z=6$. Here we assume that the temperature at reionization is $T_r = 3 \times 10^4 K$, since
atomic hydrogen line cooling should keep the temperature less than this when all three species are ionized simultaneously 
(Miralda-Escud\'e \& Ress 1994, Abel \& Haehnelt 1999, Lidz et al., in prep). The temperature after reionization depends on
the ionizing spectrum, which determines the amount of residual photoheating.
The curve here adopts a quasar like spectrum,
reprocessed by intervening absorption, to give $\alpha_{\rm HI} = 1.5$ near the HI ionization threshold and $\alpha_{\rm HeII} = 0$ near
the HeII ionization threshold (Hui \& Haiman 2003). This case is hence similar to the other $z=6$ reionization model, except with
the addition of residual HeII photoheating. 
Each of the examples considered 
gives too low a temperature in the $z=2.2$, $z=3.0$, and $z=3.4$ redshift bins, particularly at $z=3.4$ and $z=3$. 
One can further ask how hard the post-reionization ionizing spectrum would need to be to give a thermal asymptote as large as
$\sim 20,000$ K. For a power law spectrum we find, using the thermal asymptote formula of Hui \& Haiman (2003), 
that an implausibly hard spectrum with $\alpha \lesssim -0.73$ is required to match the $2-\sigma$ lower limit on the $z=3.4$ temperature.
In fact, there is evidence that galaxies rather than quasars produce most of the ionizing background at $z \gtrsim 3$ 
(e.g. Faucher-Gigu\`ere et al. 2008b), and so assuming even a quasar like spectrum likely overestimates residual photoheating for
plausible early HeII reionization models. In summary, although the errors allow the possibility of a slow 
temperature evolution and $T_0 \sim 20,000$ K,
this temperature is higher than expected from residual photoheating long after reionization.

The simplest interpretation is that HeII reionization 
heats the IGM, with the
process completing near $z \sim 3.4$, at which point there is relatively little additional heating 
and the 
Universe expands and cools. The redshift extent over which the heat input occurs is, however, not well 
constrained by our present measurement. Clearly the large error bars on the measurements still leave room for other possibilities.
For example, models in which HeII reionization completes a bit 
later at $z \sim 3$ -- or perhaps even as late as $z \sim 2.7$ as favored by a recent
analysis of HeII Ly-$\alpha$ forest data by Dixon \& Furlanetto 2009 -- or earlier at $z \sim 4$ are likely consistent with our present
measurements given the large error bars. We will consider this further in future work. Finally, other heating mechanisms may be at work
in addition to photoionization heating.

\subsection{An Inverted Temperature-Density Relation?}
\label{sec:tinverted}

Recently, Bolton et al. (2008), Becker et al. (2007) and Viel et al. (2009) have suggested that measurements
of the Ly-$\alpha$ flux PDF favor an inverted temperature density relation ($\gamma < 1$), i.e., situations 
where low
density  gas elements are hotter than overdense ones. Bolton et al. (2008) and Viel et al. (2009) construct
simulated models with inverted temperature-density relations by adding heat into the simulations in a
way that depends on the local density, i.e., on the density smoothed on the Jeans scale. This particular
case for an inverted temperature-density relation seems unphysical to us since heat input from, e.g. 
reionization, should be coherent on much larger scales. 
Nonetheless, we can consider this as a phenomenological example that the flux PDF data favor, and
examine the implications of these models for the small-scale wavelet amplitudes. 
Theoretically, Trac et al. (2008) and Furlanetto \& Oh (2009) find that hydrogen reionization does produce
a weakly inverted temperature-density relation. This effect is driven by the tendency for 
large-scale overdensities to reionize hydrogen first, coupled presumably with the small correlation between 
the overdensity on large scales and that on the Jeans scale. On the other hand, McQuinn et al. (2008) find 
that HeII
reionization leads to a {\em non-inverted} equation of state with $\gamma \sim 1.3$ in the midst and 
at the end of HeII reionization. We refer the reader to this paper for further discussion.

\begin{figure}
\bc
\includegraphics[width=9.2cm]{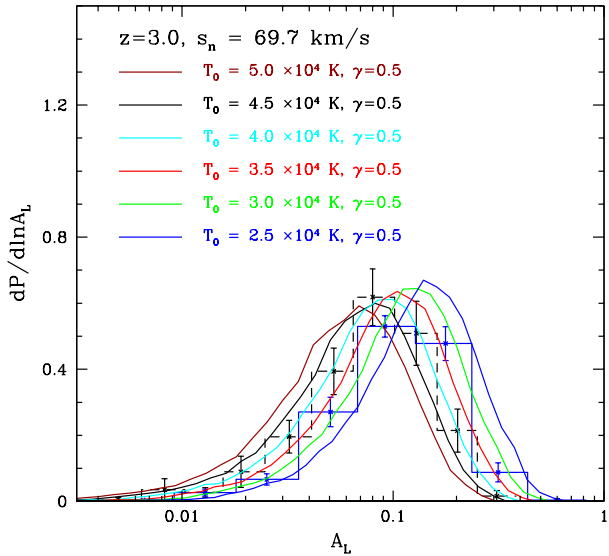}
\caption{Wavelet PDF in inverted temperature density relation models compared to measurements. The dashed
histogram shows the metal line corrected wavelet PDF at $z=3$ ($L=1,000$ km/s, $s_n=69.7$ km/s), and the blue
histogram is the same without correcting for metal line contamination. The colored lines show several
models with $\gamma=0.5$. One can fit the PDF with an inverted temperature density relation,
but this requires an extremely high temperature at mean density.
}
\label{fig:pdf_v_inverted}
\ec
\end{figure}

To explore this, we generate mock
spectra and measure wavelet amplitudes for several inverted temperature-density 
relation models and compare with our $z=3$ measurements. As before, we are considering the
impact of the temperature in a post-processing step, and so we are not accounting for differences between
the gas pressure smoothing in the inverted models and that in the simulation. 
Likewise, we incorporate thermal broadening assuming a perfect temperature density relation, and so 
the impact
of scatter in the temperature density relation is ignored in this part of the calculation. 
We consider inverted temperature density relations with a power law index 
of $\gamma=0.5$, close to the value suggested by Bolton et al. (2008) and
Viel et al. (2009) from their flux PDF measurements near $z=3$. The results of these calculations are 
shown in 
Figure \ref{fig:pdf_v_inverted}. These cases
also roughly match the observed PDF, but require a very high temperature at mean 
density of $T_0 \sim 40-45,000$ K. 
The reason for this is that the wavelet PDF measurements are sensitive mostly to the temperature around
a density of $\Delta \sim 2$ at this redshift. In the previous section we found that models with, for 
example, 
$T_0 \sim 25,000$ K
and $\gamma=1.3$ roughly match the data. A model with an inverted temperature density relation ($\gamma=0.5$) produces
the same temperature at a density of $\Delta \sim 2$ only for a much higher temperature (at mean density) 
of $T_0 \sim 45,000$ K.  The figure suggests that the expected degeneracy between $T_0$ and $\gamma$ indeed
extends to even these inverted temperature-density relations. Hence one can fit the measurements with a 
very high $T_0$, small $\gamma$ model, although the inverted cases produce slightly wider PDFs. 
While these can fit the data, the high required temperatures seem unlikely to us, and we disfavor
inverted models for this reason.   

Bolton et al. (2008) and Viel et al. (2009) found that
inverted models with substantially smaller temperature at mean density match their flux PDF measurements. 
On the other hand, Viel et al.
(2009) did a joint fit to the flux PDF and the SDSS flux power spectrum from McDonald et al. (2006). Recall that the SDSS
measurements are sensitive only to the large scale flux power spectrum ($k \lesssim 0.02$ s/km), and thus depend on IGM parameters
differently than the small-scale wavelet measurements explored here.
Their joint fit requires
high $T_0$ for cases with inverted temperature-density relations, similar to our conclusions from a different type of
measurement. There 
thus appears
to be some tension with the flux PDF measurement, which may reflect systematic errors in one or more of the 
measurements and/or the modeling. We intend to consider this further in future work.

\subsection{Inhomogeneities in the Temperature-Density Relation}
\label{sec:inhomg_sims}

Let us further consider the implications of our measurements for the presence or absence of temperature inhomogeneities
in the IGM. In most redshift bins, the measured PDF has comparable width to the simulated PDFs, which assume a perfect
temperature-density relation.\footnote{Strictly speaking, the calculations assume a perfect temperature-density relation only 
when accounting for thermal broadening since the effects of shock heating on the gas density distribution are incorporated.
We expect thermal broadening to be the most important effect of the temperature, and we are not modeling 
inhomogeneities from HeII
reionization here. It is in this sense that we assume a perfect temperature-density relation.} The possible exceptions 
are the $\bar{z}=3.4$ bin (where metal
contamination is a possible culprit) and the $\bar{z}=2.2$ bin (where scatter from shocks may be most important). One might wonder
if the widths of the wavelet PDFs are too small to be compatible with ongoing or recent HeII reionization, which is presumably
a fairly inhomogeneous process. A related question regards the precise meaning of our temperature constraints in the
presence of inhomogeneities: which temperature do we measure exactly -- the mean temperature, the minimum temperature, etc.?
We intend to
address these issues in detail in future work, but we outline a few pertinent points here. In this discussion, we draw on the results
of McQuinn et al. (2008).

The first point is that temperature inhomogeneities during HeII reionization, while likely important, are smaller 
than one might naively guess.
McQuinn et al. (2008) emphasized the 
importance of hard photons, with long mean free paths, for HeII photoheating: much of the heating during HeII
reionization by bright quasars occurs far from sources, rather than in well-defined `bubbles' around ionizing sources. This is
quite different than during HI/HeI reionization by softer stellar sources, where the ionizing photons have short mean free paths and 
heating does occur within well-defined bubbles.
Since
the hard photons have long mean free paths, and a `background' radiation field from multiple sources needs to be built up before these
photons appreciably ionize and heat the IGM, the heating is much more homogeneous than might otherwise be expected. The softer photons,
typically absorbed in bubbles around the quasar sources, only heat the IGM by $\delta T \lesssim 7,000$ K. 
Consider the temperature PDFs in Figure 11 of
McQuinn et al. (2008). This figure illustrates that by the time any gas is heated significantly, there are very few completely
cold regions left over in the IGM: the temperature field is more homogeneous than might be expected.

Simplified models with discrete
$\sim 30,000 K$ bubbles around quasar sources and a cooler IGM outside (e.g. Lai et al. 2006) are hence not realistic, and overestimate
the temperature inhomogeneities. In the McQuinn et al. (2008) simulations, the temperature inhomogeneities peak at a level
of $\sigma_T/\avg{T} \sim 0.2$, which is reached in the early phases of HeII reionization. For contrast, a toy two-phase
hot/cold IGM with hot bubbles that are $3$ times as hot as a cooler background IGM, gives a more substantial 
peak fluctuation level of $\sigma_T/\avg{T} = 0.58$, reached when the hot bubbles fill $25 \%$ of the IGM. In the midst
of HeII reionization, the McQuinn et al. (2008) simulations predict roughly $10 \%$ level temperature
fluctuations on large scales from inhomogeneous HeII heating. This level of 
scatter may be hard to discern with our existing measurement.

\begin{figure}
\bc
\includegraphics[width=9.2cm]{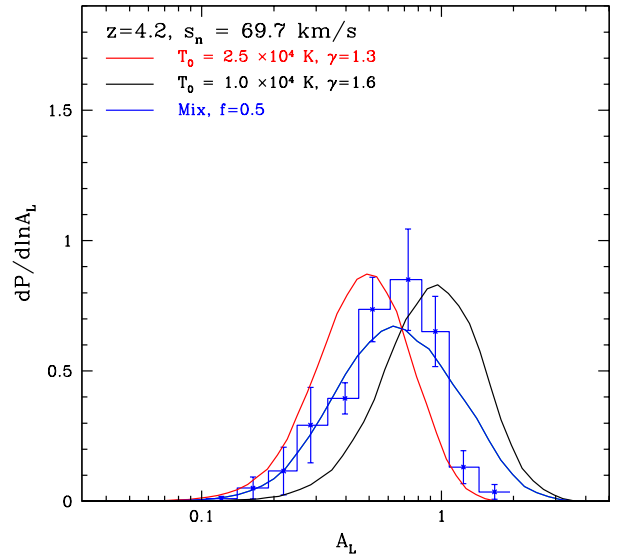}
\caption{Illustration of the challenge of detecting temperature inhomogeneities. The blue histogram with error bars
is the wavelet PDF at $\bar{z}=4.2$ and $s_n=69.7$ km/s. The curves show theoretical models: the red line is a hot
model, the black line is a cold model, while the blue curve shows a fifty-fifty mix between the hot and cold models.
This extreme model can be ruled out as it is too broad, and produces too may high amplitude regions compared
to the data, but one
can see that detecting smaller levels of inhomogeneity is challenging.}
\label{fig:pdf_twophase}
\ec
\end{figure}

To illustrate this, we compare the $\bar{z}=4.2$, $s_n=69.7$ km/s measurement to a simplified and extreme two 
phase model. This redshift
bin probes extended stretches of spectrum along just two lines of sight. Imagine a model where one line of sight passes entirely through
cold regions of the IGM with $T_0 = 10^4$ K, $\gamma=1.6$, while the other line of sight passes entirely through hot regions with
$T_0 = 2.5 \times 10^4$ K, $\gamma=1.3$. This is a contrived example since each sightline probes hundreds of co-moving Mpc,
and so each sightline should in reality probe a mix of temperatures, but this simple case nonetheless illustrates the challenge
of detecting temperature inhomogeneities. For simplicity, in this toy model we imagine that each line of sight probes an
equal stretch through the IGM so that the wavelet PDF is a fifty-fifty mix of the hot and cold models. 
In this toy scenario the mean IGM temperature is $\avg{T} = 17,500$ K and
the fluctuation level is $\sigma_T/\avg{T} \sim 0.43$, i.e., substantially larger than we expect. The wavelet PDF in
this toy model is shown in Figure \ref{fig:pdf_twophase}. This simple model clearly produces too broad a PDF, but it 
is also
apparent that smaller, likely more realistic, 
levels of inhomogeneity will be hard to distinguish with the existing data. For example, 
an inhomogeneous model with fewer cold regions than
in the toy two-phase model would agree with the measurement.
Indeed, the
data may even favor slightly inhomogeneous models, but we leave exploring this to future work. The $z=4.2$ and $z=3.4$ data, which
may be in the midst of HeII reionization, and which are sensitive to the temperature near the cosmic mean density, are the best
redshift bins for further exploration.
Provided the inhomogeneities are
relatively small, as suggested by the measurements in most redshift bins, ambiguities in which temperature we constrain precisely
are unimportant, and our temperature estimates should be accurate. 

Another possible issue, related to the discussion in \S \ref{sec:smoothing}, is that the one-dimensional 
nature of the Ly-$\alpha$ forest
may obscure detecting temperature inhomogeneities from HeII reionization. Consider the three-dimensional power spectrum
of temperature fluctuations in Figure 10 of McQuinn et al. (2008). There is a large scale peak in the three-dimensional power
spectrum, owing to inhomogeneous heating, and a prominent small-scale ramp-up that results 
from the temperature-density relation and small-scale
density inhomogeneities. The large scale peak in the power spectrum is essentially the signal we are after, while the small 
scale ramp-up
is noise as far as extracting inhomogeneities is concerned. However, the one-dimensional temperature power spectrum may be more
relevant than the three-dimensional one for absorption spectra. In the one-dimensional 
temperature power spectrum, high-$k$ transverse modes, which are dominated
by the small-scale ramp-up, will be aliased to large scales, swamping the temperature inhomogeneities. This argument
is imperfect though, since the
one-dimensional temperature power spectrum is not exactly the relevant quantity either: absorption
spectra are insensitive to the temperature of large overdensities, which regardless produce saturated absorption.
It will be interesting to consider this further in the future, and to consider the potential gains from cross-correlating the
wavelet amplitudes of pairs of absorption spectra.

A final issue, particularly relevant in the highest redshift bin, is that the temperature inhomogeneities may depend on the
timing and nature of hydrogen reionization. The temperature contrast between regions with doubly ionized helium and those
in which only HI/HeI are ionized depends on when hydrogen (and HeI) reionized.  Specifically, 
the temperature contrast between HII/HeII and HII/HeIII regions will be reduced
if hydrogen is reionized late to a high temperature, and increased if hydrogen reionizes early to a smaller temperature.
Moreover, heating from hydrogen reionization will
itself be inhomogeneous (e.g. Cen et al. 2009).
Extending the measurements in this paper to higher redshift can help disentangle the impact of hydrogen and helium
photoheating. Further modeling will also be helpful.

\subsection{The Impact of Jeans Smoothing}
\label{sec:jeans_sims}

As mentioned previously, a shortcoming of our modeling throughout is that we have run only a single simulated thermal history in describing
the gas density distribution in
the IGM: we vary the thermal state of the gas only as we construct mock absorption spectra and 
incorporate thermal broadening. Similar approximations are common in
the Ly-$\alpha$ forest literature. The gas density distribution is sensitive to the full thermal history of
the IGM (Gnedin \& Hui 1998) and so properly accounting for a range of thermal histories requires running many simulations.
This certainly deserves further exploration, but we do not expect a big impact on our present results. Thermal 
broadening directly smooths the optical depth field and results
in a roughly exponential decrease in small-scale flux power (Zaldarriaga et al. 2001), while Jeans smoothing acts on the 
three-dimensional gas distribution and has a less direct impact. Properly accounting for the impact of HeII photoheating in
the simulation run should smooth out the gas distribution a bit, and reduce the wavelet amplitudes in these models slightly.
This might reduce our favored temperatures during HeII reionization, but we expect this effect to be small compared to other
uncertainties. Observational studies of the absorption spectra of close 
quasar pairs may help disentangle the effects of thermal broadening and Jeans smoothing.

\subsection{Comparison with Previous Measurements}
\label{sec:compare_previous}

\begin{figure}
\bc
\includegraphics[width=9.5cm]{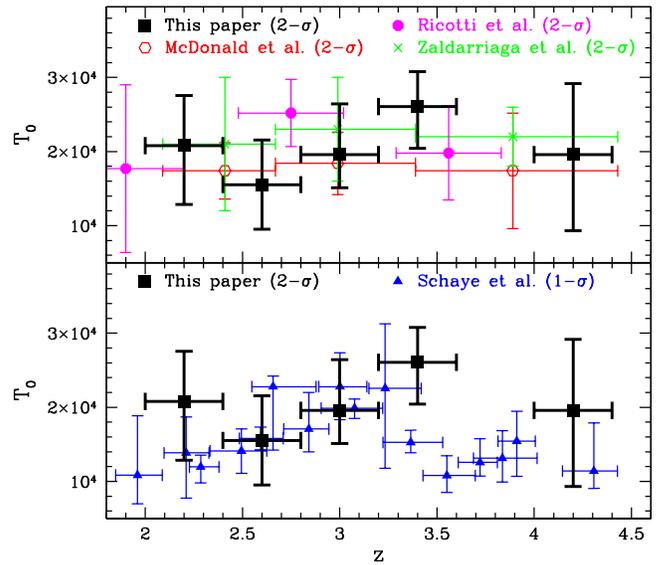}
\caption{Comparison with previous measurements from the literature. The black points with error bars show
the redshift evolution of the temperature at mean density favored by our present analysis. The other points
show various measurements from the literature.  
}
\label{fig:compare_tzero_previous}
\ec
\end{figure}

A detailed comparison with previous measurements is difficult since our methodology differs from that of most previous work.
Instead, we will simply compare the bottom line, and make a few remarks about the differences. Figure \ref{fig:compare_tzero_previous}
shows our constraints on $T_0(z)$, compared to the results of Schaye et al. (2000), Ricotti et al. (2000), McDonald et al. (2001),
\& Zaldarriaga et al. (2001). It is encouraging that some of the main trends are similar across all of the measurements: for example, all
of the measurements favor a fairly hot IGM near $z \sim 3$. In this sense, our work reinforces the previous results. 
There are differences in the details, however:
the peak temperatures in Ricotti et al. (2000), and Schaye et al. (2000) are reached at lower redshift than in our analysis. The
McDonald et al. (2001) and Zaldarriaga et al. (2001) results are, on the other hand, flat as a function of redshift, although they adopt
wide redshift bins and may average over any temperature increase. Our measurements are also fairly consistent with a flat temperature
evolution given the large error bars on our measurements.
Our results mostly favor higher temperatures than the previous measurements, particularly the high redshift points of Schaye
et al. (2000). 

One possible reason for some of the differences is related to improvements in simulations of the forest 
over the past decade or so.
In Appendix A, we found that our method -- and we suspect related methods -- require fairly large simulation volumes and high
mass and spatial resolution, particularly at high redshift (see also e.g. Bolton \& Becker 2009). The 
requisite particle number, while achievable 
today, was of course prohibitive for past
studies. Indeed, this was one of our motivations for revisiting the temperature measurements. While some of the previous studies
varied simulation resolution and boxsize, they often considered only a single additional run, which may have been 
inadequate to fully assess convergence.
Finite resolution, in particular, can bias temperature estimates low. 

It is instructive
to compare our fiducial simulation with a boxsize of $L_b = 25$ Mpc/$h$ and $N_p = 2 \times 1024^3$ particles to the main runs
of previous work. Schaye et al. (2000) used a $(L_b, N_p) = (2.5$ Mpc/$h$, $2 \times 64^3$) SPH simulation, 
Ricotti et al. (2000)'s main runs were ($2.56$ Mpc/$h$, $2 \times 256^3$) HPM calculations, McDonald et al. (2001) used
an Eulerian hydrodynamic simulation with $L_b = 10$ Mpc/$h$ and $288^3$ cells, and Zaldarriaga et al. (2001) used a dark matter only
simulation with $L_b = 16$ Mpc/$h$ and $128^3$ dark matter particles. Given the differences between methods, we will not try to estimate
the impact of systematic errors from finite boxsize and resolution on previous results. However, it is clear that increases in computing
power allow us to do a much better job with respect to boxsize and resolution than previous work. Finally, improved estimates
of the mean transmitted flux (Faucher-Gigu\`ere et al. 2008b), and improved masking of metal lines, may also contribute to some of the
differences between our results and previous work.

\section{Cross-correlating with the HeII Ly-$\alpha$ forest}
\label{sec:cross_heii} 

An interesting possibility is to cross-correlate wavelet amplitude measurements from HI Ly-$\alpha$ forest spectra
with measurements in the corresponding regions of HeII Ly-$\alpha$ forest spectra. It is timely to consider this, as larger samples
of HeII Ly-$\alpha$ forest spectra will soon be available (Syphers et al. 2009), especially given the recent
installation of the Cosmic Origins Spectrograph on the Hubble Space Telescope. 

A fundamental difficulty with HeII Ly-$\alpha$ forest observations is that the HeII 
Ly-$\alpha$ cross section is relatively large, and so even a mostly ionized (mostly HeIII)
medium may give rise to complete absorption. 
McQuinn (2009) recently emphasized, however, 
that this problem is not as acute as it is for the $z \sim 6$ HI Ly-$\alpha$ forest. 
First, the $z \sim 3$ HeII Ly-$\alpha$ optical depth is significantly smaller than the
$z \sim 6$ HI Ly-$\alpha$ optical depth owing to the lower
cosmic helium abundance, the smaller absorption cross section, and the lower mean gas density
at $z \sim 3$. 
Moreover, one can locate low density gas elements using high transmission regions from 
HI Ly-$\alpha$ 
forest observations of the same quasar: if even these low density regions manage to give 
complete absorption, these elements and surrounding gas in the absorption trough must be
significantly neutral (see McQuinn 2009 for details). As a quantitative measure, it is helpful to note that
a gas element at the $z=3$ cosmic mean density with a HeII fraction of only
$X_{\rm HeII} = 10^{-3}$ produces a significant HeII optical depth of $\tau_{\rm HeII}=3.6$
(e.g. Furlanetto 2008). A gas element at one tenth of the cosmic mean density will
give the same optical depth when it is one percent neutral.  

While constraining on their own, HeII Ly-$\alpha$ observations may 
be fruitfully combined with our
methodology to extract still more information. Specifically, we propose to measure wavelet
amplitudes from the HI Ly-$\alpha$ forest for quasar spectra with existing HeII Ly-$\alpha$ 
observations, 
contrasting the wavelet amplitudes in HeII absorption trough regions with those 
in HeII transmission regions. If the HeII troughs correspond to purely neutral HeII
regions, untouched by high energy quasar photons, we expect them to be {\em cold}, provided
that HeI and HI in the region were ionized long ago, as one expects for absorbing gas
at say $z \sim 3-4$. The temperature-density relation in the neutral HeII regions
should be at $T_0 \lesssim 10,000$ K, and $\gamma \sim 1.6$, depending on the nature of the
HI ionizing sources, and on when HI reionization occurs (Hui \& Haiman 2003). On
the other hand, if the regions instead contain mostly ionized HeII (yet are nevertheless
opaque in HeII Ly-$\alpha$ owing to the large absorption cross section), they
will be at similar temperature to the transmission regions. In this case all of the gas
will be hot, unless HeII reionization completed at much higher redshift. A final, somewhat
subtle, possibility relates to the fact that 
towards the end of HeII reionization there will likely
be very hot gas elements with neutral fractions as large as $X_{\rm HeII} \sim 0.1$
that are (partly) ionized by a heavily filtered ionizing spectrum from distant
quasars (McQuinn et al. 2008). Such regions will give rise to troughs, will generally
be {\em hotter} than more ionized regions, and occur before HeII reionization completes.
Hence, at the end of HeII reionization, we may expect the HeII troughs to be {\em hotter}
than transmission regions. Only troughs of purely neutral HeII gas, untouched by quasar
photons, should be cold. Discovering any cold regions in the HI Ly-$\alpha$ forest that correspond
to HeII troughs would also make the presence of cold regions and their connection to HeII
reionization more plausible.

\begin{figure}
\bc
\includegraphics[width=9.2cm]{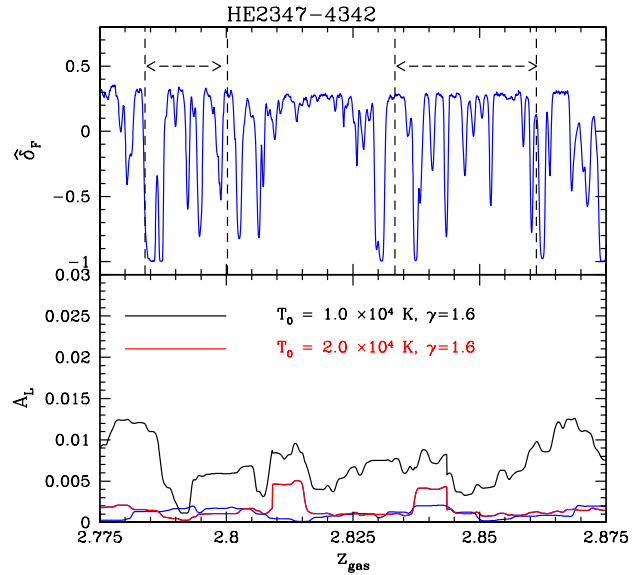}
\caption{HeII Ly-$\alpha$ troughs and the HI Ly-$\alpha$ wavelet amplitudes. {\em Top panel}: The fractional HI Ly-$\alpha$ 
transmission for the spectrum HE2347-4342. The dashed lines, demarcated by arrows, indicate redshift ranges over which
Smette et al.'s (2002) measurements are consistent with complete absorption in the HeII Ly-$\alpha$ forest. {\em Bottom
panel}: The wavelet amplitudes (with $s_n = 34.9$ km/s, $L=1,000$ km/s) for the same stretch of 
spectrum (blue line). The black and red lines 
are simulated wavelet amplitudes suggesting
that even the trough portions of the spectrum are hot and ionized.}
\label{fig:test_HI_He}
\ec
\end{figure}

A detailed study of this type will certainly await future HeII Ly-$\alpha$ observations, but we can nevertheless illustrate
the main idea with the single spectrum from our sample, HE 2347-4342, for which there is an existing HeII Ly-$\alpha$ forest
spectrum (e.g. Smette et al. 2002). These authors identify two spectral regions that are consistent with complete HeII absorption
troughs to within the signal to noise of their measurement. Specifically, an observed spectral region between 
$\lambda_{\rm obs} = 1165.00-1173.50 \AA$ (an absorption redshift of $\bar{z}_{\rm gas} = 2.849$) is estimated to have a 
mean HeII Ly-$\alpha$ transmission of $\avg{F}=-0.001 \pm 0.007$, and a region between $\lambda_{\rm obs} = 1150.00-1154.95$
($\bar{z}_{\rm gas} = 2.7938$) has $\avg{F}=0.024 \pm 0.030$ (Smette et al. 2002). 

We plot the transmission, $\hat{\delta}_F$, for the corresponding
portion of the VLT HI Ly-$\alpha$ spectrum in Figure \ref{fig:test_HI_He} (top panel). In the bottom panel of the figure, we show
the wavelet amplitudes (for $s_n=34.9$ km/s, $L=1,000$ km/s) of the corresponding stretch of spectrum, and compare it to the amplitude 
along a typical sightline drawn from a 
hot $T_0 = 20,000$ K, $\gamma=1.6$ model
and a cold $T_0 = 10,000$ K, $\gamma=1.6$ model. The cold model produces larger wavelet amplitudes than the data, and the hot
model matches more closely, (although even it has two regions of higher wavelet
amplitude than found in the observed spectrum).
Hence, our measurement suggests that the high opacity regions are already quite hot. This is unsurprising
based on the findings of the previous section that the IGM is mostly quite hot at $z \sim 3$. 
These special HeII trough
regions do not appear cooler than typical regions, and this argues against the gas in
these regions being purely neutral. In addition, the trough regions are not obviously hotter
than the transmission regions.

We tentatively suggest that the trough regions are hot and ionized. For now, 
our argument is based only on a small portion of a single spectrum, and so we caution
against drawing strong conclusions from it. We regard it as suggestive, and eagerly
await further HeII Ly-$\alpha$ spectra to perform a more complete study, hopefully
out to higher redshift. Note that there will likely be significant HeII transmission
before HeII reionization completes (Furlanetto 2008, McQuinn et al. 2008), and so we should 
be able
to contrast the temperature in trough and transmission regions even in the midst of HeII
reionization and fully exploit this method.

\section{Conclusions} \label{sec:conclusion}

In this paper, we used a method similar to that of Theuns \& Zaroubi (2000) and Zaldarriaga (2002) to quantify the amount
of small-scale structure in the Ly-$\alpha$ forest. In particular, we convolved Ly-$\alpha$ forest spectra with suitably
chosen Morlet wavelet filters, and recorded the PDF of the smoothed wavelet amplitudes. 
Using cosmological simulations, we showed that this measure of small-scale structure in the forest can
be used to extract information about the temperature of the IGM and its inhomogeneities. We then applied this methodology to $40$ VLT
spectra, spanning absorption redshifts between $z=2.2$ and $z=4.2$ and presented tables of the resulting smoothed wavelet PDFs.
The tables (Tables \ref{table:pdfz4.2}--\ref{table:pdfz2.2}) of smoothed wavelet PDFs are the main result of this paper.

In order to examine the main implications of our measurements for the thermal history of the IGM, we made an initial comparison
with high resolution cosmological simulations.
This comparison suggests that the temperature of the IGM, close to the cosmic
mean density, peaks in the redshift range studied 
near $z=3.4$, at which point it is hotter than $T_0 \gtrsim 20,000$ K at $2-\sigma$ confidence. At lower redshift, the
data appear roughly consistent with a simple adiabatic fall-off ($T_0 \propto (1+z)^2$) from the peak temperature at $z=3.4$. 
The high temperature measurements require significant amounts of late time heating, and are inconsistent with models in which
HeII reionization completes much before $z \sim 3.4$.
At 
the highest redshift considered, the
temperature in our best fit model is rather high, $T_0 \sim 15-20,000$ K but cooler $T_0 \sim 10,000$ K models are still allowed
at $2-\sigma$ confidence at this 
redshift, owing mostly to uncertainties in the mean transmitted flux. We believe that the most likely explanation for our results is
that HeII reionization completes sometime around $z \sim 3.4$, although the statistical errors are still large and 
other heating mechanisms may conceivably be at work.
In general, our analysis favors higher temperatures and higher redshift HeII reionization 
than most previous analyses in the literature (see \S \ref{sec:compare_previous}).

This work can be extended and improved upon in several ways, some theoretical and some observational. 
First, we intend to compare our measurements to more detailed theoretical
models which follow photoheating and radiative transfer during HeII 
reionization. Next, the wavelet PDF measurements can be combined with measurements
of the large scale flux power spectrum from the SDSS (McDonald et al. 2006). This should tighten our constraints, and hopefully
break some of the degeneracies present with the mean transmitted flux at high redshift. It would also be interesting to apply
our method to a larger data set, beating down the statistical error bars, and filling in the redshift gap in our present data set
around $z = 3.8$. Identifying metal line absorbers in additional spectra would help further control metal line contamination,
an important systematic for small-scale measurements. Particularly 
interesting would be to apply our methodology at higher redshifts. This would help disentangle
the effects of hydrogen and helium photoheating, and perhaps provide interesting constraints on hydrogen reionization 
(Theuns et al. 2002a, Hui \& Haiman 2003).
A similar analysis applied to the Ly-$\beta$ region of a quasar spectrum would be sensitive to the temperature of more overdense
regions, and help constrain $\gamma(z)$ (Dijkstra et al. 2004).
Finally, it would be interesting to consider the implications our our measurements for cosmological parameter constraints from
the Ly-$\alpha$ forest, for which the temperature of the IGM is an important nuisance parameter. Although challenging to extract,
the small-scale structure in the Ly-$\alpha$ forest contains a wealth of information regarding the thermal
and reionization histories of the Universe!

\section*{Acknowledgments}
We thank Jamie Bolton, Mark Dijkstra, Nick Gnedin, Lam Hui, Hy Trac, and Matteo Viel for helpful conversations.
Cosmological simulations were run on the Odyssey supercomputer at Harvard University.
CAFG acknowledges support during the course of this work from a NSERC graduate fellowship 
and the Canadian Space Agency.
Support was provided, in part, by the David and Lucile Packard Foundation, the
Alfred P. Sloan Foundation, and grants AST-0506556 and NNG05GJ40G.




\section*{Appendix A: Noise Bias} \label{sec:noise_correc}

Here we estimate the shot-noise bias introduced by random noise in the observed quasar
spectra. In order to do this, we exploit two features of the underlying signal and noise
fields: 1) the smallest measurable scales should be dominated by noise for spectra
in which the noise correction is significant, and 2) for white-noise Gaussian random
fields, one can filter the field on a very small-scale, labeled here as $s_m$, and use this to 
determine how noise contaminates the moments on larger smoothing scales, $s_n$. 
We confine our discussion here to estimates of the bias in the mean, the variance, and the wavelet amplitude
power spectrum, although we ultimately measure the full wavelet PDF.

Let us write the total filtered signal in a quasar spectrum, $a_n^{\rm tot}$, as
\beqa
a_n^{\rm tot}(x) = a_n^{\rm sig}(x) + a_n^{\rm noise}(x),
\label{eq:tot_filt}
\eeqa
where $a_n^{\rm sig}(x)$ denotes the underlying cosmic signal and $a_n^{\rm noise}(x)$ is the filtered
noise field. If the signal and noise fields are independent, it follows that
\beqa
\avg{\hat{A}(x)} = \avg{|a_n^{\rm tot}(x)|^2} = \avg{|a_n^{\rm sig}(x)|^2} + \avg{|a_n^{\rm noise}(x)|^2}.
\label{eq:tot_mean}
\eeqa
In other words, provided the signal and noise are uncorrelated, the mean wavelet amplitude
we measure, $\hat{A}(x)$, is simply the sum of that from the underlying signal, $A(x)$, and a noise contribution. 

We then require $\avg{|a_n^{\rm noise}(x)|^2}$ to estimate the noise bias for the mean wavelet amplitudes. 
One approach would be to use the pixel noise array estimates produced while performing the spectroscopic data reduction.
Here we instead estimate the noise directly from the reduced data, using 
the total wavelet amplitude (Equation \ref{eq:tot_mean}) filtered on a smaller scale $s_m$.
Recall that we normalize the wavelet filters to each have unit power (see Equation \ref{eq:filt_four}).
This means that the average wavelet amplitude for a white-noise field filtered on scale
$s_m$ is the same as when the field is instead filtered on scale 
$s_n$: $\avg{|a_m^{\rm noise}(x)|^2} = \avg{|a_n^{\rm noise}(x)|^2}$. Provided we can find
a scale $s_m$ at which the noise dominates over the signal, that the noise is white-noise, Gaussian
random, and that the noise is uncorrelated with the signal, we can construct an un-biased
estimator of the signal's mean wavelet amplitude. We simply subtract the average of the small-scale
filtered wavelet amplitudes from that on larger scales. 
Our estimate of the noise bias comes from filtering
the data on a scale $s_m = 17.4$ km/s, and assuming $\avg{|a_m^{\rm tot}(x)|^2} \sim \avg{|a_m^{\rm noise}(x)|^2}$ on this scale, after
metal excision. In a spectrum with low noise, the signal may still dominate over the noise even on this smoothing scale, and
in this case we overestimate the noise bias. However, since the signal drops off strongly with wavenumber, we conclude in
this case that the noise bias is unimportant. 

We would also like to estimate the noise bias in the wavelet amplitude power spectrum, and the bias in the
variance of the wavelet amplitudes, smoothed on length scale $L$. To begin with, we neglect any variations in the 
noise power 
spectrum, $P_N$, from sightline to sightline and assume that it is independent of scale.   
Using the notation 
$\hat{A}(x) = |a_n^{\rm sig}(x) + a_n^{\rm noise}(x)|^2$, let us consider the (configuration space) two-point
function of $\hat{A}(x)$:
\begin{eqnarray}
\avg{\hat{A}(x_1) \hat{A}(x_2)} -  \avg{\hat{A}(x_1)}\avg{\hat{A}(x_2)}  = 
\xi_A^{\rm sig}(|x_1 -x_2|) + \xi_A^{\rm noise}(|x_1 - x_2|)  
\nonumber \\ 
+ \avg{a_n^{\star \rm sig}(x_1) a_n^{\rm sig}(x_2)} \avg{a_n^{\rm noise}(x_1) a_n^{\star \rm noise}(x_2)} 
+ \avg{a_n^{\rm sig}(x_1) a_n^{\star \rm sig}(x_2)}\avg{a_n^{\star \rm noise}(x_1) a_n^{\rm noise}(x_2)} 
\nonumber \\
 + 
\avg{a_n^{\star \rm sig}(x_1) a_n^{\star \rm sig}(x_2)}\avg{a_n^{\rm noise}(x_1) a_n^{\rm noise}(x_2)}
+ \avg{a_n^{\rm sig}(x_1) a_n^{\rm sig}(x_2)}\avg{a_n^{\star \rm noise}(x_1) a_n^{\star \rm noise}(x_2)}.
\label{eq:atw}
\end{eqnarray}

Here $\xi_A^{\rm sig}(|x_1 - x_2|)$ denotes the two-point function of the underlying signal (i.e., Equation \ref{eq:atwop_dem},
although in the above expression we have not yet normalized by $\avg{A}$ in the denominator),
and $\xi_A^{\rm noise}(|x_1 - x_2|)$ is a pure noise term, while the other terms are cross-terms.  

The power spectrum of $\hat{A}(x)$ is the Fourier transform of Equation \ref{eq:atw}. 
Using the convolution theorem and Equation \ref{eq:filt_four}, the pure noise part of the 
power (i.e., the Fourier transform of $\xi_A^{\rm noise}(|x_1 - x_2|)$)
can be written as:

\beqa
P_A^{\rm noise}(k) =  B^4 P_N^2 \int \frac{dk^\prime}{2 \pi} 
\rm{exp}\left[-(k - k^\prime)^2 s^2 + k_0^2 s^2\right]  
\rm{exp}\left[-(k^\prime - k_0)^2 s^2\right].
\label{eq:power_wave_noise}
\eeqa

Here $B = \pi^{-1/4} (2 \pi s_n/\Delta u)^{1/2}$ is
a normalization factor (Equation \ref{eq:filt_four}), and we abbreviate $s_n$ as $s$. The noise contribution to the 
wavelet amplitude (squared) power spectrum
is proportional to $P_N^2$ because $A$ is a quadratic function of $\delta_F$ (Equations \ref{eq:field_filt}--\ref{eq:waveamp}). 

Next we consider the cross terms. The terms on the third line of Equation \ref{eq:atw} can be shown to be very small.
The important cross terms can be derived by again applying the convolution theorem, and using the Fourier transform of
the Morlet Wavelet filter and its complex conjugate. The result is: 
\beqa
P_A^{\rm cross}(k) = && F.T.[
\avg{a_n^{\star \rm sig}(x_1) a_n^{\rm sig}(x_2)} \avg{a_n^{\rm noise}(x_1) a_n^{\star \rm noise}(x_2)} 
+ \avg{a_n^{\rm sig}(x_1) a_n^{\star \rm sig}(x_2)}\avg{a_n^{\star \rm noise}(x_1) 
a_n^{\rm noise}(x_2)}] \nonumber \\
= && B^4 P_N \int \frac{dk^\prime}{2 \pi} 
\rm{exp}\left[-(k - k^\prime)^2 s^2 - k_0^2 s^2\right] \rm{exp}\left[-(k^\prime + k_0)^2 s^2\right] P_F(k^\prime) \nonumber \\
 + && B^4 P_N \int \frac{dk^\prime}{2 \pi} 
\rm{exp}\left[-(k - k^\prime)^2 s^2 + k_0^2 s^2\right] \rm{exp}\left[-(k^\prime - k_0)^2 s^2\right] P_F(k^\prime).
\label{eq:power_wave_cross}
\eeqa
Here $P_F(k)$ denotes the flux power spectrum. 

The power spectrum of the underlying signal, $P_A(k)$, is related to the one we measure, $P_{\hat{A}}(k)$, by 
$P_A(k) = P_{\hat{A}}(k) - P_A^{\rm cross}(k) - P_A^{\rm noise}(k)$. Note that in order to estimate the bias in the measured
power spectrum we need to first estimate the underlying flux power spectrum $P_F(k)$. The expressions also require an
estimate of the noise power spectrum which we derive from the small-scale filtered field, $P_N(k) = \avg{|a_m^{\rm tot}|^2} \Delta u$, under the assumption that $\avg{|a_m^{\rm tot}|^2} = \avg{|a_m^{\rm noise}|^2} = \avg{|a_n^{\rm noise}|^2}$. 

Finally, we want to estimate the bias on the variance of the (smoothed) wavelet amplitude squared. The variance follows
from the power spectrum by:
\beqa
\sigma_A^2(L) = \int_{-\infty}^{\infty} \frac{dk^\prime}{2 \pi} 
\left[\frac{Sin(k^\prime L/2)}{k^\prime L/2}\right]^2
P_A(k^\prime).
\label{eq:var_wave}
\eeqa
It is also useful to note that the noise contribution to the variance can be calculated analytically from Equation \ref{eq:power_wave_noise}
and Equation \ref{eq:var_wave} and is given by:
\beqa
\sigma_A^2(L)_{\rm noise} = 2 \avg{|a_n^{\rm noise}|^2}^2 \frac{s_n}{L}
\Bigg[\sqrt{\frac{\pi}{2}} Erf\left(\frac{L}{\sqrt{2}s_n}\right) + \frac{s_n}{L} \left(-1 + \rm{exp}\left(-\frac{L^2}{2 s_n^2}\right)\right)\Bigg].
\label{eq:var_noise}
\eeqa
Integrating over the power spectrum, the variance we measure, $\sigma^2_{\hat{A}}(L)$, is related to the underlying signal variance,
$\sigma^2_A(L)$, by $\sigma^2_{\hat{A}}(L) = \sigma^2_A(L)_{\rm noise} + \sigma^2_A(L)_{\rm cross} + \sigma^2_A(L)$.

This expression almost provides us with an un-biased estimate of the signal variance, but we still need to take into account 
sightline-to-sightline variations in the noise power spectrum. The above expression for $\sigma^2_{\hat{A}}(L)$ can be interpreted 
as a conditional variance 
$var(A_L|P_N)$, i.e., the variance in $A(L)$ given that the noise power is $P_N$. The unconditional variance is then
given (for uniform weighting) by 
\beqa
var(A_L) = \avg{var(A_L|P_N)}_{\rm Noise} + \avg{\avg{A_L|P_N}^2}_{\rm Noise} - \avg{A_L}^2,
\label{eq:var_final}
\eeqa
where $\avg{ }_{\rm Noise}$ denotes averaging over the ensemble of sightlines with different noise properties, and $\avg{A_L}$ is the
global average wavelet amplitude.
With these formulae in hand, we can estimate the bias in our variance estimates owing to random noise in the spectra. The cross term in Equation \ref{eq:power_wave_cross} requires an estimate of the flux power spectrum. We use here a simulated model for the flux power spectrum.

\section*{Appendix B: Simulated Metal Line Absorption}\label{sec:metals_corr}

In this Appendix, we explore the impact of metal line contamination on the wavelet PDF measurements theoretically.
Our main goal here is to build some intuition for the contamination and its relative importance at different
smoothing scales and redshifts -- i.e., we expect this investigation to be useful qualitatively but
do not expect quantitatively accurate estimates of metal line contamination.
Our strategy is to randomly populate mock spectra with metal lines in a way that roughly matches
empirical constraints on metal line absorbers, rather than attempting to directly simulate metal absorbers from first
principles. Ideally, our prescription for including metal lines would match the column density 
distribution, two-point correlation function,
b-parameter distribution, and overall opacity for many different species of metal line absorbers.
In practice, the relevant statistical properties have not been measured for all of the metal absorbers that may
contaminate the forest. We instead populate mock spectra only with lines that match the observed properties
of CIV lines, which produce the strongest contamination to the forest. To roughly account for absorption by additional
metal line species, we generate three independent sets of absorption lines, with each set of lines
drawn according to the statistical properties of CIV.
This crude approximation is adequate to the extent that the statistical properties of other metal line
absorbers are similar to those of CIV. Generating {\em three} sets of CIV-like lines is also somewhat arbitrary 
of course, and 
we find that 
even with three sets of strong CIV absorbers, we -- somewhat surprisingly -- underestimate the fractional 
contribution of 
metals to the opacity of the forest by a factor of a few (Schaye et al. 2003, Faucher-Gigu\`ere et al. 2008b).

We generate mock metal absorption lines by first generating a lognormal random field, and then Poisson sampling from 
the lognormal
field to produce random realizations of discrete metal lines. 
The measured two-point correlation function of CIV absorbers has the form (Boksenberg et al. 2003):
\beqa
\xi(\Delta v) = A_1 \rm{exp}\left(-\frac{\Delta v^2}{2 \sigma_1^2}\right) + A_2 \rm{exp}\left(-\frac{\Delta v^2}{2 \sigma_2^2}\right).
\label{eq:twop_civ}
\eeqa

We want to generate realizations of a random field with the above clustering, which we do approximately with a lognormal
model.
Specifically, we generate a Gaussian random field
$\delta_G$ and then form a lognormal field via the mapping:
\beqa
1 + \delta_{\rm CIV} = A \rm{exp}(\delta_G),
\label{eq:lognorm}
\eeqa
with the parameter $A$ chosen so that the field $\delta_{\rm CIV}$ has mean zero, $A = \rm{exp}(-\avg{\delta_G}^2/2)$. 
In order for $\delta_{\rm CIV}$ to have the correct two-point function, the Gaussian random field $\delta_G$ must
be drawn from a model with an appropriate power spectrum. By experimentation, we find that a model with 
\beqa
P_G(k) = A_G \rm{exp}\left(-\frac{k^2 \sigma_G^2}{2} \right),
\label{eq:power_gauss}
\eeqa
with $A_G = 1.11 \times 10^3$, and $\sigma_G = 135$ km/s, gives roughly the correct clustering. 

Given a line of sight realization of the random field $\delta_{\rm CIV}$, the average number of CIV lines expected
in a simulated cell of velocity width $\Delta v_{\rm cell}$, and density $\delta_{\rm CIV}$, at spatial position $x$ is:
\beqa
\avg{\mathcal{N}_{\rm CIV}}(x) = \avg{n_{\rm CIV}} \Delta v_{\rm cell} \left[1 + \delta_{\rm CIV}(x)\right]. 
\label{eq:avnum_civ} 
\eeqa
We denote the cosmic average number of lines per velocity increment, $\Delta v_{\rm cell}$, as $\avg{n_{\rm CIV}}$. 
This can be computed from the average number of lines per unit redshift, which in turn follows from the CIV column
density distribution.
The average number of lines per unit redshift above some minimum column density $N_{\rm CIV, min}$ is given by
\beqa
\frac{d\mathcal{N}}{dz} = \frac{dX}{dz} \int_{\rm N_{\rm CIV, min}}^{\infty} dN_{\rm CIV} \frac{d^2N_{\rm CIV}}{dN_{\rm CIV} dX}.
\label{eq:dndz}
\eeqa
We adopt $N_{\rm CIV, min} = 10^{12} cm^2$ throughout. 
Here $\frac{dX}{dz}$ is the absorption pathlength,
\beqa
\frac{dX}{dz} = \frac{(1+z)^2}{\left[\Omega_m (1 + z)^3 + \Omega_\Lambda\right]^{1/2}}.
\label{eq:pathl}
\eeqa

Given the average number of CIV lines in a cell, $\avg{\mathcal{N}_{\rm CIV}}(x)$, the exact number of CIV lines
to place in the cell is determined by drawing from a Poisson distribution. Each absorption line is then assigned
a column density by drawing from a power-law fit to the observed column density distribution (Scannapieco et al. 2006).
This power-law fit has $f(N) \propto (N/N_0)^{-\alpha}$, with $\alpha = 1.8$, and is normalized to 
$f=10^{12.7}$ cm$^2$ at $N_0 = 10^{13}$ cm$^{-2}$. We use this fit at all redshifts since the observed distribution 
evolves only weakly over the redshifts of interest. Since CIV is a doublet, we create a weaker partner
line for each mock absorption line generated. We give each absorption line a Gaussian profile, and approximate the
$b$-parameter distribution as a delta-function. We have experimented with delta functions 
around $b=5,10$ and $20$ km/s, comparable
to the observed values (Boksenberg et al. 2003). For reference, the stronger CIV absorption component has a rest frame
wavelength of $\lambda_r = 1548.2 \AA$, while the weaker component is at $\lambda_r = 1550.8 \AA$.
The cross section of the stronger component is $\sigma_{\rm 1, CIV} = 2.6 \times 10^{18}$ cm$^{2}$,
and is $\sigma_{\rm 2, CIV} = 1.3 \times 10^{18}$ cm$^{2}$ for the weaker component. 
It is also useful to note that the line center optical depth of the stronger component is related to the 
column density
and b-parameter of the line by: 
\beqa
\tau_0 = 1.0 \left[\frac{N_{\rm CIV}}{2.3 \times 10^{13} cm^{-2}}\right] \left[\frac{10 km/s}{b}\right],
\label{eq:taucen}
\eeqa
while the line center optical depth of the weaker component is a factor of two smaller. 

\begin{figure}
\bc
\includegraphics[width=16cm]{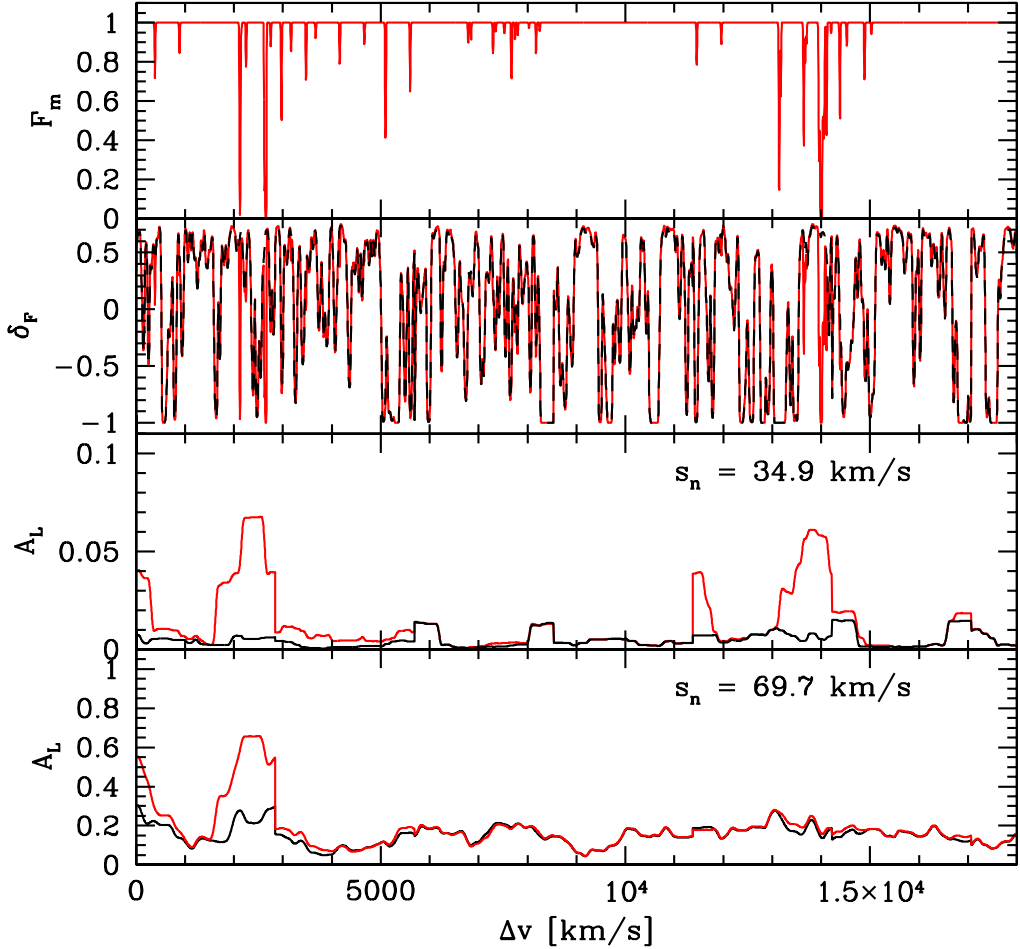}
\caption{Mock spectra with metal lines and the impact on wavelet amplitudes at $z=3.4$. {\em Top panel}: The 
transmission field
from metal line absorbers. {\em Second panel from top}: The fractional transmission, $\delta_F$, in the forest. The
black dashed line ignores metal lines while the red solid line includes metal absorbers. {\em Second panel from bottom}:
The corresponding smoothed wavelet amplitudes with $s_n=34.9$ km/s and $L=1,000$ km/s. The red lines include the
impact of metal absorbers, while the black lines ignore the metals. {\em Bottom panel}: Similar to the
previous panel for $s_n=69.7$ km/s.}
\label{fig:mock_metals}
\ec
\end{figure}

\begin{figure}
\bc
\includegraphics[width=16cm]{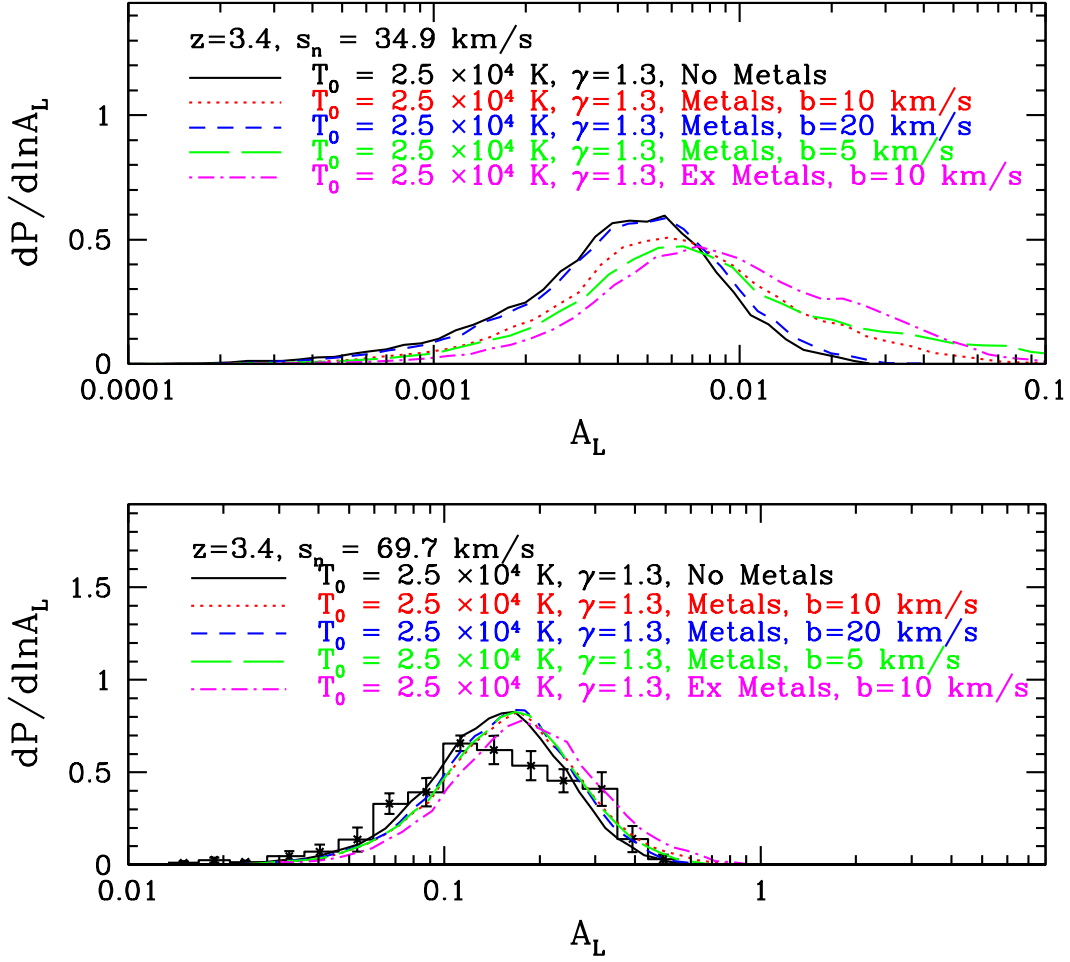}
\caption{Impact of simulated metal line absorbers on the wavelet PDF. {\em Top panel}: Simulated 
wavelet PDFs at $z=3.4$ and
$s_n = 34.9$ km/s for
a model without metals (black solid line), compared to the same model with metal lines added according to several
different prescriptions.
The magenta dot-dashed line is an extreme model that incorporates $6$ times the observed CIV abundance in metals. The
other lines each incorporate $3$ times the observed CIV abundance, and differ in the b-parameters assumed.
{\em Bottom panel}: Similar to the top panel, except for a larger filtering scale, $s_n = 69.7$ km/s. 
}
\label{fig:mock_metals_pdf}
\ec
\end{figure}

We have generated mock metal absorption lines according to the above prescription, and added them to simulated
Ly-$\alpha$ forest spectra at $z=2.2,3.0,3.4$ and $4.2$. A typical example sightline at $z=3.4$ is shown in 
Figure \ref{fig:mock_metals}, assuming $b=10$ km/s and $T_0 = 2.5 \times 10^4$ K, $\gamma=1.3$.\footnote{This sightline 
is extended
by splicing together the transmission and wavelet amplitudes from smaller segments of spectrum that are 
periodic over a box length.
This occasionally leads to slight artifacts in the associated figures. The statistics of the wavelet amplitudes are
measured before splicing and so are not impacted by these artifacts.}
This illustrates
a few key qualitative features regarding metal line contamination, and its impact on the wavelet amplitudes. The first
feature is that our mock metal absorbers do lead to prominent peaks in the wavelet amplitudes, similar to
the peaks observed and associated with metal absorbers in our observational data (\S \ref{sec:metals}).
The next feature one notices is the considerably larger impact of metal absorbers on the smaller smoothing scale, again
consistent with our previous findings from observational data.  In some cases there are peaks in the wavelet 
amplitude on
the smaller smoothing scale that are entirely absent at larger smoothing scale.
For example, the metal line absorbers
beyond $\Delta v \gtrsim 10,000$ km/s in Figure \ref{fig:mock_metals} produce peaks in the wavelet amplitude only on the
smaller filtering scale. There are also cases where metal line absorbers lead to peaks for both filters (e.g.
the lines near $\Delta v \sim 2,000$ km/s). In these cases, the fractional boost in wavelet amplitude from the metal
lines is larger for the smaller smoothing scale filter.
The metal lines are typically
narrower than the HI lines, and the fractional contamination is hence significantly larger on small scales.
Finally, a metal line that lands on a pixel where there is already significant Ly-$\alpha$ absorption is obviously 
irrelevant.
We find many examples from the mock spectra of strong, narrow metal lines that happen to overlap strong Ly-$\alpha$
lines, and have little impact as a result. The strong increase in the mean absorption with redshift,
and the corresponding boost in the amplitude of fluctuations in the forest, result in significantly less contamination
towards high redshift. For example, in our simulated models the fractional impact of metals on the mean wavelet
amplitude (for $s_n = 34.9$ km/s) is $7$ times larger at $z=2.2$ than it is at $z=4.2$. 

In order to provide a more quantitative measure of the impact of metal lines on wavelet amplitude measurements, we
measure the wavelet PDF from $1,000$ mock spectra with added metal lines. Examples at $z=3.4$ are shown in
Figure \ref{fig:mock_metals_pdf}. By comparing the top and bottom panels, one can see that the metal lines generally
have a much larger impact on the smaller filtering scale. At $s_n=34.9$ km/s, for $b=5$ and $10$ km/s, the mean wavelet amplitude is shifted significantly, and the PDF develops a long tail towards high wavelet amplitudes. There is 
relatively little impact
for lines with larger b-parameters, as demonstrated by the $b=20$ km/s curve, but most observed CIV lines have 
smaller b-parameters: $b = 20$ km/s is really at the upper end of the observed CIV 
linewidths (Boksenberg et al. 2003). 
We have also generated a more extreme model, with $6$ independent sets of CIV-like lines. Even this model produces
only a small shift in the wavelet PDF on the large smoothing scale. Although our model for metal lines is rather
crude, we expect fairly small shifts in the wavelet amplitudes on the larger smoothing scale, especially in the
higher redshift bins.

\section*{Appendix C: Convergence with Simulation Resolution and 
Boxsize}\label{sec:box_res}

In this section we assess the convergence of the simulated wavelet PDFs with
increasing simulation resolution and boxsize. It is relatively challenging
to obtain fully converged results in Ly-$\alpha$ forest simulations.
On the one-hand, one needs to simulate a large volume to: compare simulations with large scale
flux power spectrum measurements (if desired), to sample a representative
fraction of the Universe, to capture the cascade of power from large to small scales,
and to simulate peculiar velocity fields, which are coherent on rather large scales. On the
other hand, high mass and spatial resolution at the level of tens of $kpc$ (in regions of low
to moderate overdensity) are required to fully resolve the filtering (Gnedin \& Hui 1998) and 
thermal broadening scales. 

In order to examine the convergence of the wavelet PDFs with 
simulation volume, we ran a set
of cosmological SPH simulations with fixed mass and spatial resolution, yet increasing
boxsize. 
Specifically, we ran simulations with boxsize $L_b$ and particle
number $N_p$ of $(L_b, N_p) = (12.5 \rm{Mpc}/h, 2 \times 256^3)$,
$(25 \rm{Mpc}/h, 2 \times 512^3)$, and $(50 \rm{Mpc}/h, 2 \times 1024^3)$. 
To isolate resolution effects, we ran a sequence of fixed boxsize, increasing
particle number simulations 
with $(L_b, N_p) = (25 \rm{Mpc}/h, 2 \times 256^3)$, 
$(25 \rm{Mpc}/h, 2 \times 512^3)$, and $(25 \rm{Mpc}/h, 2 \times 1024^3)$.
In each simulation the force softening was set 
to $1/20$th of the mean inter-particle spacing. In general, the initial conditions in each of the fixed boxsize simulations
are drawn from the same random number seeds, so that the Fourier modes of the initial displacement field are identical
(for the wavenumbers common to each pair of simulations). Owing to imperfect planning, however, the highest resolution
simulation with $N_p = 2 \times 1024^3$ particles was run with different initial conditions, and so there are random
differences between this simulation and the lower resolution realizations, in addition to any systematic dependence on
resolution. Given that the random seed-to-seed fluctuations are fairly small, and that are results are fairly well converged,
we have not rerun the (faster) lower resolution simulations with initial conditions that match the highest resolution run.

\begin{figure}
\bc
\includegraphics[width=9.2cm]{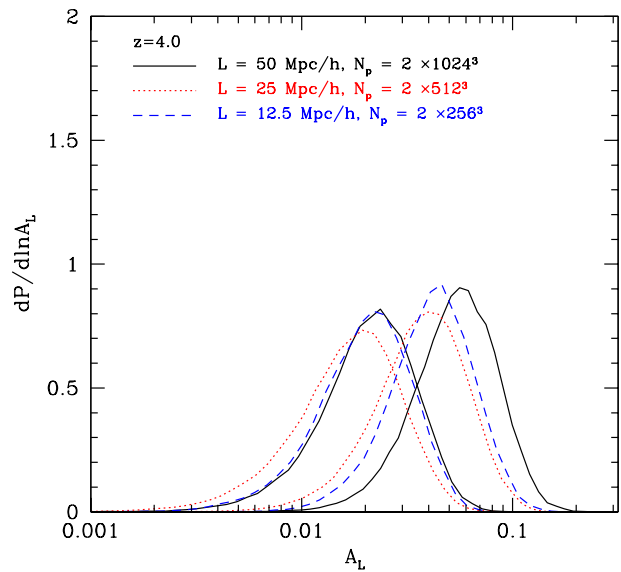}
\caption{Wavelet amplitude PDF as a function of boxsize at $z=4$.
The curves show the wavelet amplitude PDF at fixed mass resolution for
simulations of varying boxsize for each of two thermal history models. The set
of curves to the left, 
centered near $A_L = 0.02$ has $(T_0, \gamma) = (2 \times 10^4 K, 1.3)$,
while those on the right have $(T_0, \gamma) = (1 \times 10^4 K, 1.6)$.}
\label{fig:boxl_converge_z4}
\ec
\end{figure}

\begin{figure}
\bc
\includegraphics[width=9.2cm]{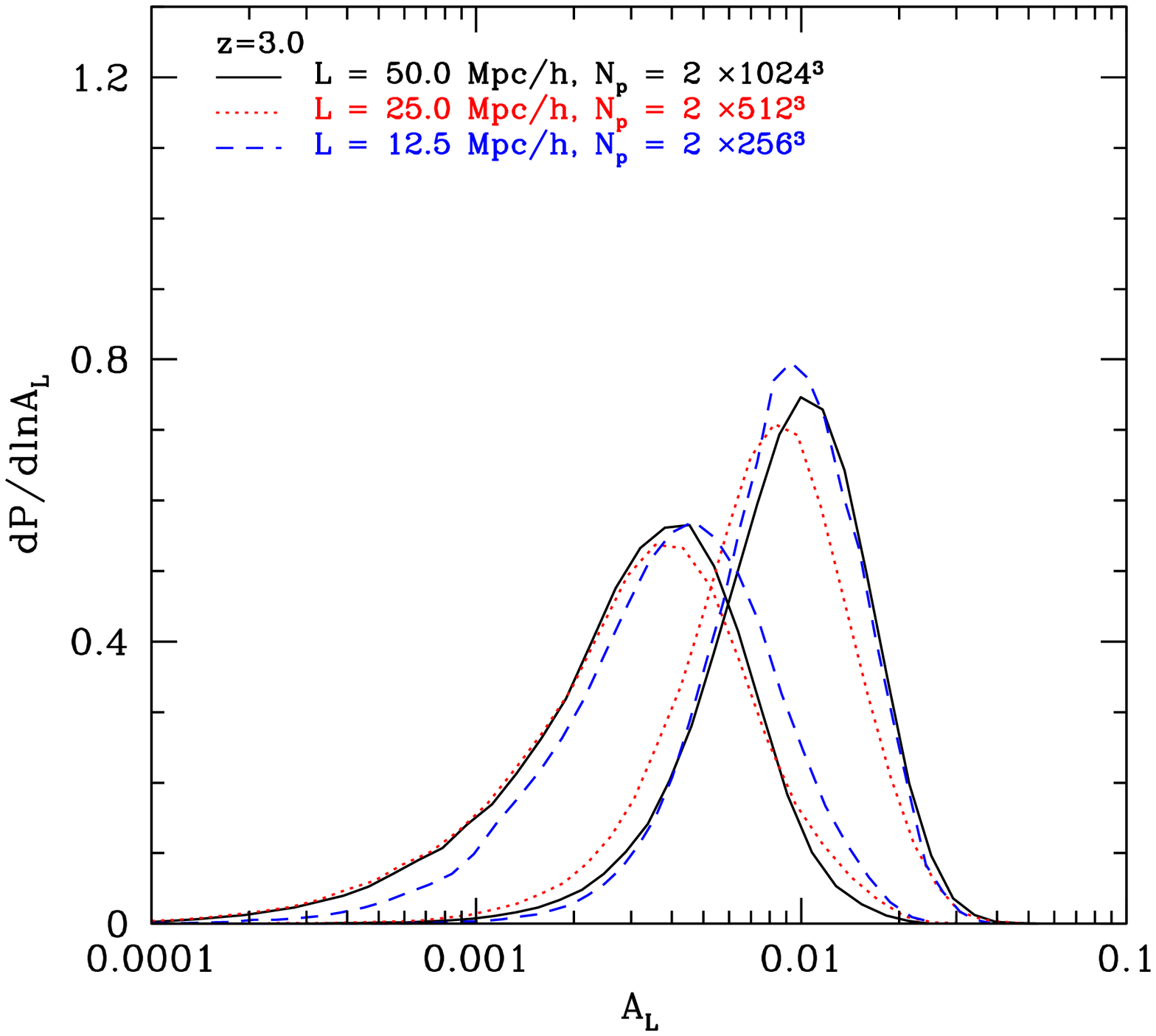}
\caption{Wavelet amplitude PDF as a function of boxsize at $z=3$. Identical
to Figure \ref{fig:boxl_converge_z4}, except at $z=3$.}
\label{fig:boxl_converge_z3}
\ec
\end{figure}

\begin{figure}
\bc
\includegraphics[width=9.2cm]{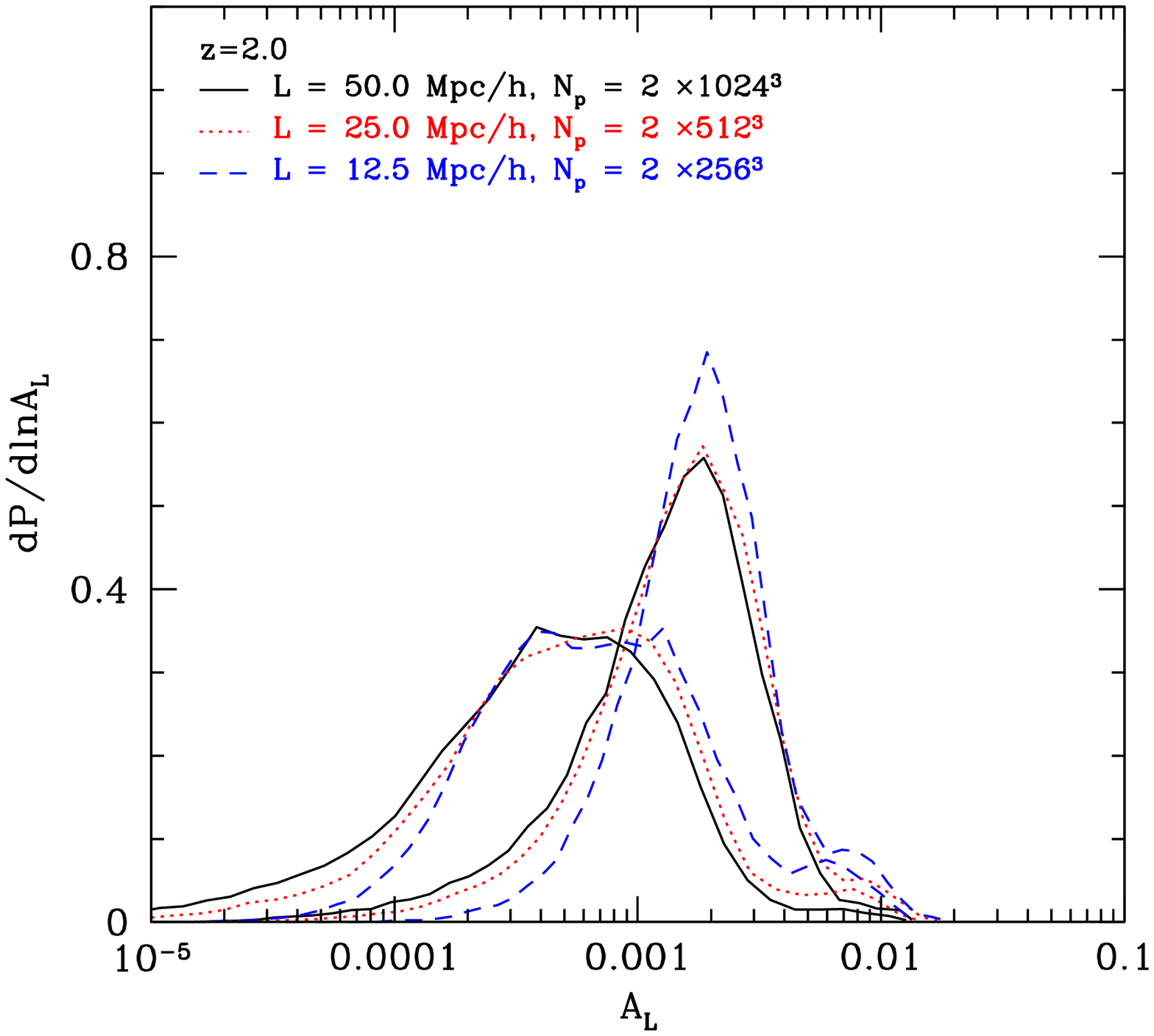}
\caption{Wavelet amplitude PDF as a function of boxsize at $z=2$.
Identical
to Figure \ref{fig:boxl_converge_z4} and Figure \ref{fig:boxl_converge_z2}, 
except at $z=2$.}
\label{fig:boxl_converge_z2}
\ec
\end{figure}

In order to test how the convergence depends on redshift (mostly owing to evolution in the mean transmitted flux)
we examine simulation outputs at $z=2, 3$, and $4$. We re-adjust the intensity of the ionizing background in
each simulation to match a given (averaged over all sightlines) mean transmitted flux. At $z=3$, we assume a mean transmitted 
flux of $\avg{F} = 0.680$.
For the tests here, we adopt $\avg{F} = 0.849$ at $z=2$, and $\avg{F}=0.393$ at $z=4$. We assume a perfect temperature
density relation when incorporating thermal broadening in the mock quasar spectra. To test whether the convergence
depends on the assumed model for the thermal state of the IGM, we consider two temperature-density relations:
$(T_0, \gamma) = (2 \times 10^4 K, \gamma=1.3)$ and $(T_0, \gamma = 1 \times 10^4 K, \gamma=1.6)$. In each case
we adopt a small-scale smoothing of $s_n = 34.9$ km/s and a large scale smoothing 
of $L=1,000$ km/s (see \S \ref{sec:smoothing}). In the text we consider $s_n=69.7$ km/s as well as $s_n=34.9$ km/s, but
the resolution requirements are more stringent on the smaller of these scales, and so we consider it throughout this
convergence study.

The results of the boxsize convergence test are shown in Figures \ref{fig:boxl_converge_z4}-\ref{fig:boxl_converge_z2}. 
The convergence with simulation boxsize is generally encouraging. In fact, the wavelet PDFs from the rather small
$L_b=12.5$ Mpc/$h$ box are similar to those in the larger $L_b=25$ Mpc/$h$ and $L_b=50$ Mpc/$h$ volumes. 
The $z=2$ results, however, suggest that the $L_b=12.5$ Mpc/$h$ box is a bit small: the wavelet PDF looks
systematically narrow compared to the PDF in the larger volume simulations, although the differences are
fairly small. It is not particularly
surprising that this small volume run is inadequate at $z=2$, even for the relatively undemanding task of 
characterizing the distribution of 
small-scale power. For one, the amplitude of the linear power spectrum at the fundamental mode of this simulation 
box is $\Delta^2(k_F) \sim 0.4$ in our adopted cosmology at this redshift,
and so one does expect to start seeing systematic errors from missing large scale modes. 
In some of the $z=3$ and
$z=4$ models the trend with boxsize appears to be non-monotonic. This may suggest that some of the differences
are random, rather than systematic: i.e., a different choice of random number seed in the initial conditions
can shift the PDF around a little bit in the smaller volumes. This scatter can be reduced
by running several different realizations of each model and averaging, but the effects are small
and so we do not pursue this here. It may also be that some of the non-monotonic trends result from two 
competing systematic effects.
For present purposes, bear in mind that our main goal
is to distinguish hotter $T_0 \sim 2 \times 10^4 K, \gamma=1.3$ models from cooler 
$T_0 \sim 1 \times 10^4 K, \gamma=1.6$
models: the differences between simulations of different boxsize are mostly quite small compared
to the model differences. The one possible exception appears to be for the cooler model at $z=4$, where the
peak of the PDF appears at surprisingly large amplitude in the large volume simulation, although the boxsize
shift is still relatively small compared to the difference between the hot and cold models. 
Since we focus on small-scale
fluctuations in this paper, and we find that the resolution requirements are fairly
stringent at high redshift (see below), we sacrifice simulation volume slightly for resolution and 
adopt $L=25$ Mpc/$h$ as our fiducial boxsize.

\begin{figure}
\bc
\includegraphics[width=9.2cm]{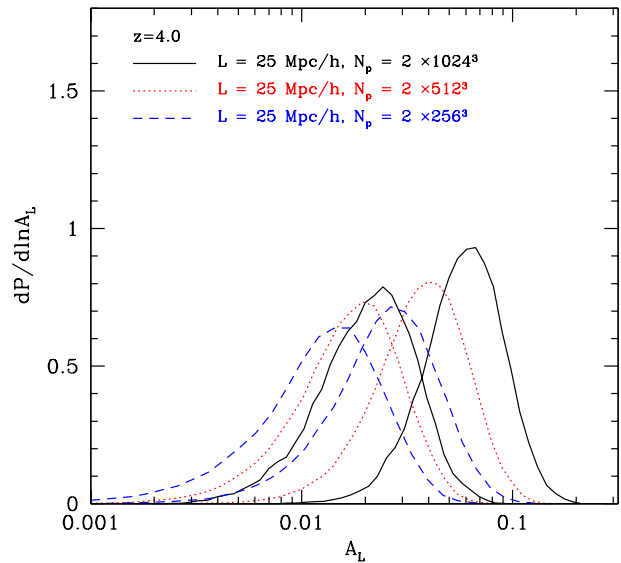}
\caption{Wavelet amplitude PDF as a function of resolution at $z=4$.
The curves show the wavelet amplitude PDF at fixed boxsize in
simulations of varying mass and spatial resolution for each of two thermal history models. The set
of curves to the left have $(T_0, \gamma) = (2 \times 10^4 K, 1.3)$,
while those on the right have $(T_0, \gamma) = (1 \times 10^4 K, 1.6)$.}
\label{fig:res_converge_z4}
\ec
\end{figure}

\begin{figure}
\bc
\includegraphics[width=9.2cm]{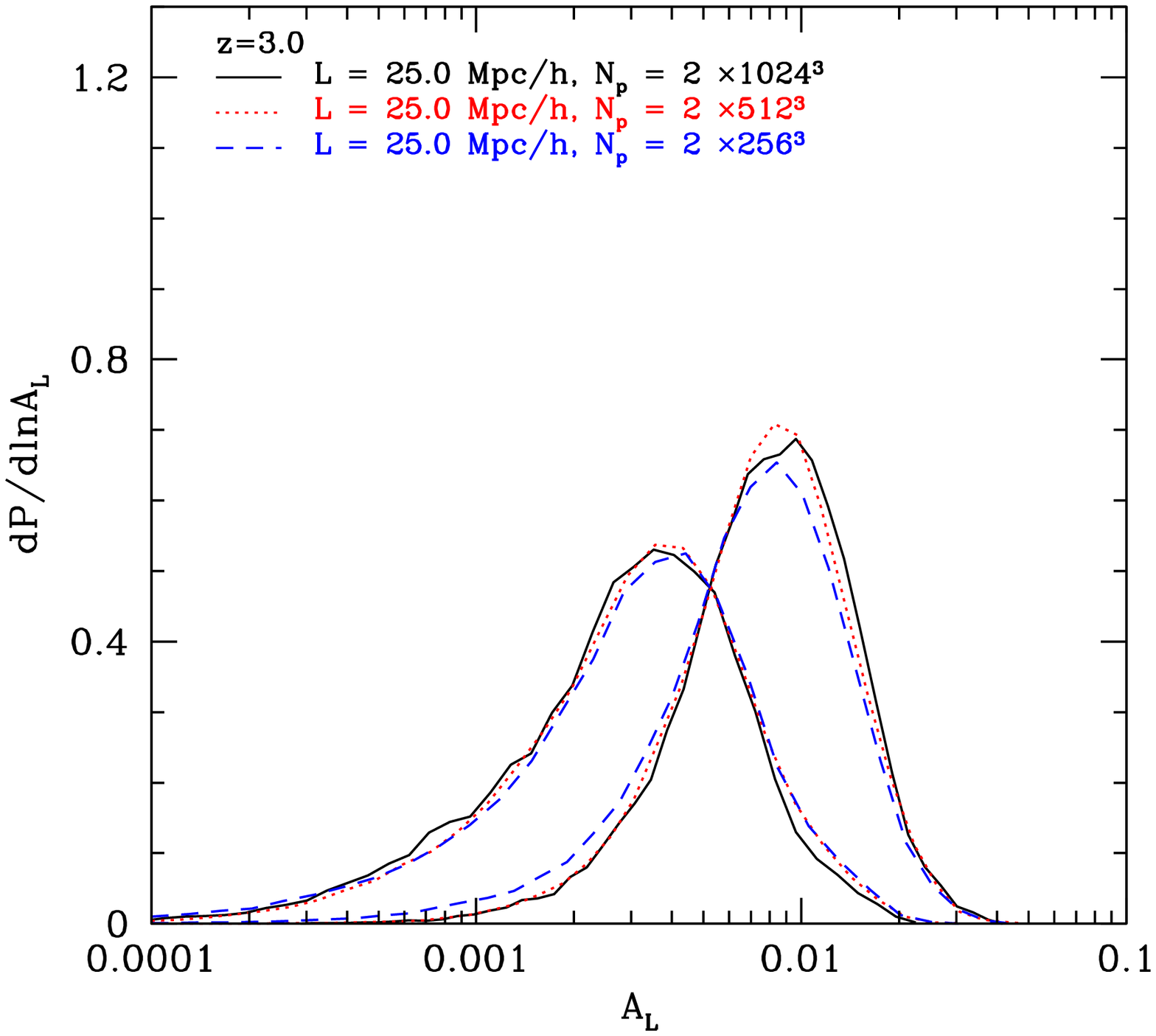}
\caption{Wavelet amplitude PDF as a function of resolution at $z=3$. Identical
to Figure \ref{fig:res_converge_z4}, except at $z=3$.}
\label{fig:res_converge_z3}
\ec
\end{figure}

\begin{figure}
\bc
\includegraphics[width=9.2cm]{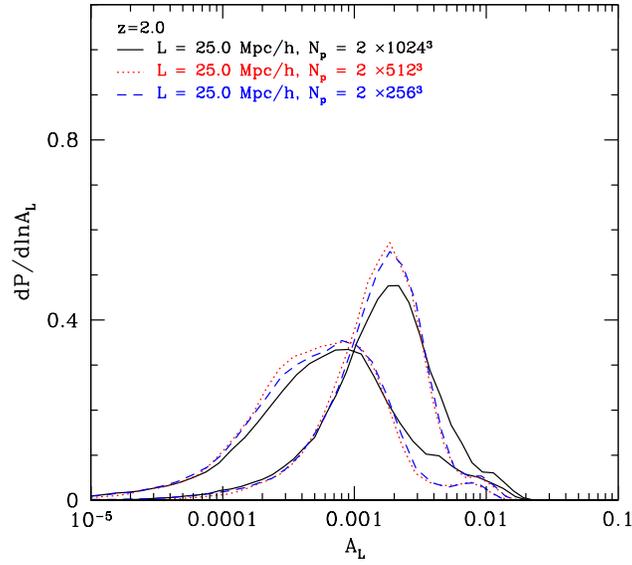}
\caption{Wavelet amplitude PDF as a function of resolution at $z=2$.
Identical
to Figure \ref{fig:res_converge_z4} and Figure \ref{fig:res_converge_z2}, 
except at $z=2$.}
\label{fig:res_converge_z2}
\ec
\end{figure}

Next we show the results of varying the spatial and mass resolution at fixed simulation volume
(Figures \ref{fig:res_converge_z4}-\ref{fig:res_converge_z2}). 
At $z=2$ and $z=3$, the results of the $N_p = 2 \times 256^3$, $L_b = 25$ Mpc/$h$ and the $N_p = 2 \times 512^3$, $L_b=25$ Mpc/$h$
simulations are quite similar. This gives us confidence that even the $N_p = 2 \times 512^3$, $L_b = 25$ Mpc/$h$ simulation is adequately converged at these redshifts for measurements of the wavelet PDF. 
At $z=4$, however, there are noticeable 
differences, suggesting
that higher spatial resolution is required. Note that the convergence with resolution is better for the hotter model. Since the
data appear to favor this model over the cooler model, its convergence properties may be more relevant. It is clear, however, that
resolution requirements are rather stringent at high redshift and so we use the $L_b = 25$ Mpc/$h$, $N_p = 2 \times 1024^3$ simulation
as our main simulation run throughout.
Note also that any bias from limited simulation resolution causes us to systematically underestimate the temperature of 
the IGM, and strengthens the argument for a hot IGM.

\tabletypesize{\normalsize}

\begin{deluxetable*}{c  c  c  c}
\tablecaption{The probability distribution of $A_L$ at $\bar{z}=4.2$.\label{table:pdfz4.2}}
\tablehead{\colhead{Bin No.} & \colhead{$A_L$} & \colhead{$dP/dlnA_L$} & \colhead{$\sigma_P$}}
\startdata
 1 &    0.121E+00 &     0.102E-01 &     0.958E-02\\
 2 &    0.164E+00 &     0.511E-01 &     0.421E-01\\
 3 &    0.220E+00 &     0.117E+00 &     0.913E-01\\
 4 &    0.285E+00 &     0.292E+00 &     0.145E+00\\
 5 &    0.396E+00 &     0.395E+00 &     0.598E-01\\
 6 &    0.516E+00 &     0.736E+00 &     0.124E+00\\
 7 &    0.726E+00 &     0.850E+00 &     0.195E+00\\
 8 &    0.943E+00 &     0.652E+00 &     0.135E+00\\
 9 &    0.124E+01 &     0.131E+00 &     0.631E-01\\
10 &    0.167E+01 &     0.362E-01 &     0.281E-01\\
\enddata
\tablecomments{Here the Morlet filter scale is $s_n=69.7$ km/s.
The first column is the bin number, the second column is the average
wavelet amplitude in the bin, the third column is the differential PDF (per ln $A_L$) in the bin, and
the fourth column is the $1-\sigma$ error on the differential PDF. The measurements have not been corrected for metal line contamination.}
\end{deluxetable*}

\begin{deluxetable*}{cccc}
\tablecaption{The probability distribution of $A_L$ at $\bar{z}=3.4$.
\label{table:pdfz3.4}}
\tablehead{\colhead{Bin No.} & \colhead{$A_L$} & \colhead{$dP/dlnA_L$} & \colhead{$\sigma_P$}}
\startdata
 1 &    0.155E-01 &     0.108E-01 &     0.800E-02\\
 2 &    0.207E-01 &     0.226E-01 &     0.113E-01\\
 3 &    0.334E-01 &     0.420E-01 &     0.223E-01\\
 4 &    0.493E-01 &     0.115E+00 &     0.504E-01\\
 5 &    0.710E-01 &     0.321E+00 &     0.452E-01\\
 6 &    0.108E+00 &     0.601E+00 &     0.526E-01\\
 7 &    0.156E+00 &     0.577E+00 &     0.641E-01\\
 8 &    0.231E+00 &     0.478E+00 &     0.601E-01\\
 9 &    0.335E+00 &     0.289E+00 &     0.835E-01\\
10 &    0.466E+00 &     0.404E-01 &     0.220E-01\\
\enddata
\tablecomments{Similar to Table \ref{table:pdfz4.2} except at $\bar{z}=3.4$.}
\end{deluxetable*}

\begin{deluxetable*}{cccc}
\tablecaption{The probability distribution of $A_L$ at $\bar{z}=3.0$.
\label{table:pdfz3.0}}
\tablehead{\colhead{Bin No.} & \colhead{$A_L$} & \colhead{$dP/dlnA_L$} & \colhead{$\sigma_P$}}
\startdata
 1 & 0.498E-02 &     0.591E-02 &     0.617E-02\\
 2 & 0.825E-02 &     0.351E-01 &     0.339E-01\\
 3 & 0.129E-01 &     0.271E-01 &     0.146E-01\\
 4 & 0.190E-01 &     0.897E-01 &     0.474E-01\\
 5 & 0.322E-01 &     0.196E+00 &     0.495E-01\\
 6 & 0.523E-01 &     0.394E+00 &     0.708E-01\\
 7 & 0.800E-01 &     0.618E+00 &     0.858E-01\\
 8 & 0.129E+00 &     0.509E+00 &     0.970E-01\\
 9 & 0.202E+00 &     0.214E+00 &     0.650E-01\\
10 & 0.310E+00 &     0.159E-01 &     0.166E-01\\ 
\enddata
\tablecomments{Similar to Table \ref{table:pdfz4.2} except corrected for metal line contamination, and at $\bar{z}=3.0$.}
\end{deluxetable*}

\begin{deluxetable*}{cccc}
\tablecaption{The probability distribution of $A_L$ at $\bar{z}=2.6$.
\label{table:pdfz2.6}}
\tablehead{\colhead{Bin No.} & \colhead{$A_L$} & \colhead{$dP/dlnA_L$} & \colhead{$\sigma_P$}}
\startdata
 1 &    0.135E-02 &     0.315E-02 &     0.327E-02\\
 2 &    0.221E-02 &     0.159E-01 &     0.165E-01\\
 3 &    0.445E-02 &     0.188E-01 &     0.151E-01\\
 4 &    0.786E-02 &     0.523E-01 &     0.195E-01\\
 5 &    0.149E-01 &     0.887E-01 &     0.418E-01\\
 6 &    0.242E-01 &     0.126E+00 &     0.494E-01\\
 7 &    0.479E-01 &     0.367E+00 &     0.579E-01\\
 8 &    0.829E-01 &     0.595E+00 &     0.832E-01\\
 9 &    0.142E+00 &     0.328E+00 &     0.578E-01\\
10 &    0.237E+00 &     0.764E-01 &     0.427E-01\\
\enddata
\tablecomments{Similar to Table \ref{table:pdfz3.0} except at $\bar{z}=2.6$.}
\end{deluxetable*}

\begin{deluxetable*}{cccc}
\tablecaption{The probability distribution of $A_L$ at $\bar{z}=2.2$.
\label{table:pdfz2.2}}
\tablehead{\colhead{Bin No.} & \colhead{$A_L$} & \colhead{$dP/dlnA_L$} & \colhead{$\sigma_P$}}
\startdata
 1 &    0.846E-03 &     0.154E-01 &     0.160E-01\\
 2 &    0.129E-02 &     0.331E-01 &     0.246E-01\\
 3 &    0.230E-02 &     0.605E-01 &     0.304E-01\\
 4 &    0.464E-02 &     0.129E+00 &     0.620E-01\\
 5 &    0.834E-02 &     0.191E+00 &     0.706E-01\\
 6 &    0.154E-01 &     0.149E+00 &     0.511E-01\\
 7 &    0.285E-01 &     0.248E+00 &     0.567E-01\\
 8 &    0.507E-01 &     0.554E+00 &     0.749E-01\\
 9 &    0.820E-01 &     0.216E+00 &     0.758E-01\\
10 &    0.150E+00 &     0.531E-01 &     0.399E-01\\
\enddata
\tablecomments{Similar to Table \ref{table:pdfz3.0} except at $\bar{z}=2.2$.}
\end{deluxetable*}


\end{document}